\documentclass[aps,amsmath,amssymb,prd,notitlepage,superscriptaddress,twocolumn,eqsecnum,nofootinbib]{revtex4-1}

\usepackage{esdiff}
\usepackage{graphicx}
\usepackage{hyperref}
\usepackage[usenames,dvipsnames,svgnames,table]{xcolor}
\usepackage{multirow}
\usepackage{float}

\newcommand{\go}{\mathring{g}}
\newcommand{\sigmab}{\bar{\sigma}}

\DeclareMathOperator{\sgn}{sgn}

\def\simless{\mathbin{\lower 3pt\hbox
   {$\rlap{\raise 5pt\hbox{$\char'074$}}\mathchar"7218$}}} %< or of order
\def\simgreat{\mathbin{\lower 3pt\hbox
   {$\rlap{\raise 5pt\hbox{$\char'076$}}\mathchar"7218$}}} %> or of order

\newcommand{\lta}[0]{\simless}

\begin{document}

\title{Gravitational backreaction on a cosmic string: Formalism}
\author{David F.~Chernoff}
\author{\'Eanna \'E.~Flanagan}
\affiliation{Department of Astronomy, Cornell University, Ithaca, NY 14853, USA}
\author{Barry Wardell}
\affiliation{Department of Astronomy, Cornell University, Ithaca, NY 14853, USA}
\affiliation{School of Mathematics and Statistics, University College Dublin, Belfield, Dublin 4, Ireland}

\begin{abstract}
We develop a method for computing the linearized gravitational backreaction for Nambu-Goto strings
using a fully covariant formalism. We work with equations of motion expressed in terms of a higher
dimensional analog of the geodesic equation subject to self-generated forcing terms. The approach
allows arbitrary spacetime and worldsheet gauge choices for the background and perturbation.

The perturbed spacetime metric may be expressed as an integral over a distributional stress-energy
tensor supported on the string worldsheet. By formally integrating out the distribution, this quantity
may be re-expressed in terms of an integral over the retarded image of the string. In
doing so, one must pay particular attention to contributions that arise from the field point and
from non-smooth regions of the string. Then, the gradient of the perturbed metric decomposes
into a sum of boundary and bulk terms. The decomposition depends upon the worldsheet coordinates
used to describe the string, but the total is independent of those considerations.

We illustrate the method with numerical calculations of the self-force at every point on the
worldsheet for loops with kinks, cusps and self-intersections using a variety of different
coordinate choices. For field points on smooth parts of the worldsheet the self-force is finite. As
the field point approaches a kink or cusp the self-force diverges, but is integrable in the sense
that the displacement of the worldsheet remains finite. As a consistency check, we verify that the
period-averaged flux of energy-momentum at infinity matches the direct work the self-force performs
on the string.

The methodology can be applied to address many fundamental questions for string loop evolution.
\end{abstract}

\date{\today}

\maketitle

%%%%%%%%%%%%%%%%%%%%%%%
\section{Introduction}%
%%%%%%%%%%%%%%%%%%%%%%%

\subsection{Cosmological superstrings}

Cosmic superstrings \cite{HenryTye:2006uv} are the strings of string
theory stretched to macroscopic length scales by the universe's early
phase of exponential, inflationary growth
\cite{Guth:1980zm,Linde:1981mu,Albrecht:1982wi}.  During subsequent
epochs when the scale factor grows as a more leisurely power law of
time a complicated network of various string elements forms
\cite{Albrecht:1984xv,Bennett:1987vf,Allen:1990tv}. Long,
horizon-crossing strings stretch, short curved pieces accelerate and
attempt to straighten, and, occasionally, individual segments
intercommute (collide, break and reattach) chopping out loops and
forming new, connected string pathways.  Analytic and numerical calculations
demonstrate that these processes rapidly drive the network to a
self-similar evolution with statistical properties largely determined
by the string tension
\cite{Vanchurin:2005yb,Ringeval:2005kr,Martins:2005es,BlancoPillado:2011dq,Blanco-Pillado:2013qja}. The
energy densities in long strings, in loops, and in gravitational
radiation divided by the critical energy density are all independent of time.
The distribution of loops of a given size relative to the
horizon scale is also fixed.

An understanding of this evolution is informed by previous
studies of one dimensional defects
in the context of symmetry breaking in
grand unified theories (GUTs; \cite{Kibble:1976sj};
for a general review see \cite{Vilenkin:2000jqa}).
One important difference for superstrings is the expected
value of the string tension.  In GUT theories the string tension
$G \mu/c^2 \sim \Lambda_{GUT}^2/M_p^2 \sim 10^{-6}$ is fixed by the
GUT energy scale $\Lambda_{GUT}$. Observations of the microwave sky have ruled out GUT strings as
the source of the cosmological perturbations \cite{Smoot:1992td,Bennett:1996ce,Spergel:2006hy} and led to upper bounds
on the tension. Currently, broadly model-independent limits from
lensing  \cite{Vilenkin:1981zs,Hogan:1984unknown,Vilenkin:1984ea,deLaix:1997dj,Bernardeau:2000xu,Sazhin:2003cp,Sazhin:2006fe,Christiansen:2008vi},
CMB studies \cite{Smoot:1992td,Bennett:1996ce,Pogosian:2003mz,Pogosian:2004ny,Tye:2005fn,Wyman:2005tu,Pogosian:2006hg,Seljak:2006bg,Spergel:2006hy,Bevis:2007qz,Fraisse:2006xc,Pogosian:2008am,Ade:2013xla}
and gravitational wave background and bursts \cite{Vachaspati:1984gt,Economou:1991bc,Battye:1997ji,Damour:2000wa,Damour:2001bk,Damour:2004kw,Siemens:2006vk,Hogan:2006we,Siemens:2006yp,Abbott:2006vg,Abbott:2009rr,Abbott:2009ws,Aasi:2013vna,TheLIGOScientific:2016dpb}
give $G \mu/c^2 \lta 10^{-7}$. More stringent but somewhat more model-dependent limits from pulsar timing
\cite{Bouchet:1989ck,Caldwell:1991jj,Kaspi:1994hp,Jenet:2006sv,DePies:2007bm}
have regularly appeared. Currently, the
strongest inferred limit
is $G \mu/c^2 \lta 10^{-11}$ \cite{Blanco-Pillado:2017oxo,Blanco-Pillado:2017rnf}.

Low tension strings are natural in string theory and have little difficulty in this regard. In the most
well-studied compactifications the standard model physics is located
at the bottom of a warped throat where all energy scales are
exponentially diminished compared to the string scale. Superstrings
have tensions that are reduced by exactly this effect and can
correspond to energies as small as TeV (see \cite{HenryTye:2006uv,Chernoff:2014cba} for reviews).

The magnitude of $G\mu/c^2$
influences many properties of the strings and loops that make up the
network. A loop with characteristic size $\ell$ and energy
$\propto \mu \ell$ will completely dissipate by gravitational wave
emission in times $t \sim \ell/(\Gamma G \mu/c)$ where $\Gamma \sim
50$ is a loop-dependent pure number
\cite{Vachaspati:1984gt,Burden:1985md,Garfinkle:1987yw,Durrer:1989zi,Allen:1994bs,Allen:1994iq,Allen:1994ev,Casper:1995ub}.
If $G\mu/c^2 \ll 10^{-6}$ then
superstring loops evaporate by gravitational wave emission much less
rapidly than GUT string loops. The characteristic size of loops that
evaporate gravitationally in $t_H$, the age of the universe, is $\ell_g =
t_H \Gamma G \mu/c $. These turn out to dominate the distribution
of loop sizes found in the universe today.

Current simulations report that about 10-20\% of
the string network that is chopped out ends up in the form of large
loops, with sizes within a few orders of magnitude of the horizon
scale at birth \cite{BlancoPillado:2011dq,Blanco-Pillado:2013qja}.
The rest forms very small loops with size scale
relative to the horizon set by a power of $G \mu/c^2$
\cite{Polchinski:2004hb,Polchinski:2006ee,Polchinski:2007qc,Polchinski:2007rg,Dubath:2007mf,Polchinski:2008hu}. These
rapidly evaporate.
Today, the string network's energy density is dominated by the large loops formed
at an early epoch. If
$G \mu/c^2 < 7 \times 10^{-9}$
it is before matter-radiation
equality. Today's size distribution increases
as $\ell \to \ell_g$ from above (the universe was denser at earlier
times and formed more smaller loops); the distribution is cutoff by the
evaporation process at $\ell < \ell_g$.

Long gravitational lifetimes have another important effect:
the center of mass
velocity of the old loops is small and they cluster like cold dark
matter \cite{Chernoff:2009tp}. This opens the way to experimental
tests of string
theory that are based upon direct detection of gravitational wave
emission and observation of string microlensing of background stellar sources
\cite{Chernoff:2017xxx}

\subsection{Gravitational backreaction in the string network}

The most numerous loops are close
to the characteristic size $\ell_g$, set by
gravitational backreaction.
An understanding of string gravitational backreaction is crucial for
making forecasts of experimental studies and
planning future observational campaigns. The emission of gravitational
radiation and the associated dissipative forces shrink the size of the
loop (energy loss) and impart a recoil (momentum and angular momentum
loss). These may change the character of the loop oscillation over
long timescales.  The radiative emission processes have been
well-studied assuming that the loop is a long-lived periodic
oscillator \cite{HoganRees:1984unknown,Vachaspati:1984gt,
  Burden:1985md,Hogan:1987unknown,Garfinkle:1987yw,Durrer:1989zi,Allen:1994bs,Allen:1994iq,Allen:1994ev,Casper:1995ub}.
The secular effects of gravitational backreaction on the loop
oscillation are relatively unexplored. Two important aspects are the
propensity of loops to self-intersection and the evolution of
discontinuous features on the loops.

Self-intersections are important because they can lead to the rapid
demise of the long-lived loops which are of greatest observational
interest. The reason is simple: isolated, dissipationless loops are
exactly periodic.  If a loop can self-intersect it will do so over and
over again eventually leading to intercommutation and breakage. This
process shatters the loop into many small looplets \cite{Casper:1995ub}
moving apart at
relativistic speeds, each of which will evaporate in only a
fraction of the time required by the original loop.
Self-intersections have the potential to radically depress the number
of old loops of size $\ell_g$ that would otherwise exist throughout the
universe.  The loop distribution will be cutoff at scale $> \ell_g$; the number density at that cutoff will be substantially smaller.
Furthermore, the intercommutation
process evicts the shattered progeny from
being bound to the galaxy. Backreaction
can significantly alter experimental
forecasts.

Another important aspect of gravitational backreaction is the presence
of kinks and cusps on loops.  Typically when a new loop is formed from
a smooth segment of string the orbit of the new loop will contain an
infinitesimal element of string that moves at the speed of light for
an infinitesimal time, repeating once per period. This is a cusp, a
well-characterized, periodic strong source of gravitational wave
emission.  Cusp emission is the principle target of gravitational wave
searches from string loops because it is strong, beamed and has a
well-understood signal form
\cite{Vachaspati:1984gt,Economou:1991bc,Battye:1997ji,Damour:2000wa,Damour:2001bk,Damour:2004kw,Siemens:2006vk,Hogan:2006we,Siemens:2006yp,Abbott:2006vg,Abbott:2009rr,Abbott:2009ws,Aasi:2013vna,TheLIGOScientific:2016dpb}.
Ref.~\cite{Polchinski:2008hu} has argued that a scaling
network may be inefficient at forming loops with cusps for the
following reason.  Scaling requires chopping out a significant
fraction of the long strings' length each time the universe doubles in
size. The chopping removes loops and inevitably adds kinks (derivative
discontinuities) to the remaining long string segments. Smooth long
strings accumulate kinks and grow dense with small scale structure as
the universe ages.  New loops inherit the small scale structure. The
first time that the loop begins to form a
cusp-like structure the kinky string reconnects, effectively excising
the part of the loop responsible for the cusp.  Such a loop is left with
nothing but kinks. Kinks may also be detected by gravitational wave
searches but are not as strong or as unidirectional.  Recent
cosmological network simulations support this
theoretical prediction
\cite{BlancoPillado:2011dq,Blanco-Pillado:2013qja}. In particular,
they show that loops with kinks are
formed preferentially and there are few cusps\footnote{It must be noted
that it isn't clear whether the string substructure in even the biggest simulations has entered a
scaling regime or is still in the process of evolving.}.

This general evolutionary outline
prompts a number of questions related to how gravitational
backreaction influences the evolution of derivative discontinuities on
loops and long strings. Qualitatively, we understand that
gravitational backreaction will smooth kink discontinuities (lessening
the size of the jump in the tangent vector from one side to the other) and
theoretically allow new cusps to form. There is a competition between
the rate at which the discontinuity diminishes and the rate at which the loop
shrinks.
One question is whether the loop fully evaporates before the
cusp reforms. Another question
is whether
a reformed cusp has the same scale
as the loop itself or an intrinsically
smaller scale.
These can be answered by calculating the dynamical evolution of
a string loop with backreaction for many orbits.

Another aspect that requires a full treatment of backreaction is how a
loop with many kinks evolves (since the scaling solution suggests the
ubiquity of kinks). If the total rate of gravitational
wave emission scales linearly with the number of kinks \cite{Bohe:2011rk}
then the loop's lifetime is shortened.  However, the backreaction of many
closely spaced radiating kinks may qualitatively effect the evolution
predicted on the basis of a single kink. It is therefore of interest to
understand how backreaction operates when there is a high density of
kinks on long strings and loops.

\subsection{Theory and simulation}

In this paper we develop a complete formalism for computing the gravitational backreaction on
cosmic string loops, and demonstrate the method by computing the gravitational self-force for
several specific cosmic string configurations. Some similar studies were previously done in
Refs.~\cite{Quashnock:1990wv,Scherrer:1990pj}, but these were limited in scope and did not include
many of the details considered here.

Quashnock and Spergel (QS) \cite{Quashnock:1990wv} derived linearized equations of motion for a
string interacting with its own gravitational field (in this context, linearized means first order
in $G \mu/c^2$ expanded about flat spacetime). They worked with particular coordinates and gauge
choices that were chosen to simplify many aspects of the calculation. The weak field approximation
breaks down at kinks, cusps and self-intersections, but these freely moving line singularities were
treated in a perturbative sense.

QS computed the self-force at a field point as sourced by elements
of the retarded, distant string image. They concluded that only finite divergence-free backreaction
forces existed for field points with smooth sources, and that the contribution to the backreaction
forces tended to zero as the source point approached the field point. This situation stands in
contrast to the analogous point particle case studied by Dirac \cite{Dirac:1938nz}, in which
self-interaction leads to a renormalized mass. Carter and Battye \cite{Carter:1998ix} and Buonanno
and Damour \cite{Buonanno:1998is} showed that while a general string has a local divergent part to
its perturbed metric, the Nambu-Goto string is special and the total force density due to all the
local divergent pieces exactly vanishes. The remaining force is given by long-range interactions.

Kinks and cusps are examples where smoothness in the vicinity of
the field point fails to hold. QS did not explicitly
discuss the limiting behavior near a kink but did argue on general
grounds that the backreaction force per source coordinate interval at
a cusp would be infinite, but integrable. They also solved numerically for
the evolution of the loop represented both as a continuous function
and as a set of kinks (straight line segments with small tangent
vector discontinuities) by integrating the backreaction over a full
period.  The simulations showed that cusps survive backreaction but are
deformed and delayed. Longer integrations suggested that the amplitude
of the cusp and the associated asymmetric rocket effects were
suppressed by backreaction. Finally, QS also showed that small (compared to
the size of the loop) kinks decay more rapidly than the string as a
whole. The magnitude of the discontinuity at a kink (change in tangent
vectors) lessens but the discontinuity itself is not
smoothed out by dissipation.

It is some measure of the complexity of the problem that most work since the QS investigation has
dealt with specific issues and not attempted such an ambitious numerical treatment. Anderson
\cite{Anderson:2005qu} analytically calculated the gravitational backreaction forces for the
Allen-Casper-Ottewill (ACO) loop \cite{Allen:1994bs}, a rotating loop configuration with a pair of
kinks (one tangent vector is continuous and the other is discontinuous). The coordinates and gauge
conditions used were equivalent to those of \cite{Quashnock:1990wv}. Anderson demonstrated
explicitly that all the components of 4-vector acceleration diverged near the kink. The calculated
forces were, however, integrable so that the equations of motion in the weak field limit were
integrable too\footnote{\cite{Anderson:2005qu} did not evaluate forces at the kink itself where the
metric is ill-determined.}.

In this paper we do not evolve the string configuration (that
will be for a followup) but study in detail the method of
calculation of the first order self-force.  Certain intermediate quantities in our calculations exhibit divergences.
The occurence of these calculational divergences is tied to three interrelated factors:
the choice of worldsheet gauge (eg conformal or other), the specification of residual gauge freedom
in the choice of worldsheet coordinates (eg null or non-null coordinates), and the existence of discontinuous
sources anywhere on the loop's retarded image (the intersection of the
worldsheet with the past lightcone of the field point).  However, the
total integrated self-force at any point
on a smooth region of the worldsheet is always finite due to cancelations of divergences, and is independent of these choices.
This finiteness is consistent with the lack of renormalization of the string tension discussed in \cite{Buonanno:Damour:1998}
and with the the general conclusions of smoothness of \cite{Quashnock:1990wv}.

While the self-force is finite in smooth regions of the worldsheet,
it diverges in the limit when the field point approaches cusps or kinks on the
worldsheet.  However, when one solves
the linearized equation of motion for the perturbation to the
worldsheet, the linearized displacement of the worldsheet is finite.
Going beyond this treatment will involve critically examining
the linearized approximation and the
distributional representation of features such as
kinks and cusps. The question of whether
physical divergences occur in a fully self-consistent evolution
is beyond the scope of this paper.  Nevertheless, the methodology we
develop in this paper should allow adressing certain aspects of
the question in the
future.  Our methodology will allow us to refine the gauge during the
course of a self-consistent evolution (continuing to use
linearization with distributional models)
to separate invariant physical divergences from
calculational divergences. In the case of the cusp, for example, we
would need to step carefully through a single period of oscillation
to handle the occurrence of the divergence at a single spacetime
point.

As a result of the work in this paper
there is evidence that any such singularity is weak in the
``physical'' sense. In particular, period-averaged changes are given
by simple quadratures over the worldsheet. Orbit-averaging does not require
instant-by-instant evolution but presumes the metric and string
are only mildly perturbed in some average sense.
We find that over an oscillation both
the kink and cusp lead to finite displacements of the
worldsheet and finite small changes in energy, momentum and angular
momentum.  All period-averaged physical divergences are small and bounded in the
sense of being proportional to $G \mu$. This is quite mild compared
to the character of the singular behavior of point masses in general relativity,
for example.

Recently, Wachter and Olum \cite{Wachter:2016hgi,Wachter:2016rwc}
have studied the evolution of
loops composed of linear pieces (both right and left moving modes are
given by a set of fixed tangent vectors which generate
kinks). Using the methodology of
\cite{Quashnock:1990wv} they found
the metric perturbations and the loop's acceleration and analytically
evaluated the backreaction
for a planar rectangular loop \cite{Garfinkle:1987yw}. They deduced the energy loss, changes to the left and
right moving modes and the kink smoothing (diminishing the tangent
vector jumps). Small angle kinks (acute angles) disappeared more
quickly than large angle kinks (of order $\pi/4$).  This observation
is complementary to that of
\cite{Quashnock:1990wv} which
reported small
sized kinks (length small compared to the loop size) disappeared more
quickly than large sized kinks (length a fraction of the full loop
size). Refs.~\cite{Wachter:2016hgi,Wachter:2016rwc} compared the loop evaporation time to the kink
smoothing time and found that the loop angle was a key
parameter. For small angles, kinks disappeared rapidly.  For large
angles, the loop evaporated first. Finally, the analysis of the
piecewise loops showed that the straight line segments begin to curve
after a short period for all except loops with special symmetry.

\subsection{Lagrangian Methodology}

Carter pioneered the treatment of perturbations in an arbitrarily curved
spacetime background with relativistic string, membrane or other brane
models where $p$, the spatial dimension of the brane, is less than $n$,
the spatial dimension of spacetime
(\cite{Battye:1995hv,Carter:1993wy,Battye:1998zk};
see \cite{Carter:1997pb} for a review). The action in such models is
\begin{eqnarray}
  {\cal I} & = & \int {\cal L} d{\bar \Sigma} \\
  d{\bar \Sigma} & = & | \gamma |^{1/2} d^{p+1} \zeta
\end{eqnarray}
where $d{\bar \Sigma}$ is the surface measure element induced on the
timelike world sheet by the background metric, $\gamma$ is the determinant
of the induced metric and $\zeta$ stands for the $(p+1)$ internal
coordinates. We may assume a constant scalar Lagrangian ${\cal L} = -m^{p+1}$
where $m$ is a characteristic mass scale and $\hbar=c=1$. For $p<n$
the brane and the Lagrangian are distributional in spacetime. The
brane is concentrated on lower dimensional world sheets
in the higher dimensional spacetime. The case $p=1$ and $n=3$ is
an effective low-energy description of minimally coupled F- and D-strings
with 2 dimensional world sheets.

In this work we apply the formalism to compute the metric perturbation generated
by a cosmic string. Two important considerations guide our efforts.  First, the
distributional nature of the strings motivates a Lagrangian approach.
Second, we work as much as possible in terms of
tensorial quantities of the background spacetime and avoiding the use
of specific systems of intrinsic coordinates for the brane
submanifolds. We develop a fully covariant
formalism and apply it in a variety
of circumstances.

\subsection{Conventions used in this paper}

Throughout this paper we follow the conventions of Ref.~\cite{MTW}; we use a ``mostly positive''
metric signature, $(-,+,+,+)$ for the spacetime metric and $(-,+)$ for the worldsheet metric, the
connection coefficients are defined by
$\Gamma^{\lambda}_{\mu\nu}=\frac{1}{2}g^{\lambda\sigma}(g_{\sigma\mu,\nu}
+g_{\sigma\nu,\mu}-g_{\mu\nu,\sigma}$), the Riemann tensor is
$R^{\alpha}{}_{\!\lambda\mu\nu}=\Gamma^{\alpha}_{\lambda\nu,\mu}
-\Gamma^{\alpha}_{\lambda\mu,\nu}+\Gamma^{\alpha}_{\sigma\mu}\Gamma^{\sigma}_{\lambda\nu}
-\Gamma^{\alpha}_{\sigma\nu}\Gamma^{\sigma}_{\lambda\mu}$, the Ricci tensor and scalar are
$R_{\alpha\beta}=R^{\mu}{}_{\!\alpha\mu\beta}$ and $R=R_{\alpha}{}^{\!\alpha}$, and the Einstein
equations are $G_{\alpha\beta}=R_{\alpha\beta}-\frac{1}{2}g_{\alpha\beta}R=8\pi T_{\alpha\beta}$.
We use standard geometrized units, with $c=G=1$, Latin indices for worldsheet components and Greek indices for four-dimensional spacetime components.

%%%%%%%%%%%%%%%%%%%%%%%%%%%%%%%%%%%%%%%%%%%%%%%%%%%%%%%%%%%%%%%%%%%%%%%%%%%%
\section{Covariant equations of motion for a Nambu-Goto cosmic string loop}%
%%%%%%%%%%%%%%%%%%%%%%%%%%%%%%%%%%%%%%%%%%%%%%%%%%%%%%%%%%%%%%%%%%%%%%%%%%%%
\label{sec:eom}

We begin by considering a Nambu-Goto cosmic string tracing out a two-dimensional worldsheet in
spacetime. We identify a point on the string by a pair of worldsheet coordinates $\{\zeta^1,
\zeta^2\}$ and denote the spacetime coordinate of that point by $z^\alpha(\zeta^a)$.

Given the full spacetime metric, $g_{\alpha \beta}$, the induced metric on the worldsheet is defined
by
\begin{equation}
  \gamma_{ab} = g_{\alpha \beta} \partial_a z^\alpha \partial_b z^\beta.
\end{equation}
The worldsheet-tangent projection tensor is defined as
\begin{equation}
\label{eq:projection}
  P^{\alpha \beta} = \gamma^{ab} \partial_a z^\alpha \partial_b z^\beta
\end{equation}
where $\gamma^{ab} \gamma_{bc} = \delta^a_c$. The corresponding worldsheet-orthogonal projection
tensor is
\begin{equation}
  \perp^{\alpha \beta} = g^{\alpha \beta} - P^{\alpha \beta}.
\end{equation}
For tensor fields with support confined to the world sheet,
the tangentially projected covariant derivative is
\begin{equation}
{\bar \nabla}_\alpha = P_{\alpha}{}^{\mu} \nabla_\mu.
\end{equation}
Finally, defining the extrinsic curvature (or second fundamental tensor) and its trace,
\begin{equation}
\label{eq:Kdef}
 K_{\alpha \beta}{}^{\gamma} \equiv P_{\mu \beta} \bar{\nabla}_\alpha P^{\mu \gamma},
 \quad
 K^{\gamma} \equiv g^{\alpha \beta} K_{\alpha \beta}{}^{\gamma},
\end{equation}
Battye and Carter \cite{Battye:Carter:1995} showed that the equation of motion of the string
may be written in the compact form
\begin{equation}
\label{eq:zeroorder}
  K^\rho = 0.
\end{equation}
This can be expanded explicitly as
\begin{equation}
\label{eq:coordK}
  K^\gamma = \frac{1}{\sqrt{-\gamma}} \partial_a(\sqrt{-\gamma}\gamma^{ab}\partial_b z^\gamma) + P^{\alpha \beta} \Gamma^\gamma_{\alpha \beta},
\end{equation}
where $\gamma = \det(\gamma_{ab})$
and $\Gamma^\gamma_{\alpha\beta}$ is
evaluated at the spacetime coordinate
of the worldsheet point
$z^\mu(\zeta^a)$.

%%%%%%%%%%%%%%%%%%%%%%%%%%%%%%%%%%%%%%%%%%%%%%%%%%%%%%%%%%%%%
\section{Gravitational perturbations of cosmic string loops}%
%%%%%%%%%%%%%%%%%%%%%%%%%%%%%%%%%%%%%%%%%%%%%%%%%%%%%%%%%%%%%
\label{ser:pert}

We now wish to specialize to the case where the string tension is small and the problem may be
treated perturbatively in $G\mu/c^2$. Then, the string can be considered to be moving in a
perturbed spacetime with metric
$$
g_{\alpha\beta} = \go_{\alpha\beta} + h_{\alpha\beta},
$$
where the background, unperturbed metric is $\go_{\alpha\beta}$ and
the perturbation $h_{\alpha \beta}$ is sourced by the string's own stress energy. We can
likewise parameterise the worldsheet in terms of a background piece plus a perturbation,
\begin{equation}
z^\alpha = z_{(0)}^\alpha + z_{(1)}^\alpha,
\end{equation}
and will work to first order\footnote{For notational simplicity, from here on we will always make explicit the dependence on the perturbed quantities $h_{\alpha\beta}$ and $z_{(1)}^\alpha$, but will implicitly define everything else in terms of background quantities. So, for example, we will have $\gamma_{ab} = \go_{\alpha \beta} \partial_a z_{(0)}^\alpha \partial_b z_{(0)}^\beta$ and likewise for all of the other quantities defined in Sec.~\ref{sec:eom}.} in both the metric perturbation, $h_{\alpha\beta}$, and in the worldsheet
perturbation, $z_{(1)}^\alpha$.

\subsection{Zeroth order equation of motion}
For the case where backreaction is ignored, we may treat the string as moving in a fixed background
spacetime with equation of motion
\begin{equation}
  \frac{1}{\sqrt{-\gamma}} \partial_a(\sqrt{-\gamma}\gamma^{ab}\partial_b
  z_{(0)}^\gamma) + P^{\alpha \beta} \Gamma^\gamma_{\alpha \beta} = 0.
\label{eq:general-eom}
\end{equation}
If the background is Minkowski spacetime, the second term vanishes and the equation of motion is
just the standard two-dimensional scalar wave equation for each component of the worldsheet
coordinate vector.

Further simplification can be obtained by considering the gauge freedom in defining the worldsheet
coordinates. A common class of choices invokes the conformal gauge condition, whereby the
worldsheet metric is required to be conformally flat:
\begin{equation}
\label{eq:conformal}
  \gamma_{ab} = \phi \eta_{ab},
\end{equation}
where $\eta_{ab}$ is a two-dimensional Minkowski metric and $\phi > 0$ is a conformal factor. A
consequence of this choice is that the worldsheet derivatives, $\partial_{\zeta^1} z^\mu$ and
$\partial_{\zeta^2} z^\mu$, must satisfy certain orthogonality conditions (the details of which
depend on the particular choice of worldsheet coordinates) and that the equation of motion is given
by
\begin{equation}
\label{eq:conformal-eom}
  \phi^{-1} \eta^{ab}\partial_a \partial_b z_{(0)}^\gamma = 0.
\end{equation}
This is just the $1$+$1$D flat space scalar wave equation for each spacetime component of the
string worldsheet vector $z_{(0)}^\alpha$. The solutions to this equation are periodic in both $\zeta^1$ and $\zeta^2$ in
the sense that for a loop of length $L$ we have $z^\alpha(\zeta^1, \zeta^2) = z^\alpha(\zeta^1+L/2,
\zeta^2+L/2)$.

Weak solutions of equations \eqref{eq:general-eom} and \eqref{eq:conformal-eom}
allow derivative discontinuities, so generic solutions are not smooth.
The tangent sphere representation provides a description
of the derivatives of the two components of a solution \cite{Vilenkin:2000jqa}. Perfectly smooth
string loop solutions have two continuous paths on the tangent sphere. However,
there may be long-lived kinks (corresponding to gaps in the
tangent sphere) that propagate around the string along null worldsheet directions, and cusps
(corresponding to intersections in the tangent sphere) that only exist instantaneously. There may
also be self-intersections, where the string crosses over on itself.

%%%%%%%%%%%%%%%%%%%%%%%%%%%%%%%%%%%%%%%%%%%%%%%%%%%%%%%%%%%%%%%%%%%%%
\subsection{First order equation of motion for the string worldsheet}%
%%%%%%%%%%%%%%%%%%%%%%%%%%%%%%%%%%%%%%%%%%%%%%%%%%%%%%%%%%%%%%%%%%%%%%
We now return to the general case (no specialization of gauge or metric) to write down the
perturbed equation of motion. Demanding that the perturbed trace of the extrinsic curvature vanish as
in Eq.~\eqref{eq:zeroorder}, and assuming that the zeroth order equation of motion is satisfied
gives \cite{Battye:Carter:1995,Battye:1998zk}
\begin{align}
&&\perp^\rho_{\ \,\chi} {\bar \nabla}_\mu {\bar \nabla}^\mu z_{(1)}^\chi - 2
{\bar \nabla}_\mu z_{(1)}^\alpha K^{\mu\ \, \rho}_{\ \,\alpha} +
\perp^{\beta\rho} P^{\mu\nu} R_{\mu\varepsilon\nu\beta}
z_{(1)}^\varepsilon \nonumber \\
&&= K^{\alpha\beta\rho} h_{\alpha\beta} -  \perp^\rho_{\
  \beta} P^{\lambda\tau} \left( \nabla_\lambda h^\beta_{\ \,\tau} - {1
    \over 2} \nabla^\beta h_{\lambda\tau} \right).
\label{eq:final}
\end{align}
The homogeneous version of this equation is a higher dimensional analog of the geodesic deviation
equation.

Identifying the term on the right hand side as a self-force, it is convenient to split this force into separate contributions, one involving the metric perturbation and the other involving its derivative,
\begin{subequations}
  \label{eq:self-force}
\begin{eqnarray}
  F^\rho &=&\, F_1^\rho + F_2^\rho, \\
  F_1^\rho &\equiv&\, - \perp^\rho{}_\lambda P^{\mu\nu} \Big(\nabla_\mu h_\nu{}^\lambda - \frac12 \nabla^\lambda h_{\mu\nu}\Big), \\
  F_2^\rho &\equiv&\, K^{\mu \nu \rho} h_{\mu \nu}.
\end{eqnarray}
\end{subequations}
Using the definition \eqref{eq:Kdef} of $K^{\mu\nu\rho}$ we can write
$F_2^\rho$ in terms of $H_{ab} \equiv h_{\mu\nu} \partial_a
z^\mu \partial_b z^\nu$ (the projection of $h_{\mu\nu}$ along the
worldsheet),
\begin{align}
        \label{eq:F2-hproj}
  F_2^\rho &= \left(\gamma^{ac} \gamma^{bd} \partial_c z^{\sigma} \partial_d z^\lambda \nabla_\lambda P_\sigma{}^\rho \right) H_{ab}\nonumber \\
         &= \perp^\rho{}_\sigma \gamma^{ac} \gamma^{bd} \left(\partial_c \partial_d z^\sigma+ \Gamma^\sigma_{\lambda\tau} \partial_c z^\lambda \partial_d z^\tau\right) H_{ab}.
\end{align}
We can also write the first term as
\begin{equation}
F_1^\rho = - \frac{1}{\sqrt{\gamma}} \perp^\rho{}_\lambda {\cal
  F}^\lambda_{\rm conf},
\end{equation}
where\footnote{We use a caligraphic font for ${\cal F}^\mu_{\rm conf}$
  since it is not a gauge-specialized version of the general self
  force $F^\mu$, because the left hand side of Eq.~\eqref{eq:simple} is not
  obtained from the left hand side of \eqref{eq:final} by a gauge
  specialization.}
\begin{align}
  \label{eq:trad-self-force}
  {\cal F}_{\rm conf}^\rho \equiv&\, \sqrt{-\gamma} P^{\mu\nu} \Big(\nabla_\mu h_\nu{}^\rho - \frac12 \nabla^\rho h_{\mu\nu}\Big).
\end{align}
is the quantity that appears on
the right hand side of the conformal gauge equation of motion
\eqref{eq:simple} below.
Battye and Carter \cite{Battye:Carter:1995} showed that for a general choice of
gauge it is crucial to both project orthogonal to the worldsheet
and to include the additional term involving $K^{\mu\nu\rho}$ in order to get the correct
gravitational self-force\footnote{In fact, a sequence of papers provided derivations of the
fundamental equations of motion with increasing degrees of rigor. Following on from
Ref.~\cite{Battye:Carter:1995}, Battye and Carter \cite{Battye:1998zk} performed a more
careful analysis using a second order Lagrangian variational treatment to derive the first order
equations of motion for the displacement vector of the world sheet and for the metric
perturbations. When restricted to the linearized backreaction regime, their final results (given in
Eqs.~(30), (31) and (33) of \cite{Battye:1998zk} with terms involving $K_\rho$ identically zero for linearized backreaction) are consistent with their earlier results and with the
expressions above.}. In a subsequent work \cite{Carter:Battye:1998} they showed that, despite the
presence of divergences in the metric perturbation, the
gravitational self-force \eqref{eq:self-force} is finite for strings in four spacetime
dimensions with smooth worldsheets.

The very general form for the equations of motion given by Eq.~\eqref{eq:final} allows arbitrary
choice of gauge for the background, both for the spacetime coordinates and for the worldsheet
coordinates. It also allows separate arbitrary gauge transformations for the perturbations, and is
invariant under two different types of linearized gauge transformations:
\begin{itemize}
\item Linearized coordinate transformations in spacetime, which induce changes in the worldsheet and
      metric perturbations, $z_{(1)}^\alpha \to z_{(1)}^\alpha + \xi^\alpha$,
      $h_{\alpha\beta} \to h_{\alpha\beta} - 2 \nabla_{(\alpha} \xi_{\beta)}.$
\item Linearized coordinate transformations on the worldsheet, which induce the changes
      \begin{equation}
      z_{(1)}^\alpha \to z_{(1)}^\alpha + \partial_a z^\alpha \xi^a.
      \label{eq:gaugec}
      \end{equation}
      This gauge freedom shows that only the component of $z_{(1)}^\alpha$ that is perpendicular to
      the worldsheet contains physical information.
\end{itemize}

%%%%%%%%%%%%%%%%%%%%%%%%%%%%%%
\subsection{Choices of Gauge}%
%%%%%%%%%%%%%%%%%%%%%%%%%%%%%%

%%%%%%%%%%%%%%%%%%%%%%%%%%%%%%%%%%%%%%%%%%%%%
\subsubsection{Gauge choice to zeroth order}%
%%%%%%%%%%%%%%%%%%%%%%%%%%%%%%%%%%%%%%%%%%%%%

We now once again specialize to Minkowski spacetime in Lorentzian coordinates at zeroth order. Then,
the third term on the left hand side of Eq.~\eqref{eq:final} vanishes identically. The first term
simplifies to
\begin{equation}
\perp^\rho{}_{\chi}  {1 \over \sqrt{-\gamma}} \partial_a \left( \sqrt{-\gamma} \gamma^{ab} \partial_b z_{(1)}^\chi \right),
\end{equation}
and the second term is
\begin{equation}
-2 \partial_a z^{(1)}_{\alpha} \gamma^{ab} z^\sigma_{(0),bd} z^\alpha_{(0),c}
\gamma^{cd} \perp_\sigma{}^{\rho}.
\end{equation}
The first two terms simplify further if we use the conformal gauge [Eq.~\eqref{eq:conformal}] to
zeroth order, in which case the left hand side becomes
\begin{equation}
\perp^\rho{}_{\chi}  {1 \over \sqrt{-\gamma}} \eta^{ab} \partial_a \partial_b z_{(1)}^\chi.
\end{equation}

%%%%%%%%%%%%%%%%%%%%%%%%%%%%%%%%%%%%%%%%%%%%
\subsubsection{Gauge choice to first order}%
%%%%%%%%%%%%%%%%%%%%%%%%%%%%%%%%%%%%%%%%%%%%

At first order we adopt Lorenz gauge\footnote{This gauge condition is often referred to as Lorentz gauge but is actually due to Lorenz \cite{5672647}.} for the spacetime coordinates. For the worldsheet coordinates
there are several natural choices. We focus here on the conformal gauge as it is computationally
the most convenient, and direct the reader to Appendix \ref{app:gauge} for a discussion of other
possible choices.

The choice of conformal gauge at first order amounts to choosing the worldsheet coordinates so that
the conformal flatness condition \eqref{eq:conformal} holds to first order as well as zeroth order.
Anderson \cite{Anderson:2005qu} showed that in this gauge the equation of motion,
Eq.~\eqref{eq:final} takes the simple form
\begin{equation}
  \eta^{ab} \partial_a \partial_b z_{(1)}^\chi =
  - {\cal F}^\chi_{\rm conf}
\label{eq:simple}
\end{equation}
When our sign convention for the metric is taken into account, this form is consistent with that
used by Buonanno and Damour \cite{Buonanno:Damour:1998}.

Comparing with the covariant equation, Eq.~\eqref{eq:final}, we see a number of differences due to
the gauge specialization:
\begin{itemize}
\item The right hand side of Eq.~\eqref{eq:simple} corresponds to the second term on the right
      hand side of Eq.~\eqref{eq:final}, but with the projection tensor dropped.
\item The left hand side of Eq.~\eqref{eq:simple} corresponds to the first term on the left hand
      side of Eq.~\eqref{eq:final}, but again with the projection tensor dropped.
\item The remaining two terms in Eq.~\eqref{eq:final} involving couplings to the extrinsic
      curvature tensor have been dropped -- they cancel against the effect of dropping the
      projection tensors in this gauge. (We have already dropped the term involving the Riemann
      tensor since we are working in flat spacetime.)
\end{itemize}

A simple proof of this can be obtained by starting with the general coordinate expression
\eqref{eq:coordK} for $K^\rho$ before considering perturbations, and applying the conformal gauge
condition \eqref{eq:conformal}. We have
\begin{equation}
K^\rho = {1 \over \sqrt{-\gamma}} \eta^{ab} \partial_a  \partial_b z^\rho
+ \gamma^{ab} z^\lambda_{,a} z^\mu_{,b} \Gamma^\rho_{\lambda\mu}.
\label{eq:expr}
\end{equation}
without approximation ($z^\rho$, $\gamma_{ab}$, $g_{\alpha\beta}$, $\Gamma^\rho_{\lambda\mu}$ exact). Now consider evaluating this expression with the metric $
g_{\alpha\beta} \to \eta_{\alpha\beta} + h_{\alpha\beta}$ and
worldsheet $z^\alpha \to z_{(0)}^\alpha + z_{(1)}^\alpha$. The zeroth order term vanishes by assumption. The
variation of the first term in Eq.~\eqref{eq:expr} comes from replacing $z^\rho$ with
$z_{(0)}^\rho + z_{(1)}^\rho$, since the zeroth order quantity $\eta^{ab} \partial_a \partial_b
z_{(0)}^\rho$ vanishes. Therefore this term yields the left hand side of Eq.~\eqref{eq:simple}.
Similarly, the variation of the second term in Eq.~\eqref{eq:expr} comes from the variation
in $\Gamma^\rho_{\lambda\mu}$, since this quantity vanishes in the background by assumption (we are
working in Lorentzian coordinates in Minkowski spacetime). Using expression
\eqref{eq:projection} for the projection tensor we see that the variation of this term yields the
right hand side of Eq.~\eqref{eq:simple}.

For the specific choice of gauge in this section ${\cal F}^\rho_{\rm conf}$
naturally appears in the balance laws for energy and momentum relating
the flux of radiation at infinity to the local dissipation forces (see
Appendix \ref{sec:energymomentumlossdiscussion}).

\subsection{First order metric perturbation}
The stress tensor for a Nambu-Goto cosmic string is given by \cite{Vilenkin:2000jqa}
\begin{equation}
  T^{\alpha \beta} (x)=
  - G \mu \iint P^{\alpha \beta} \delta_4 (x,z) \sqrt{-\gamma} \, d \zeta^{1'} d\zeta^{2'}
  \label{exactstressenergy}
\end{equation}
where $\delta_4 (x,z) = \frac{\delta_4 (x-z)}{\sqrt{-g}}$ is the four-dimensional invariant Dirac
delta distribution and $z$, $P^{\alpha \beta}$ and $\gamma$ are all functions of $\zeta^{a'}$. A
coupling of the string to gravity leads to deviations of the spacetime from the background. For
sufficiently small string tensions, $G \mu/c^2 \ll 1$, this deviation may be treated perturbatively
by expanding the metric about the background spacetime,
\begin{equation}
  g_{\alpha \beta} = \go_{\alpha \beta} + h_{\alpha \beta}.
\end{equation}
The perturbation satisfies the linearized Einstein equation, which in Lorenz gauge is just
the wave equation,
\begin{equation}
  \Box \bar{h}_{\alpha \beta} + 2 R_\alpha{}^\gamma{}_\beta{}^\delta h_{\gamma \delta} = -16 \pi T_{\alpha \beta}
\end{equation}
where $\bar{h}_{\alpha \beta} \equiv h_{\alpha \beta} - \tfrac12 \go_{\alpha \beta} \go^{\gamma \delta} h_{\gamma \delta}$
is the trace-reversed metric perturbation.
We can invert this equation using the retarded Green function, which satisfies the wave equation,
\begin{equation}
  \Box G_{\alpha \beta}{}^{\alpha' \beta'}
    + 2 R_\alpha{}^\gamma{}_\beta{}^\delta G_{\gamma \delta}{}^{\alpha' \beta'} =
    - g_{\alpha}{}^{\alpha'} g_{\beta}{}^{\beta'} \delta^4 (x,x').
\end{equation}
In a four-dimensional Minkowski background ($\go_{\alpha \beta} = \eta_{\alpha \beta}$) the solution is
\begin{equation}
G^{\rm ret}_{\alpha \beta}{}^{\alpha' \beta'} (x,x') = \tfrac{1}{4\pi} \Theta_{-}(x,x') \delta_{(\alpha}^{\alpha'} \delta_{\beta)}^{\beta'} \delta[\sigma(x,x')].
\end{equation}
Here, $\sigma(x,x')$ is the Synge world-function, defined to be one-half of the square of the
geodesic distance between $x$ and $x'$, so that the Dirac delta function is non-zero only when
$x$ and $x'$ are null-separated. In Minkowski spacetime, we have the closed form
\begin{equation}
  \label{eq:Synge}
  \sigma (x, x') = \frac12 \eta_{\alpha \beta} (x^\alpha - x^{\alpha'}) (x^\beta - x^{\beta'}).
\end{equation}
The metric perturbation is then given by convolving the Green function with the source,
\begin{align}
\label{eq:hbar-convolution}
  \bar{h}_{\alpha \beta}(x) =&\, 16 \pi \int G^{\rm ret}_{\alpha \beta}{}^{\alpha' \beta'} (x,x') T_{\alpha' \beta'} (x') \sqrt{-g(x')} d^4 x'
\nonumber \\
   =& \, - 4 \,G \mu \iint P_{\alpha \beta} \delta [\sigma(x, z)] \sqrt{-\gamma} d \zeta^{1'} d\zeta^{2'}.
\end{align}
where $P_{\alpha \beta}$, $z^\alpha$ and $\gamma$ are all functions of $\zeta^{1'}$ and
$\zeta^{2'}$.

In practical calculations it is convenient to perform one of the integrals immediately using the
identity
\begin{equation}
  \delta\Big[\sigma\big(x,z(\zeta^1, \zeta^2)\big)\Big]
    = \frac{\delta\big[\zeta^1 - \zeta^1_{\rm ret}(x, \zeta^2)\big]}{|r_1|}
\end{equation}
where $r_1 \equiv \partial_{\zeta^{1'}} \sigma = (\partial_{\zeta^1}
z^{\alpha'}) (\partial_{\alpha'} \sigma)$ and $\zeta^1_{\rm ret}(x,
\zeta^2)$ parameterizes the retarded image, defined by
\begin{equation}
\sigma[ x, z(\zeta^1_{\rm ret}, \zeta^2) ] =0.
\end{equation}
This gives
\begin{equation}
\label{eq:hbar-zeta2-convolution}
  \bar{h}_{\alpha \beta}(x) = \, - 4 G \mu \oint \Bigg[\frac{\sqrt{-\gamma} P_{\alpha \beta}}{|r_1|}\Bigg]_{\zeta^{1'}_{\rm ret}} d\zeta^{2'},
\end{equation}
where the quantity in square brackets is evaluated at
$\zeta^{1'} = \zeta^{1'}_{\rm ret}(\zeta^{2'})$.
The one-dimensional integration traces exactly one period of the loop's
retarded image and there is no boundary; it is a closed loop.
Equivalently, the non-trace-reversed metric
perturbation is given by
\begin{equation}
\label{eq:h-zeta2-convolution}
  h_{\alpha \beta}(x) = \, - 4 G \mu \oint \Bigg[\frac{\sqrt{-\gamma}}{|r_1|} \Sigma_{\alpha \beta} P \Bigg]_{\zeta^{1'}_{\rm ret}} d\zeta^{2'},
\end{equation}
where $\Sigma_{\alpha \beta} \equiv P_{\alpha \beta}-\tfrac12 \eta_{\alpha \beta} P$ with $P \equiv
P^{\gamma}{}_{\gamma}$. Note that the integral does not converge when $x$ is
a point on the worldsheet; this is because the integrand diverges whenever $r_1 = 0$,
which occurs when source and field points coincide, i.e. $x = z$.

Derivatives of the first order metric perturbation may be computed in a similar manner to
$h_{\alpha \beta}$ itself, with the caveat that care must be taken in non-smooth regions of the
string. These non-smooth regions occur at kinks and cusps, and also in the vicinity of the field
point, $x$, if it is on the worldsheet.

Ignoring the issue of smoothness for now, and differentiating Eq.~\eqref{eq:hbar-convolution} with
respect to the field point, $x$, we get
\begin{align}
\label{eq:dh-convolution}
  \partial_\gamma & \bar{h}_{\alpha \beta}(x)
    = - 4 \,G \mu \iint P_{\alpha \beta} \, \partial_\gamma \big( \delta [\sigma(x, z)] \big) \sqrt{-\gamma} \, d \zeta^{1'} d\zeta^{2'}
\nonumber \\ &
    = - 4 \,G \mu \iint P_{\alpha \beta} \,\partial_\gamma \sigma \, \delta'[\sigma(x, z)] \sqrt{-\gamma}\, d \zeta^{1'} d\zeta^{2'}
\nonumber \\ &
    = - 4 \,G \mu \iint P_{\alpha \beta} \frac{\partial_\gamma \sigma}{\partial_{\zeta^{1'}} \sigma} \partial_{\zeta^{1'}} \big( \delta [\sigma(x, z)] \big) \sqrt{-\gamma}\, d \zeta^{1'} d\zeta^{2'}
\nonumber \\ &
    = - 4 \,G \mu \iint \frac{P_{\alpha \beta} \Omega_\gamma}{r_1} \partial_{\zeta^{1'}} \big( \delta [\sigma(x, z)] \big) \sqrt{-\gamma}\, d \zeta^{1'} d\zeta^{2'},
\end{align}
where $\Omega^\alpha \equiv x^\alpha - x^{\alpha'}$ is the coordinate distance
between $x$ and $x'$. On a smooth worldsheet, this may be integrated by parts to give
\begin{align}
\label{eq:dhbar-zeta2-convolution}
  \partial_\gamma & \bar{h}_{\alpha \beta}(x)
\nonumber \\
    & = 4 \,G \mu \iint \partial_{\zeta^{1'}} \bigg[ \frac{\sqrt{-\gamma} P_{\alpha \beta} \Omega_\gamma}{r_1} \bigg] \delta [\sigma(x, z)] d \zeta^{1'} d\zeta^{2'}
\nonumber \\
    & = 4 \,G \mu \oint \Bigg[\frac{1}{|r_1|} \partial_{\zeta^{1'}} \bigg( \frac{\sqrt{-\gamma} P_{\alpha \beta} \Omega_\gamma}{r_1} \bigg) \Bigg]_{\zeta^{1'}_{\rm ret}} d\zeta^{2'}.
\end{align}
Note that there are no boundary terms introduced in the integration by parts as the integration is
over a closed loop. Additionally, note that we can also arrive at the same equation by
differentiating Eq.~\eqref{eq:hbar-zeta2-convolution} and accounting for the fact that the
dependence on $x$ appears both through $r_1$ and through $\zeta^{1'}_{\rm ret}$, along with the
relation $\partial_\gamma \zeta^{1'}_{\rm ret} = -\Omega_\gamma / r_1$ (see Sec.~10 of
\cite{Poisson-review}). Again, we may write this in the non-trace-reversed form,
\begin{align}
\label{eq:dh-zeta2-convolution}
  \partial_\gamma & h_{\alpha \beta}(x) =
\nonumber \\
    & 4 \,G \mu \oint \Bigg[\frac{1}{|r_1|} \partial_{\zeta^{1'}} \bigg( \frac{\sqrt{-\gamma} \Sigma_{\alpha \beta} \Omega_\gamma}{r_1} \bigg) \Bigg]_{\zeta^{1'}_{\rm ret}} d\zeta^{2'}.
\end{align}

\subsection{First order self-force}
With the results from the previous section at hand, it is straightforward to obtain an integral
expression for the first-order gravitational self-force. Substituting
Eqs.~\eqref{eq:h-zeta2-convolution} and \eqref{eq:dh-zeta2-convolution} into \eqref{eq:self-force}
we obtain
\begin{align}
\label{eq:F1-convolution}
  F_1^{\mu}(z) &= - 4 \,G \mu \perp^{\mu \gamma} P^{\alpha \beta} \times \nonumber \\
		&  \hspace*{-0.5cm} \oint \Bigg[
     \frac{1}{|r_1|} \partial_{\zeta^{1'}} \Bigg( \frac{\sqrt{-\gamma}\big(\Sigma_{\beta \gamma} \Omega_\alpha
     - \tfrac12 \Sigma_{\alpha \beta} \Omega_\gamma\big)}{r_1} \Bigg) \Bigg]_{\zeta^{1'}_{\rm ret}} d\zeta^{2'},
\\
\label{eq:F2-convolution}
  F_2^{\mu}(z) &= \, - 4 G \mu K^{\beta \alpha \mu} \oint \Bigg[\frac{\sqrt{-\gamma}\Sigma_{\alpha \beta}}{|r_1|}  \Bigg]_{\zeta^{1'}_{\rm ret}} d\zeta^{2'}.
\end{align}
Here, it is understood that the $\perp^{\mu \gamma}$, $P^{\alpha \beta}$ and $K^{\beta \alpha \mu}$
appearing outside the integral are to be evaluated at $z$, whereas the $P^{\alpha \beta}$ and
$\gamma$ appearing inside the integral are to be evaluated at the retarded point $z'$.

One may expect a difficulty to arise from the fact that $\bar{h}_{\alpha\beta}$ diverges
logarithmically (and $\partial_\gamma \bar{h}_{\alpha\beta}$ is even more divergent) when the
source and field points coincide. This would appear to be a major obstacle for computing the
self-force since the integral expressions for $\bar{h}_{\alpha\beta}$ and $\partial_\gamma
\bar{h}_{\alpha\beta}$ will not converge when the field point, $x$, is on the worldsheet.
Fortunately, it turns out that for field points on smooth parts of the worldsheet, some miraculous
cancellations in the particular combination appearing in the equation of motion [and hence the
self-force, Eq.~\eqref{eq:self-force}] lead to many of the divergent terms canceling. The result
is that one obtains a convergent integral and a finite self-force. This was shown to hold in
\cite{Buonanno:Damour:1998} for the conformal gauge and in \cite{Carter:Battye:1998} for an
arbitrary gauge. However, both cases implicitly assumed a smooth string worldsheet. It turns out
that the conclusions continue to hold for a non-smooth worldsheet provided the field point is on
a smooth part of the worldsheet. As a field point approaches a
non-smooth point on the worldsheet the total self-force diverges.% At a non-smooth point it is truly divergent.

Despite this latter divergence, there is one further important consideration, namely the physical
significance of the self-force itself. It is possible that a divergence in the self-force is a
spurious artifact arising from, for example, an unfortunate choice of gauge or from a distributional
treatment of non-smooth worldsheet features. Indeed,
Anderson \cite{Anderson:2005qu} computed explicit closed form expressions for the self-force in the
case of the ACO string. His expressions diverge logarithmically and as negative
powers in the vicinity of the kink. However, this divergence is integrable and he was able to solve
the equations of motion to compute finite deviations in both the position and velocity of the
string\footnote{More precisely, the derivative along the direction orthogonal to the kink's
propagation direction was divergent at the kink, however Anderson was able to obtain a gauge
transformation which eliminated this divergence and so it can be attributed to nonphysical
coordinate effects.}. Similar conclusions have also been drawn in other work \cite{Quashnock:1990wv,Wachter:2016rwc,Wachter:2016hgi}.

In this work, we empirically find results that are consistent with these previous conclusions;
although the equation of motion has a divergent self-force term it turns out to give a finite
change to the worldsheet. Any physical measurement must be consistent
with the inferred finite displacement. With a distributional description of
kinks and cusps as adopted here finite displacements can lead to singular
changes in derivative quantities such as tangent
vectors on the worldsheet.\footnote{Divergent behavior of this sort
  (changes of order $G \mu/c^2$ in the tangent vector direction over a single
  period of oscillation) has recently been reported by
  Blanco-Pillado, Olum and Wachter [see acknowledgements]. 
  In our treatment here we emphasize that
  we ignore the possibility of additional contributions coming from the kink
itself. It is difficult to validate this assumption within a distributional approach. It is
likely that a matched asymptotic approach along the lines of Ref.~\cite{pound:2010} for point
particles would be required to provide a definitive answer to the question
of whether the distributional treatment omits any important physical
effects. We anticipate that such a treatment would also regularize singular
tangent vector derivatives so that all physical measurements
are finite, not merely consistent with the inferred finite worldsheet
displacement.}
Optimistically, we can expect the finiteness of worldsheet displacements
to carry through to more general scenarios, and hope that the divergences in the force are
always integrable. A proof of this fact can likely be obtained from a local expansion of the type
given in Sec.~\ref{sec:local-expansion} below, adapted to allow for a kink or cusp within the
``local'' region. Since there are considerable subtle details in this calculation, we will leave
its exploration for future work.

\section{Evaluating the Gradient of the Retarded Metric Perturbation}
\label{sec:dh}

In the previous section, we obtained integral expressions for the metric perturbation, its
derivative, and the gravitational self-force. The latter two are valid provided the retarded image
of the worldsheet is smooth. In reality, we do not have the luxury of a smooth worldsheet for at
least two reasons:
\begin{enumerate}
  \item We are interested in studying strings with kinks and cusps, and the worldsheet is non-smooth at the
        location of any kink or cusp;
  \item We are interested in computing the self-force, which requires us to evaluate the metric
        perturbation and its derivative in the limit $x \to z$. In that case, if one considers the
        retarded image of a point directly on the string, $x = z$, one finds that it is not in
        general smooth at the field point, $\zeta^{2'} = \zeta^{2'}(x)$.
\end{enumerate}
These can lead to important distributional-type contributions
to the integrand in the expression for the self force
which are easily missed.
In the following subsections, we extend Eqs.~\eqref{eq:dh-zeta2-convolution} and
\eqref{eq:F1-convolution} above to allow for these non-smooth
features. We begin with a general covariant derivation of the integral
to explain how coordinate-dependent divergences arise,
 and follow up with an explicit
treatment of both issues mentioned above.

\subsection{Covariant evaluation of the worldsheet integral}
\label{sec:coord-depend-integral}

The expression for the gradient of the metric perturbation at a point $x^\alpha$ is of the form (dropping spacetime tensor indices)
\begin{equation}
I = \int_{\cal W} \omega_{ab} \delta'(\sigma).
\label{dds}
\end{equation}
Here ${\cal W}$ is the worldsheet defined by $x^\alpha = z^\alpha(\zeta^a)$, $\omega_{ab}$ is some
given smooth two-form on the worldsheet, and the function $\sigma$ is as defined in
Eq.~\eqref{eq:Synge}. In this subsection we will derive some identities for integrals of the form
\eqref{dds} for arbitrary $\omega_{ab}$ and arbitrary smooth $\sigma$, and in the next subsection
we will specialize to the specific form \eqref{eq:Synge} of $\sigma$ for our application here.

As a warm up, let us first consider a simpler version of the integral \eqref{dds}, namely
\begin{equation}
J = \int_{\cal W} \omega_{ab} \delta(\sigma).
\label{Jdef}
\end{equation}
Let ${\cal C}$ be the curve given by $\sigma = 0$.  We would like to derive an expression for $J$ of the form
\begin{equation}
J = \int_{\cal C} \theta_a
\label{ans22}
\end{equation}
where $\theta_a$ is a one-form on the worldsheet. The result for $\theta_a$ is
\begin{equation}
\theta_a = \frac{D_a h}{ \omega^{bc} D_b \sigma D_c h}
\label{oneforma}
\end{equation}
Here $D_a$ can be taken to be either a covariant or a partial derivative on the worldsheet, and
$\omega^{ab}$ is the inverse of $\omega_{ab}$. Finally $h$ can be taken to be any smooth function
on the worldsheet which has the property that $dh \wedge d\sigma \ne 0$.

Note that the expression \eqref{oneforma} for the one-form, when pulled back onto the curve ${\cal
C}$, is independent of the choice of $h$. To see this, suppose we replace $h$ with a function $H$
of $h$ and $\sigma$,
\begin{equation}
h \to H(h,\sigma).
\label{hchange}
\end{equation}
Under this transformation
\begin{equation}
D_a h \to H_{,h} D_a h + H_{,\sigma} D_a \sigma.
\end{equation}
When this expression is inserted into the one-form \eqref{oneforma}, the contribution from the
second term to the denominator vanishes because of antisymmetrization, and the contribution to the
numerator vanishes when the one form is pulled back to ${\cal C}$, since $\sigma=0$ on ${\cal C}$.
The factors of $H_{,h}$ cancel between the numerator and the denominator, and so we see that the
pullback of $\theta_a$ to ${\cal C}$ is invariant under the transformation. Thus it is independent
of the choice of $h$.

We now turn to the derivation of the formula \eqref{oneforma}. We specialize to coordinates
$\zeta^{\bar 0} = \sigma$, $\zeta^{\bar 1} = h$. The integral \eqref{Jdef} becomes
\begin{equation}
J = \int d\sigma \int dh \, \omega_{\sigma h}(\sigma,h) \delta(\sigma)
\end{equation}
where $\omega_{\sigma h} = \omega_{{\bar 0}{\bar 1}}$. Evaluating the integral using the delta
function gives
\begin{equation}
J = \int dh \, \omega_{\sigma h}(0,h).
\label{b9}
\end{equation}
We now rewrite this in a form which is valid in arbitrary coordinate systems.  The factor $\int dh$ can be written as the integral over ${\cal C}$ of the one-form $D_a h$.  The factor $\omega_{\sigma h}$ can be written as
\begin{equation}
\omega_{\sigma h}  = \frac{1}{\omega^{\sigma h}}.
\label{b10}
\end{equation}
Using the tensor transformation law we have
\begin{equation}
\omega^{\sigma h} = \omega^{{\bar 0}{\bar 1}} = \omega^{ab} \frac{\partial \zeta^{\bar 0}}{\partial \zeta^a}
\frac{\partial \zeta^{\bar 1}}{\partial \zeta^b}
= \omega^{ab} \frac{\partial \sigma}{\partial \zeta^a}
\frac{\partial h}{\partial \zeta^b}.
\label{b11}
\end{equation}
Combining Eqs.~\eqref{b9}, \eqref{b10}, and \eqref{b11} now yields the result given by
Eqs.~\eqref{ans22} and \eqref{oneforma}.

Turn next to the corresponding analysis for the integral \eqref{dds}. Suppose that instead of
integrating over the entire worldsheet, we integrate over a region $\Delta {\cal W}$ of it. The
intersection of the boundary $\partial \Delta {\cal W}$ of this region with the curve ${\cal C}$
will consist of a set of discrete points ${\cal P}_i$. The formula for the integral is
\begin{equation}
I = \int_{\Delta {\cal W}} \omega_{ab} \delta'(\sigma) = I_{\rm boundary} + I_{\rm bulk}
\label{int5}
\end{equation}
where the contribution from the boundary is
\begin{equation}
I_{\rm boundary} = \sum_i \pm \frac{1}{\varphi} \frac{k^a D_a h}{ k^b D_b \sigma}.
\label{Ib}
\end{equation}
Here $k^a$ is the tangent to the boundary $\delta \Delta {\cal W}$
and
\begin{equation}
\varphi = \omega^{ab} D_a \sigma D_b h.
\label{varphidef}
\end{equation}
The contribution from the bulk is
\begin{equation}
I_{\rm bulk} = \int_{\cal C} \theta_a
\end{equation}
where the one-form $\theta_a$ is
\begin{equation}
\theta_a = \frac{1}{\varphi^3} \left( \omega^{bc} D_b \varphi D_c h \right) D_a h.
\label{xyy}
\end{equation}

Under a change of the function $h$ of the form \eqref{hchange}, the one-form $\theta_a$ is no
longer invariant. Instead, it transforms by an exact form\footnote{This formula is valid when
pulled back to the curve ${\cal C}$.}
\begin{equation}
\theta_a \to \theta_a + D_a \lambda,
\label{ccc}
\end{equation}
where $\lambda = H_{,\sigma} / (\varphi H_{,h})$.  The change in the boundary integral is
\begin{equation}
\sum_i \pm \frac{H_{,\sigma}}{\varphi H_{,h}},
\end{equation}
which cancels against the change \eqref{ccc} in the one-form. Thus we make the important
observation that the integral \eqref{int5} is independent of choice of $h$, but the split into
boundary and integral terms is not.

We now turn to the derivation of the formula \eqref{int5}. As before we initially specialize to
coordinates $(\zeta^{\bar 0}, \zeta^{\bar 1}) = (\sigma,h)$. Inserting the identity
\begin{equation}
\omega_{\sigma h} \delta'(\sigma ) d\sigma  \wedge dh = d \left[ \omega_{\sigma h} \delta(\sigma ) dh \right] - \delta(\sigma ) d\omega_{\sigma h} \wedge dh
\end{equation}
into the integral \eqref{int5}and using Stokes's theorem gives a result of the form of the right
hand side of \eqref{int5}, with
\begin{equation}
I_{\rm boundary} = \int_{\partial \Delta {\cal W}} \omega_{\sigma h} \delta(\sigma ) dh
\end{equation}
and
\begin{equation}
I_{\rm bulk}  = -\int_{\Delta {\cal W}} \delta(\sigma ) \, \omega_{\sigma h,\sigma } \, d\sigma dh.
\label{bulkf}
\end{equation}
We evaluate the first term by taking the parameter along the boundary $\delta \Delta {\cal W}$ to
be $\sigma$ and using $dh = d\sigma (dh/d\sigma)$. This gives
\begin{equation}
I_{\rm boundary} = \sum_i \, \omega_{\sigma h} \, \frac{dh}{d\sigma},
\end{equation}
Using Eqs.~\eqref{b11} and \eqref{varphidef} this reduces to the formula \eqref{Ib}.

For the bulk contribution, from the formula (\ref{bulkf}) and using arguments similar to those given for the integral $J$, we find
\begin{equation}
\theta_a = - \partial_\sigma (1/\varphi) D_a h = \frac{\varphi_{,\sigma }}{\varphi^2} D_a h,
\label{xy0}
\end{equation}
where we have used $\varphi = 1/\omega_{\sigma h}$.  We evaluate the $\sigma$ derivative using
\begin{equation}
\varphi_{,\sigma } = \frac{\partial  \varphi}{\partial \zeta^{\bar 0}} = \frac{\partial  \varphi}{\partial \zeta^a}  \, \frac{\partial \zeta^a}{\partial \zeta^{\bar 0}}.
\label{xy}
\end{equation}
We express the Jacobian matrix in terms of its inverse using
\begin{equation}
\frac{\partial \zeta^a}{\partial \zeta^{\bar a}}
= \frac{2}
{\left[ \omega^{cd} \omega_{{\bar c}{\bar d}}
\frac{\partial \zeta^{\bar c}}{\partial \zeta^c}
\frac{\partial \zeta^{\bar d}}{\partial \zeta^d}\right]}\, \omega^{ab} \omega_{{\bar a}{\bar b}} \frac{\partial \zeta^{\bar b}}{\partial \zeta^b}.
\end{equation}
This formula is specific to two dimensions, and is valid for any choice of two-form.  Specializing to ${\bar a} = {\bar 0}$ gives
\begin{equation}
\frac{\partial \zeta^a}{\partial \zeta^{\bar 0}} = \frac{ \omega^{ab} D_b h}{\omega^{cd} D_c \sigma \, D_d h}.
\end{equation}
Inserting this into \eqref{xy} and then into \eqref{xy0} finally gives the result \eqref{xyy}.

Finally, although the results derived in this subsection are
covariant, they do depend on a choice of arbitrary function $h$ on the
worldsheet.  While the complete final result (\ref{int5}) does not
depend on $h$, the integrand (\ref{xyy}) of the bulk integral, as well as
the splitting into bulk and boundary terms, do depend on $h$.
Elsewhere in this paper, we choose to identify $h$ with one of the
worldsheet coordinates, which explains the coordinate dependence of the
integrand and of the splitting.

\subsection{Worldsheets with kinks}
\label{sec:kinks}
We may now consider how our 1-D integral expressions \eqref{eq:h-zeta2-convolution} and \eqref{eq:dh-zeta2-convolution}
for the retarded metric perturbation and its gradient must be modified to allow for the presence of
a kink. A cosmic string with a kink may be treated as piecewise smooth, with discontinuities in
certain tangent vectors whenever a kink is crossed. To obtain an expression allowing for these
discontinuities we assume that the retarded image on the worldsheet is non-smooth at $\zeta^2 = k$, where $k$ may
depend on the field point $x$.\footnote{Although there are worldsheet coordinates where one of the
coordinates is a constant along a kink, we will not restrict ourselves to only that case here.}
Then, one way to achieve the desired result is to break up the integration in
Eq.~\eqref{eq:hbar-zeta2-convolution} at the discontinuity,
\begin{equation}
\label{eq:h-zeta2-discontinuous-convolution}
  \bar{h}_{\alpha \beta}(x) = \, - 4 G \mu \int_{k^+}^{k^- + L} \Bigg[\frac{\sqrt{-\gamma} P_{\alpha \beta}}{|r_1|} \Bigg]_{\zeta^{1'}_{\rm ret}} d\zeta^{2'},
\end{equation}
where $k^+$ ($k^-$) is a point just to the right (left) of the kink. Now, when we differentiate
this expression we have to take account of the possible dependence of the end points on $x$. If the
discontinuity in the string is at a fixed value of $\zeta^2$ (i.e. in the case of a kink
propagating along the $\zeta^1$ direction), then $k$ does not depend on $x$ and the boundary terms
vanish. If, instead, the discontinuity is at a fixed value of $\zeta^1$ (i.e. in the case of a kink
propagating along the $\zeta^2$ direction), then $k$ does depend on $x$. Then, using
$\partial_\gamma \zeta^{2}(\zeta^{1}_{\rm ret}) = -\Omega_\gamma / r_2$, where $r_2 \equiv
\partial_{\zeta^{2'}} \sigma = (\partial_{\zeta^2} z^{\alpha'}) (\partial_{\alpha'} \sigma)$, we get
\begin{align}
\label{eq:dh-zeta2-discontinuous-convolution}
  \partial_\gamma & \bar{h}_{\alpha \beta}(x)
\nonumber \\
    & = 4 \,G \mu \Bigg\{\int_{k^+}^{k^- + L} \Bigg[\frac{1}{|r_1|} \partial_{\zeta^{1'}} \bigg( \frac{\sqrt{-\gamma} P_{\alpha \beta} \Omega_\gamma}{r_1} \bigg) \Bigg]_{\zeta^{1'}_{\rm ret}} d\zeta^{2'}
\nonumber \\
    & \qquad + \bigg[\frac{\sqrt{-\gamma} P_{\alpha \beta} \Omega_\gamma}{|r_1| r_2}\bigg]_{k^-}
      - \bigg[\frac{\sqrt{-\gamma} P_{\alpha \beta} \Omega_\gamma}{|r_1| r_2}\bigg]_{k^+}
    \Bigg\}.
\end{align}
It is easy to check that one can arrive at the same expression
by appropriately including the boundary
terms from the integration by parts described in Sec.~\ref{sec:coord-depend-integral} above.
The presence (or lack thereof) of boundary terms is then manifestly dependent on the particular
choice of worldsheet coordinates. Importantly, this apparent worldsheet coordinate dependence only
appears in the split between boundary and bulk terms; the sum of the two does
\emph{not} have any worldsheet coordinate dependence.

In the case of a smooth string, the two boundary terms are identical and cancel, so we recover the
same formula as we had before. In the presence of a kink, however, the boundary terms in the two
limits $k^+$ and $k^-$ yield different values and so we pick up an overall
contribution from the kink in addition to the integral over the smooth portion of the string.

\subsection{Worldsheets with cusps}

We have already seen that care must be taken in computing the self-force for cosmic strings with
kinks. Since cusps also introduce non-smoothness in the worldsheet, one may expect similar care to
be required for cuspy strings. However, there is one crucial difference between a string with
kinks and one with cusps: cusps typically occur at a single \emph{point} on a worldsheet while kinks
occur along a one-dimensional curve. The result is that, in the case of kinks, all points on the
string ``see'' a kink at some point in their retarded image, and hence the integrand in Eq.~\eqref{eq:dh-zeta2-convolution}
will always be supplemented by a boundary term somewhere. Conversely, there is
only a one-dimensional set of points on the string which ``see'' a cusp in their retarded image;
everywhere else the integrand does not encounter a discontinuity.

This suggests that strings with cusps may not need the same careful treatment as those with kinks.
This appears to be the case in our test case in Sec.~\ref{sec:results} below, where we probe the
region around the one-dimensional cusp-seeing curve and find no evidence of unusual behavior. This
is, of course, merely empirical evidence, and should be followed with a more formal treatment; it
is likely that the local expansions developed in Sec.~\ref{sec:local-expansion} will prove useful
in such an analysis.

\subsection{Contribution from the field point}
\label{sec:local-expansion}

The final place where we must take care is in the case where the field point itself is on the
string. Then, just as in the case of a kink, the retarded image may have a discontinuity at the
field point. While it may be possible in such cases to use a similar treatment to what we have done
for kinks, there is subtlety in taking the limit of the field point to the worldsheet which makes
such a treatment difficult. Instead we choose a more robust approach, by using a local expansion of
the integrand for field points nearby\footnote{Here, we use the term ``nearby'' loosely as such a
notion is obviously dependent on the choice of worldsheets, and in particular on the choice of
coordinate which is used as the variable of integration. Not surprisingly, we will find that the
conclusions we draw will depend on the choice of worldsheet coordinates. Nevertheless, just as in
Sec.~\ref{sec:coord-depend-integral}, this apparent coordinate dependence is merely an artifact of
how we choose to split up the self-force into contributions from various integrals and boundary
terms. In reality, the total self-force obtained by combining all of these contributions is
independent of the choice of worldsheet coordinates.} the string and then analytically taking the
limit of the field point to the worldsheet.

The purpose of the following subsections is to develop the pieces required for such an expansion. In doing so we make some assumptions:
\begin{enumerate}
  \item We will study the contribution to the self-force integral nearby where the force is to be
        computed and will ultimately shrink the size of this region down to zero;
  \item We will assume that the worldsheet is smooth in this region. This is true everywhere except
        when the field point exactly lies on a kink or cusp; points arbitrarily close to a kink or
        cusp will, however, be perfectly acceptable.
  \item We will assume that the induced metric does not diverge (or vanish) on the string. This will be true
        everywhere except where a field point lies exactly on a cusp.
  \item We will assume conformal gauge for the background worldsheet, in particular Eq.~\eqref{eq:conformal} and the orthogonality relations for $\partial_{\zeta^1} z^\mu$ and
$\partial_{\zeta^2} z^\mu$ which follow from it. This step is not a strict requirement of the approach, but does significantly simplify the tensor algebra in the calculation.
\end{enumerate}

Before we proceed with the derivation of the local expansion, we point out one interesting feature,
namely that the divergence in the self-force that arises on kinks and cusps comes purely from the
short-distance portion of the self-field, i.e. the contribution to the integral from nearby
points. It is therefore likely that a more careful treatment of what happens to the self-force
exactly on a kink or cusp may be obtained from a local expansion of the kind given here. We leave
the exploration of this issue for future work.

\subsubsection{Setup of the local expansion}
We wish to compute the contribution to the self-force for points near the field point. To do so,
we will construct a local expansion of the self-force integrand about a point on the worldsheet
which is assumed to be nearby the field point, $x^\alpha$, and to lie on its retarded image,
$z^\alpha[\zeta^1_{\rm ret}(x, \zeta^{2}), \zeta^{2}]$. We denote this expansion point by
$\bar{z}^\alpha \equiv z^{\alpha}[\bar{\zeta}^{1}, \bar{\zeta}^{2}]$ with $\bar{\zeta^1} \equiv
\zeta^1_{\rm ret}(x, \bar{\zeta}^{2})$ for a particular choice of $\bar{\zeta}^2$.
The conformal factor at this point is $\bar{\phi} \equiv \phi(\bar{\zeta}^1,\bar{\zeta}^2)$ and we assume
the expansion has a radius of convergence that includes
part of the image. We can then simplify the evaluation of the
local integration over that part of the image utilizing the approximate
expansion.

We will now seek an expansion of the self-force integrand \eqref{eq:F1-convolution} (note that there is no contribution to $F_2^\mu$ from the field point since it does not involve derivatives of $h_{\alpha \beta}$) in $\Delta \zeta^2 \equiv \zeta^{2} -
\bar{\zeta}^{2}$.\footnote{Notationally, the integration in \eqref{eq:dh-zeta2-convolution} is over the dummy variable $\zeta^{2'}$ but we suppress these primes for clarity.} The first stage in our calculation is to find an expansion of the retarded
coordinate $\zeta^1_{\rm ret}(x, \zeta^{2})$ about $\bar{\zeta}^1 =
{\zeta}^1_{\rm ret}(x, \bar{\zeta}^{2})$.
We denote the difference between these two quantities $\Delta \zeta^1$ and will seek an expansion
of it in powers of $\Delta \zeta^2$. In doing so, we will need to be careful about what our
particular choice of worldsheet coordinate is. We will also need to separately consider the cases
where $\Delta \zeta^2$ is positive or negative, as in some instances the expansion has a different
form in the two cases.

\subsubsection{Expansion of the light-cone condition: space-time coordinates}

In this section we focus on a pair of spacelike and timelike type coordinates which we will denote by $\zeta$
(for space) and $\tau$ (for time), i.e. $(\zeta^1,\zeta^2) = (\tau,\zeta)$. The important defining feature of these coordinates are the conformal gauge orthogonality relations
\begin{align}
  g_{\alpha \beta} \partial_\tau z^\alpha \partial_\tau z^\beta &= - \phi, \\
  g_{\alpha \beta} \partial_\tau z^\alpha \partial_\zeta z^\beta &= 0, \\
  g_{\alpha \beta} \partial_\zeta z^\alpha \partial_\zeta z^\beta &= \phi.
\end{align}
We can also obtain similar relations involving higher derivatives (with respect to $\tau$ and/or
$\zeta$) of $z^\alpha$ by differentiating these fundamental relations. We additionally have the conformal gauge equation of motion, Eq.~\eqref{eq:conformal-eom}, which in $\tau-\zeta$ coordinates gives us a relation between second $\tau$ and second $\zeta$ derivatives of $z^\alpha$:
\begin{equation}
  \partial_{\tau\tau} z^\alpha = \partial_{\zeta\zeta} z^\alpha.
\end{equation}
We will use these identities throughout the following calculation to simplify the results we obtain.

We will start from the fact that the (retarded) source point $z^{\alpha'}$ and the field point
$z^{\alpha}$ are null-separated, $\sigma(z^\alpha, z^{\alpha'}) = 0$. Expanding this about $\sigmab
\equiv \sigma(z^\alpha, \bar{z}^{\alpha})$ we obtain a power series in $\Delta \tau$ and $\Delta
\zeta$,
\begin{align}
  \label{eq:sigma-expansion-tau-zeta}
  \sigma &= \sigmab + \sigmab_{,\tau} \Delta \tau + \sigmab_{,\zeta} \Delta \zeta \nonumber \\
  &\quad+ \tfrac12 (\sigmab_{,\tau\tau} \Delta \tau^2 + 2\sigmab_{,\tau\zeta} \Delta \tau \Delta \zeta + \sigmab_{,\zeta\zeta} \Delta \zeta^2) \nonumber \\
  & \quad+\tfrac16 (\sigmab_{,\tau\tau\tau} \Delta \tau^3 + 3 \sigmab_{,\tau\tau\zeta} \Delta \tau^2 \Delta \zeta + 3 \sigmab_{,\tau\zeta\zeta} \Delta \tau \Delta \zeta^2 \nonumber \\
  & \qquad + \sigmab_{,\zeta\zeta\zeta} \Delta \zeta^3) + \cdots
\end{align}
Using $\partial_a = (\partial_a z^\alpha) \nabla_\alpha$ (acting upon the second
argument of $\bar{\sigma}$) along with the identities above and the
fact that $\nabla_\alpha \nabla_\beta \sigma = g_{\alpha \beta}$ for Minkowski spacetime, it is
straightforward to rewrite the coefficients in terms of worldsheet derivatives of $\bar{z}^\alpha$
and $\bar{\phi}$:
\begin{align}
  \bar{\sigma}_{,\tau\tau} &= \bar{z}^\alpha_{,\zeta\zeta} \bar{\sigma}_\alpha - \phi, \\
  \bar{\sigma}_{,\tau\zeta} &= \bar{z}^\alpha_{,\tau\zeta} \bar{\sigma}_\alpha, \\
  \bar{\sigma}_{,\zeta\zeta} &= \bar{z}^\alpha_{,\zeta\zeta} \bar{\sigma}_\alpha + \phi, \\
  \bar{\sigma}_{,\tau\tau\tau} &= \bar{z}^\alpha_{,\tau\zeta\zeta} \bar{\sigma}_\alpha - \tfrac32 \phi_{,\tau}, \\
  \bar{\sigma}_{,\tau\tau\zeta} &= \bar{z}^\alpha_{,\zeta\zeta\zeta} \bar{\sigma}_\alpha - \tfrac12 \phi_{,\zeta}, \\
  \bar{\sigma}_{,\tau\zeta\zeta} &= \bar{z}^\alpha_{,\tau\zeta\zeta} \bar{\sigma}_\alpha + \tfrac12 \phi_{,\tau}, \\
  \bar{\sigma}_{,\zeta\zeta\zeta} &= \bar{z}^\alpha_{,\zeta\zeta\zeta} \bar{\sigma}_\alpha + \tfrac32 \phi_{,\zeta},
\end{align}
and likewise for higher order terms (for the current calculation of the contribution to the self-force from the field point it is only necessary to
go to the cubic order given here).

\subsubsection{Expansion of the retarded time}
In order to obtain the desired expansion of $\Delta \tau(\Delta \zeta)$, we now make the ansatz
that $\Delta \tau$ has an expansion in integer powers of an order counting parameter $\epsilon \sim
\Delta \zeta$ and that $\bar{\sigma} = \mathcal{O}(\epsilon^2)$. Substituting our ansatz into
Eq.~\eqref{eq:sigma-expansion-tau-zeta} and solving order by order in $\epsilon$ then yields the
desired expansion of $\Delta \tau$ in terms of $\epsilon$,
\begin{equation}
  \label{eq:tret-spatial-expansion}
  \Delta \tau = \frac{1}{\phi}\bigg[\bar{\sigma}_{,\tau} - \sqrt{\bar{\sigma}_{,\tau}^2 + \phi(2\Delta \zeta \bar{\sigma}_{,\zeta} + 2 \bar{\sigma} + \phi \Delta \zeta^2)}\bigg]  \epsilon + \mathcal{O}(\epsilon^2).
\end{equation}
The expressions for the higher order coefficients are somewhat cumbersome, but are fortunately not
required for the current calculation.

\subsubsection{Expansion of quantities appearing in the integrand for the self-force}

We now expand each of the quantities appearing in the integrand of $F^\mu_1$ [Eq.~\eqref{eq:F1-convolution}]:
\begin{equation}
  \Sigma_{\mu \nu} = \Sigma_{\mu\nu}^{(0,0)} + \Sigma_{\mu\nu}^{(1,0)} \Delta \tau + \Sigma_{\mu\nu}^{(0,1)} \Delta \zeta + \cdots,
\end{equation}
where
\begin{align}
  \Sigma_{\mu\nu}^{(0,0)} &= \partial_\zeta z_\mu \partial_\zeta z_\nu - \partial_\tau z_\mu \partial_\tau z_\nu-\phi g_{\mu\nu}, \\
  \Sigma_{\mu\nu}^{(1,0)} &= \partial_\tau \partial_\zeta z_\mu \partial_\zeta z_\nu +  \partial_\zeta z_\mu \partial_\tau \partial_\zeta z_\nu - \partial_\tau z_\mu \partial_\zeta \partial_\zeta z_\nu \nonumber \\
    &\quad- \partial_\zeta \partial_\zeta z_\mu \partial_\tau z_\nu - \partial_\tau\phi g_{\mu\nu}, \\
  \Sigma_{\mu\nu}^{(0,1)} &= \partial_\zeta \partial_\zeta z_\mu \partial_\zeta z_\nu +  \partial_\zeta z_\mu \partial_\zeta \partial_\zeta z_\nu - \partial_\tau z_\mu \partial_\tau \partial_\zeta z_\nu \nonumber \\
    &\quad- \partial_\tau \partial_\zeta z_\mu \partial_\tau z_\nu - \partial_\zeta \phi g_{\mu\nu}.
\end{align}
We also have
\begin{align}
  r &= \bar{\sigma}_{,\tau} + \sigmab_{,\tau\tau} \Delta \tau + \sigmab_{,\tau\zeta} \Delta \zeta \nonumber \\
  &\quad + \tfrac12 (\sigmab_{,\tau\tau\tau} \Delta \tau^2 + 2\sigmab_{,\tau\tau\zeta} \Delta \tau \Delta \zeta + \sigmab_{,\tau\zeta\zeta} \Delta \zeta^2) + \cdots,
\end{align}
and
\begin{align}
\Omega_\mu &= \bar{\Omega}_\mu - z_{\mu, \tau} \Delta \tau - z_{\mu,\zeta} \Delta \zeta  - \tfrac12 z_{\mu, \tau\tau} \Delta \tau^2 \nonumber \\
& \quad - z_{\mu,\tau\zeta} \Delta \tau \Delta \zeta - \tfrac12 z_{\mu,\zeta\zeta} \Delta \zeta^2 + \cdots.
\end{align}
Note that there are three potentially small parameters in these expansions: $\Delta \tau$,
$\Delta \zeta$ and the distance of the field point from the string, which we will denote $\Delta x$.
In the above, the dependence on $\Delta \tau$ and $\Delta \zeta$ appears explicitly; the dependence
on $\Delta x$ appears through $\bar{\Omega}_\mu \sim \Delta x$ and $\bar{\sigma}_\alpha \sim \Delta x$.

To make further progress, we will assume that all three are of the same order, $\Delta \tau \sim
\epsilon$, $\Delta \zeta \sim \epsilon$ and $\Delta x \sim \epsilon$. Now, substituting the
expansions into the integral equation for the derivative of the metric perturbation,
Eq.~\eqref{eq:dh-zeta2-convolution} and expanding out in powers of $\epsilon$, we find that the
integrand has a contribution at order $\epsilon^{-2}$ and at order $\epsilon^{-1}$, plus higher
order terms. More explicitly, the $\mathcal{O}\left(\epsilon^{-2}\right)$ piece is given by
\begin{widetext}
\begin{align}
\partial_\gamma h_{\alpha \beta} \approx -4 \int
  \frac{
  [\sigmab_{,\tau\tau}] \Sigma_{\alpha\beta}^{(0,0)} \bar{\Omega}_\gamma
  - [\sigmab_{,\tau\tau}] \Sigma_{\alpha\beta}^{(0,0)} z_{\gamma,\zeta}\Delta \zeta
  + \Sigma_{\alpha\beta}^{(0,0)} z_{\gamma,\tau} \bar{\sigma}_{,\tau}
  }{\left(\bar{\sigma}_{,\tau} + [\sigmab_{,\tau\tau}] \Delta \tau\right)^3} d\zeta + \mathcal{O}(\epsilon^{-1})
\end{align}
where square brackets denote a coincidence limit, $[\sigmab_{,\tau\tau}] \equiv \lim_{\Delta x \to 0} \sigmab_{,\tau\tau}$. Now, it is
immediately apparent that if we instead substitute our expansions into the integral expression for
$F_1^\mu$ this leading order piece identically vanishes since $P^{\mu \nu} \Sigma_{\mu \nu} = 0$
\footnote{Strictly speaking, this depends on how we extend the definition of $P^{\mu \nu}$ off the
worldsheet. However, since we are in the end only interested in taking the limit to the worldsheet
the particular choice of extension is irrelevant and does not change the result.} Likewise, since
$\Sigma_{\mu\nu}^{(0,1)} z_{\mu, \zeta} = \Sigma_{\mu\nu}^{(1,0)} z_{\mu, \tau}$, many other terms
either identically vanish or simplify significantly. Then, the only remaining piece of the
$\mathcal{O}(\epsilon^{-1})$ contribution to the derivative of the metric perturbation \emph{which does not
vanish upon substitution into the self-force} is given by
\begin{align}
 & -\int \tfrac{4}{\left(\bar{\sigma}_{,\tau} + [\sigmab_{,\tau\tau}] \Delta \tau\right)^3} \bigg(
  [\sigmab_{,\tau\tau}] \Sigma_{\alpha \beta}^{(0,1)} \bar{\Omega}_{\gamma} \Delta \zeta -
   \Sigma_{\alpha \beta}^{(1,0)} \bar{\Omega}_{\gamma} \bar{\sigma}_{,\tau} +
   \Sigma_{\alpha \beta}^{(1,0)} z_{\gamma,\zeta} \bar{\sigma}_{,\tau} \Delta \zeta +
   \Sigma_{\alpha \beta}^{(0,1)} z_{\gamma,\tau} \bar{\sigma}_{,\tau} \Delta \zeta \nonumber \\ & \qquad \qquad \qquad \qquad \qquad -
   [\sigmab_{,\tau\tau}] \Sigma_{\alpha \beta}^{(1,0)} z_{\gamma,\tau} \Delta \zeta^2 +
   2 \Sigma_{\alpha \beta}^{(1,0)} z_{\gamma,\tau} \bar{\sigma}_{,\tau} \Delta \tau +
   [\sigmab_{,\tau\tau}] \Sigma_{\alpha \beta}^{(1,0)} z_{\gamma,\tau} \Delta \tau^2 \bigg)
  d\zeta .
\end{align}
\end{widetext}

Our final step is to substitute in the expansion of the retarded time, rescale our integration range
by $\epsilon$ and integrate from $\Delta \zeta / \epsilon = -\infty$ to $+\infty$.
The factor of $\epsilon$ in the integral weight cancels with the $1/\epsilon$ in the integrand
and so the result is ultimately independent of $\epsilon$.

\subsubsection{Expansion of the self-force}

Performing the integral explicitly in the limit where the field field point tends to the
worldsheet, we finally arrive at a surprisingly simple expression for the field point contribution to the
self-force. In $\tau-\zeta$ coordinates, this is given by
\begin{equation}
  \label{eq:F-local-ST}
 F^\mu_{\rm{field,ST}} = 4 \, \phi^{-2} \perp^\mu{}_\alpha \left(z^\alpha_{,\zeta} \phi_{,\zeta} + z^\alpha_{,\tau} \phi_{,\tau} - 2 z^\alpha_{,\zeta\zeta} \phi\right).
\end{equation}

One can go through a similar procedure in the null case (see Appendix
\ref{sec:local-expansion-null} for details of the retarded time expansion in null coordinates).
Then, if we use $\zeta^-$ as our integration variable, the
equivalent expression for the field point contribution to the self-force is
\begin{equation}
  \label{eq:F-local-N}
  F^\alpha_{\rm{field,N}} = 4 \, \phi^{-2} \perp^\mu{}_\alpha \left(z^\alpha_{,\zeta^+} \phi_{,\zeta^+} -  z^\alpha_{,\zeta^+\zeta^+} \phi\right) .
\end{equation}
Likewise, one can change $+ \to -$ when $\zeta^+$ is used as the integration variable.
The expressions \eqref{eq:F-local-ST} or \eqref{eq:F-local-N} must be
added to the previous results given by Eq.~\eqref{eq:F1-convolution} to obtain the total
contribution to $F_1^\alpha$.

\section{Numerical Methods and Regularization}
For this work have developed several different techniques to
evaluate the self-force on the string by completely finite, numerical
calculations. In the next section, we will compare these calculations to validate
the exact methods we have discussed. Before doing so, here we will
schematically outline the different approaches. The
abbreviation for the methods are given in square brackets.

\subsection*{2D, smoothed kink or cusp [2D]}
The most general approach is to do the 2D integration over the
worldsheet in Eq.~\eqref{eq:hbar-convolution}. This circumvents having
to eliminate one worldsheet coordinate in terms of another
(e.g. solving for the retarded time in $\tau-\zeta$ coordinates) and possibly
having to patch different coordinate systems (e.g. two different null
coordinate systems either side of the field point). The worldsheet integration
produces manifestly coordinate invariant results.

Schematically, we replace the singular retarded Green function with a finite
approximation. For a source at $x_s$ and field at $x_f$
\begin{eqnarray}
  {\cal G}(x_f,x_s) & = & \Theta(x_s,x_f) \delta(\sigma)
\end{eqnarray}
where the $\Theta=1$ when the time of the source $t_s$ precedes the
time of the field point $t_f$ and $0$ otherwise. We transform
\begin{eqnarray}
  \delta(\sigma) & \to & \frac{e^{-\sigma^2/(2 w_1^2)}}{\sqrt{2 \pi} w_1} \\
  \Theta & \to &\frac{1-\tanh((t_s-t_f)/w_2)}{2}
\end{eqnarray}
to generate a smooth, finite integrand. The parameters $w_1$ and $w_2$
describe the width of the smoothed delta function and the width of the causal
function. (We use $w_i$ schematically in this discussion. In
Appendix \ref{sec:2D} we introduce unique symbols.)

Source points are over-retarded and appear slightly
inside the field point's backwards light cone. Over-retardation
\cite{1975NCimB..26..157D} is a covariant method for classical renormalization.
%\cite{Damour:1975uj}
We modify the Synge function
\begin{eqnarray}
  \sigma(x,z) & = & \frac{1}{2} (x-z)^\alpha g_{\alpha\beta}(x-z)^\beta + w_3
\end{eqnarray}
where $w_3 \ge 0$ is the parameter. Over-retardation disallows the
source-field point coincidence.

Finally, we round off discontinuous features on the string.  For kinks
the transition from one derivative value to another is smoothed.  For
cusps a small patch of the worldsheet near the cusp is excised.
We introduce a parameter $w_4$ that yields the discontinuous
solution when $w_4 \to 0$. Smoothing must be implemented separately
for each loop of interest. In the 2D approach any discontinuity, even if it
were not on the field point's exact light cone, must be smoothed
because all worldsheet points are sampled by the smoothed delta function.

The 2D calculation does not require any special treatment for
boundaries, any special choice of coordinates or any special handling
of the field point. The discontinuities in the source must be
smoothed. We let $\{w_1,w_2,w_3,w_4\} \to 0$ in lockstep together.
We have found that the limit is not impacted if we
set $w_2=0$ (the smoothing of the causal step function)
and $w_3=0$ (the over-retardation) from the beginning. Using the Gaussian
approximation to the delta function and smoothing the discontinuities
on the string are sufficient to regulate the calculation.

\subsection*{1D, over-retarded, smoothed kink or cusp [1DOS]}
\label{sec:1DOS}

The 1D calculations in which the Green function has been integrated
out must handle the field equals source point, the string
discontinuities and coordinate changes along the retarded loop
image.

In the [1DOS] method we use over-retardation and smooth the discontinuities on the
string if they are visible on the field point's exact light cone.
We integrate Eq.~\eqref{eq:F1-convolution} over the
image if coordinate $\zeta^2$ covers the entire image;
we additionally include boundary terms of the type given in
Eq.~\eqref{eq:dh-zeta2-discontinuous-convolution} if multiple
coordinate systems are utilized. Here a boundary term
arises not because of a string discontinuity
but because of the coordinate change. We let $\{w_3,w_4\} \to 0$.

\subsection*{1D, over-retarded, discontinuous kink or cusp [1DO]}
\label{sec:1DO}

As above we use over-retardation for the [1DO] method, but we do not smooth
the kink. We numerically locate the kink and
use boundary terms of the type given in Eq.~\eqref{eq:dh-zeta2-discontinuous-convolution}
to handle both jumps in the string source and coordinate changes.
We can evaluate the force for the cusp as long as the cusp is not on the
light cone (almost all worldsheet points). We let $\{w_3\} \to 0$.

\subsection*{1D, discontinuous kink or cusp [1D]}
\label{sec:1D}

For the [1D] method we use the analytic results \eqref{eq:F-local-ST}
for the source equals field point
and boundary terms of the form given in Eq.~\eqref{eq:dh-zeta2-discontinuous-convolution} for
jumps in the string source and coordinate changes. As above
we can evaluate the force for strings with cusps as long as the cusp is not
on the light cone. This is the computationally most efficient
method and the one we are primarily interested in validating
for future calculations of loops evolving under the effect of
gravitational backreaction.  It does not require any regularization
parameters $w_i$.

There are many related questions that we address using
these techniques. For example, we compare the self force
calculated utilizing different coordinate systems (this
is possible for all the methods, but we
concentrate on the [1D] case). We also consider a limiting
process in which the [1DO] method is used for a field point off the worldsheet, and verify
the correct behavior is recovered as the field point approaches the worldsheet.

\section{Numerical results}
\label{sec:results}

We now apply the derivations of the previous sections to some specific examples, numerically
computing the self-force for a range of nontrivial string configurations that feature kinks, cusps
and self-intersections. We perform several consistency checks in the process:
\begin{enumerate}
  \item For strings with a particularly simple structure we compare against existing calculations in the literature;
  \item For more non-trivial strings we compare different versions of the [1D] integration done with different choices of worldsheet coordinates;
  \item We compare against the smoothed approaches [1DOS] and [1DO] for handling kinks and field point contributions. The field point contribution is recovered by evaluating the integral for the force with a small over-retardation of the retarded time and numerically taking the limit as this over-retardation vanishes. The kink contribution is similarly recovered by introducing a small smoothing to the kink and taking the limit of the smoothing parameter going to zero.
  \item We further compare against our other entirely independent [2D] approach, whereby the force is directly determined from a full 2D integration over the worldsheet, approximating the Dirac $\delta$ distribution in the Green function by a narrow Gaussian.
  \item We verify that the flux of radiation to infinity (as computed using standard frequency domain methods
\cite{2001PhRvD..63f3507A}) appropriately balances the local self-force.
\end{enumerate}

There are infinitely many possible cosmic string loops which satisfy \eqref{eq:conformal-eom}. The
examples which have typically been studied in the literature are those with a low number of
harmonics. As a demonstration of our prescription for computing the self-force, we will compute the
self-force for several of these strings. Our goal is not to be exhaustive, but rather to select a
set of test cases that cover all scenarios (kinks, cusps, self-intersections, and strings without
too much symmetry). In all cases below, we define the worldsheet in terms of two functions
$a^{\alpha} (\zeta^+)$ and $b^{\alpha} (\zeta^-)$, where $\zeta^+ \equiv \tau + \zeta$ and $\zeta^-
\equiv \tau - \zeta$ are null worldsheet coordinates. Then, the spacetime position of the string is
$z^\mu = (1/2)[a^\mu(\zeta^+) + b^\mu(\zeta^-)]$. Throughout the discussion, we will also refer to
the three-vectors $\mathbf{a}$ and $\mathbf{b}$, which are defined to be the spatial projections of
$a^\alpha(\zeta^+)$ and $b^\alpha(\zeta^-)$. Finally, we will specialize to the specific case
$t=\tau$ within the class of conformal gauges.

\subsection{Allen, Casper and Ottewill self-similar string}
\label{sec:ACO}
Allen, Casper and Ottewill (ACO) \cite{Allen:1994bs} identified a particularly simple class of
strings for which the average power radiated is easily calculated in closed form. All strings in
the class have a pair of kinks, each propagating along lines of constant $\zeta^+_{k_1} = 0$ and
$\zeta^+_{k_2} = L/2$, respectively. ACO's motivation was to find the string which radiates most
slowly and is thus most long-lived. Our motivation for studying the ACO string\footnote{We will
study just one case in the class of ACO strings, the one which is simplest and which radiates power
most slowly. ACO call this particular string ``case (1) with $M=1$''. We will simply refer to it as
\emph{the} ACO string.} stems from a different consequence of the simplicity of the ACO solution.
Anderson \cite{Anderson:2005qu} showed that the description of the ACO string worldsheet is
sufficiently simple that it is possible to determine the self-force analytically.\footnote{In fact,
in \cite{Anderson:2008wa} Anderson was able to go one step further and analytically
self-consistently evolve the string under the influence of gravitational backreaction.} This
provides a valuable reference point against which we can check our numerical approach.

\begin{figure}[H]
\centering
\includegraphics[width=0.6\linewidth]{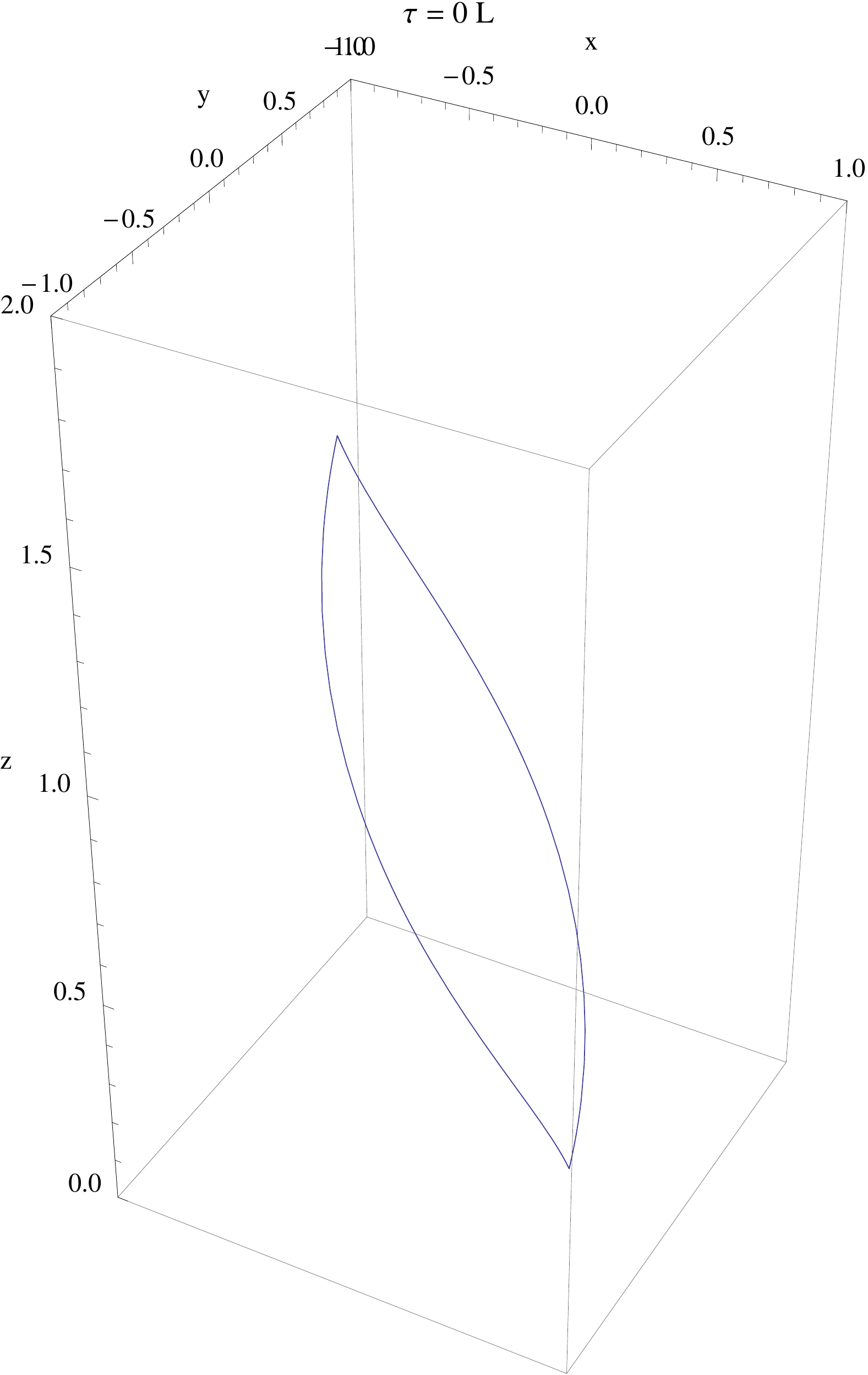}
\caption{
\label{fig:ACOloopspacetime}
Snapshot of the ACO string loop configuration in spacetime at time $\tau=0$. At later times the
configuration can be obtained by a rigid rotation about the $z$-axis.}
~\\
\includegraphics[width=0.6\linewidth]{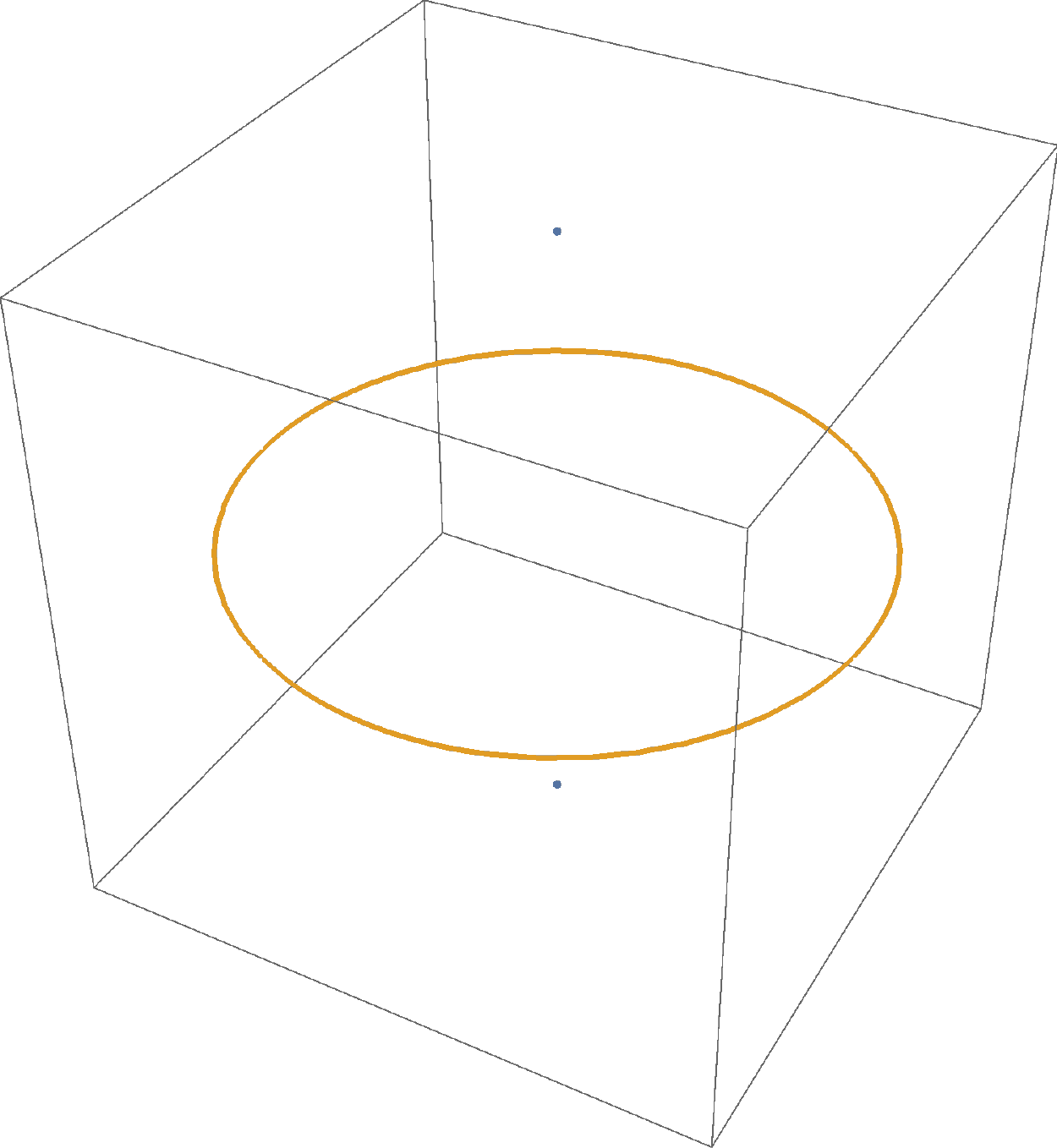}
\caption{
\label{fig:ACO-tangent-sphere}
Tangent sphere representation of the ACO string loop configuration with $\mathbf{a}'(\zeta^+)$ denoted by the two blue dots and $\mathbf{b}'(\zeta^-)$ by the orange circle.}
\end{figure}

The ACO string worldsheet is given in Cartesian coordinates by
\begin{align}
  a^{\alpha} (\zeta^+) =& A [\zeta^+/A, 0, 0, |\zeta^+|],
\nonumber \\
  b^{\alpha} (\zeta^-) =& A [\zeta^- /A, \cos (\zeta^- / A), \sin (\zeta^- / A), 0],
\end{align}
where $A \equiv \tfrac{L}{2\pi}$ and $L$ is the length of the string. For $\zeta^+<-\tfrac{L}{2}$
or $\zeta^+>\tfrac{L}{2}$ the periodic extension of $a^z$ is used, i.e. $a^z$ is the triangle
function centered about the origin and with period $L$. The ACO string can be visualised as shown
in Fig.~\ref{fig:ACOloopspacetime}; its evolution is a rigid rotation of this shape about the
$z$-axis.\footnote{In \cite{Anderson:2008wa} Anderson showed that this shape is preserved when
backreaction is taken into account, in which case the string evolves (shrinks) self-similarly.} We
characterize the ACO string in terms of its tangent-sphere representation, as shown in
Fig.~\ref{fig:ACO-tangent-sphere}.

Adopting conformal gauge to first order, Anderson \cite{Anderson:2005qu} was able to compute the self-force (which, in the conformal gauge case is defined to be the right-hand side of Eq.~\eqref{eq:simple}) by analytically determining the first-order metric perturbation generated by an ACO string. Factoring out the rigid rotation using the matrix
\begin{equation}
M^\alpha{}_\beta = \left(
\begin{array}{cccc}
 1 & 0 & 0 & 0 \\
 0 & \cos (2 \pi  \zeta^-) & \sin (2 \pi  \zeta^-) & 0 \\
 0 & -\sin (2 \pi  \zeta^-) & \cos (2 \pi  \zeta^-) & 0 \\
 0 & 0 & 0 & 1 \\
\end{array}
\right),
\end{equation}
the conformal gauge self-force in a co-rotating frame is given by
$f^\mu = M^\mu{}_\alpha {\cal F}_{\rm conf}^\alpha$, where $f^\mu = [f^t(\zeta^+),
f^L(\zeta^+), f^N(\zeta^+), \sgn(\zeta^+) f^t(\zeta^+)]$. We can interpret $f^N$ and $f^L$ as
the normal and longitudinal components of the force in the $x$--$y$ plane, respectively. Note that
this factorized
form is quite convenient as the dependence on $\zeta^+$ is entirely contained within $f^\mu$,
while the dependence on $\zeta^-$ is entirely in $M^\chi{}_\mu$.

\begin{figure}[H]
  \center
  \includegraphics[width=8cm]{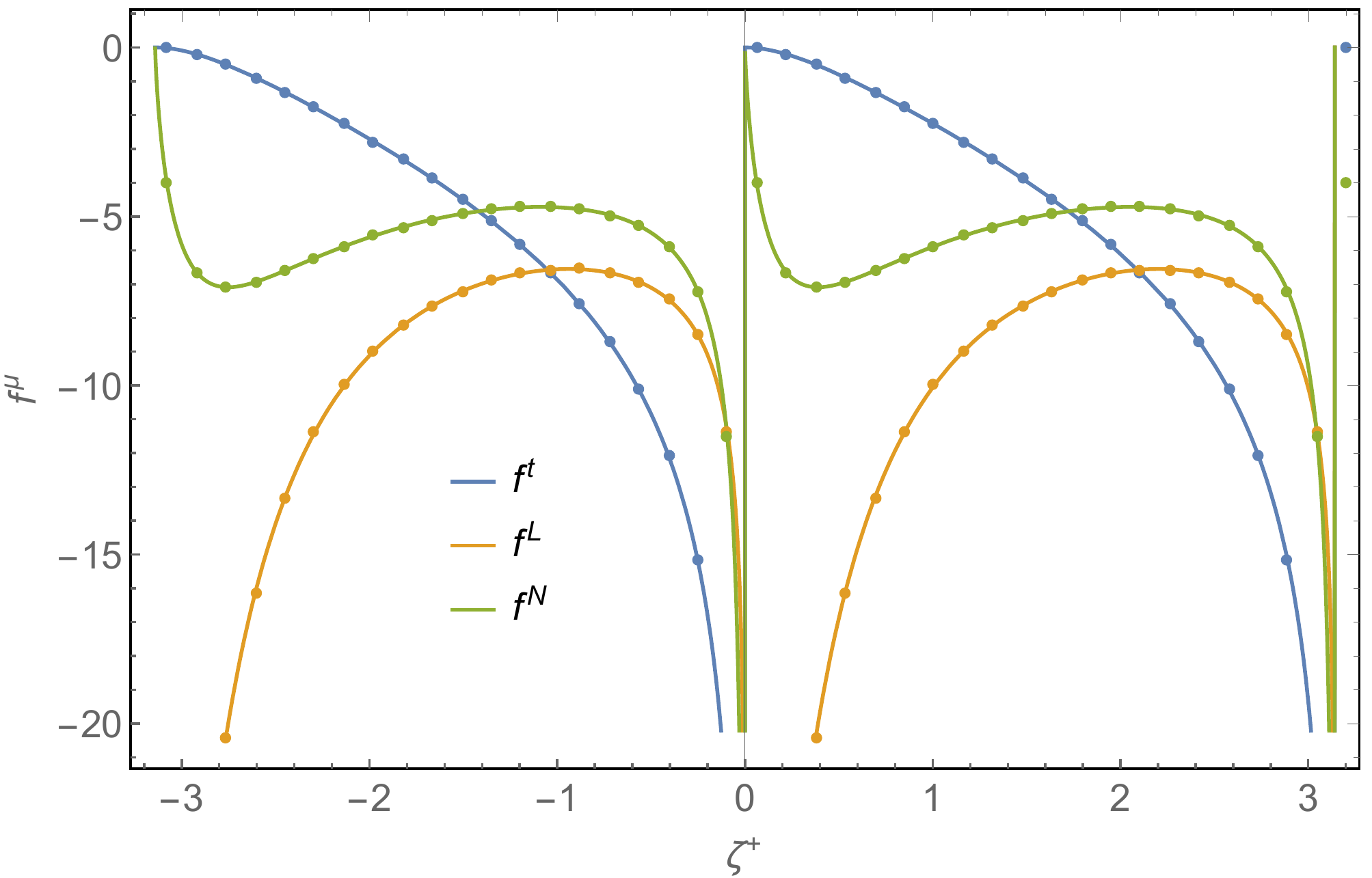}
  \caption{Co-rotating self-force for the ACO string.}
\label{fig:ACOForce}
\end{figure}
As an important consistency check on our work, we have verified that our numerical approach exactly
reproduces the analytic result derived by Anderson. Figure \ref{fig:ACOForce} shows the factored
components of the force as a function of $\zeta^+$, with Anderson's expressions plotted as solid
lines and our numerical values (computed using Eq.~\eqref{eq:F1-convolution} plus boundary terms of
the type given in Eq.~\eqref{eq:dh-zeta2-discontinuous-convolution} at the kinks and
Eq.~\eqref{eq:F-local-ST} for the field point contribution) shown as dots.

One interesting feature is the divergence of the force components as a kink is approached. Although one may be concerned about the physical implications of this divergence, for the ACO string it turns out that it is a spurious gauge artifact, and that the string worldsheet itself only ever picks up a small perturbation from the self-force. The simplicity of the ACO solution makes it straightforward to see this explicitly: as shown by Anderson \cite{Anderson:2005qu}, the explicit form of the divergence near the kink can be written as
\begin{align}
f^t \approx&\,\{-32 (\tfrac{1}{6}\pi^2)^{1/3} \mu /|\zeta^+|^{1/3}, -128 \pi^2 \mu (\zeta^+)^2\}, \nonumber \\
f^L \approx&\,32 \pi \mu \ln |\zeta^+| \{\tfrac{1}{3}, 1\}, \nonumber \\
f^N \approx&\,\{-32 (\tfrac{1}{6}\pi^2)^{1/3} \mu /|\zeta^+|^{1/3}, 128 \pi^2 \mu \zeta^+ \ln |\zeta^+|\},
\end{align}
depending on whether the limit $\zeta^+ \to 0$ is taken from the left or the right. Anderson goes on to show that integrating up the equation of motion, the physical (non-gauge) displacement of the string due to this divergent force is finite.

\subsection{Kibble and Turok strings with cusps and self-intersections}
A simple family of string loop solutions of the zeroth order equations
of motion was written down by Kibble
and Turok~\cite{Kibble:1982cb,Turok:1984cn}. The gravitational
radiation of representative examples was
calculated by Vachaspati and Vilenkin~\cite{Vachaspati:1984gt}.
We will refer to the family as KT strings. The family is described by the general form
\begin{align}
  a^{\alpha} (\zeta^+) =& A \Big[\zeta^+ / A,
     (1-\alpha) \sin (\zeta^+ / A) + \tfrac{\alpha}{3} \sin (3 \zeta^+/A),
\nonumber \\ & \quad
     (\alpha-1) \cos (\zeta^+ / A) - \tfrac{\alpha}{3} \cos (3 \zeta^+/A),
\nonumber \\ & \quad
     - 2 \sqrt{\alpha(1-\alpha)} \cos (\zeta^+ / A)\Big],
\nonumber \\
  b^{\alpha} (\zeta^-) =& A \Big[\zeta^- / A,
    \sin (\zeta^- / A),
\nonumber \\ & \quad
    -\cos \phi \cos (\zeta^- / A),
    -\sin \phi \cos (\zeta^- / A)\Big],
\end{align}
where $0 \le \alpha \le 1$ and $-\pi \le \phi \le \pi$ are two
parameters.

\begin{figure}[H]
\centering
\includegraphics[width=0.9\linewidth]{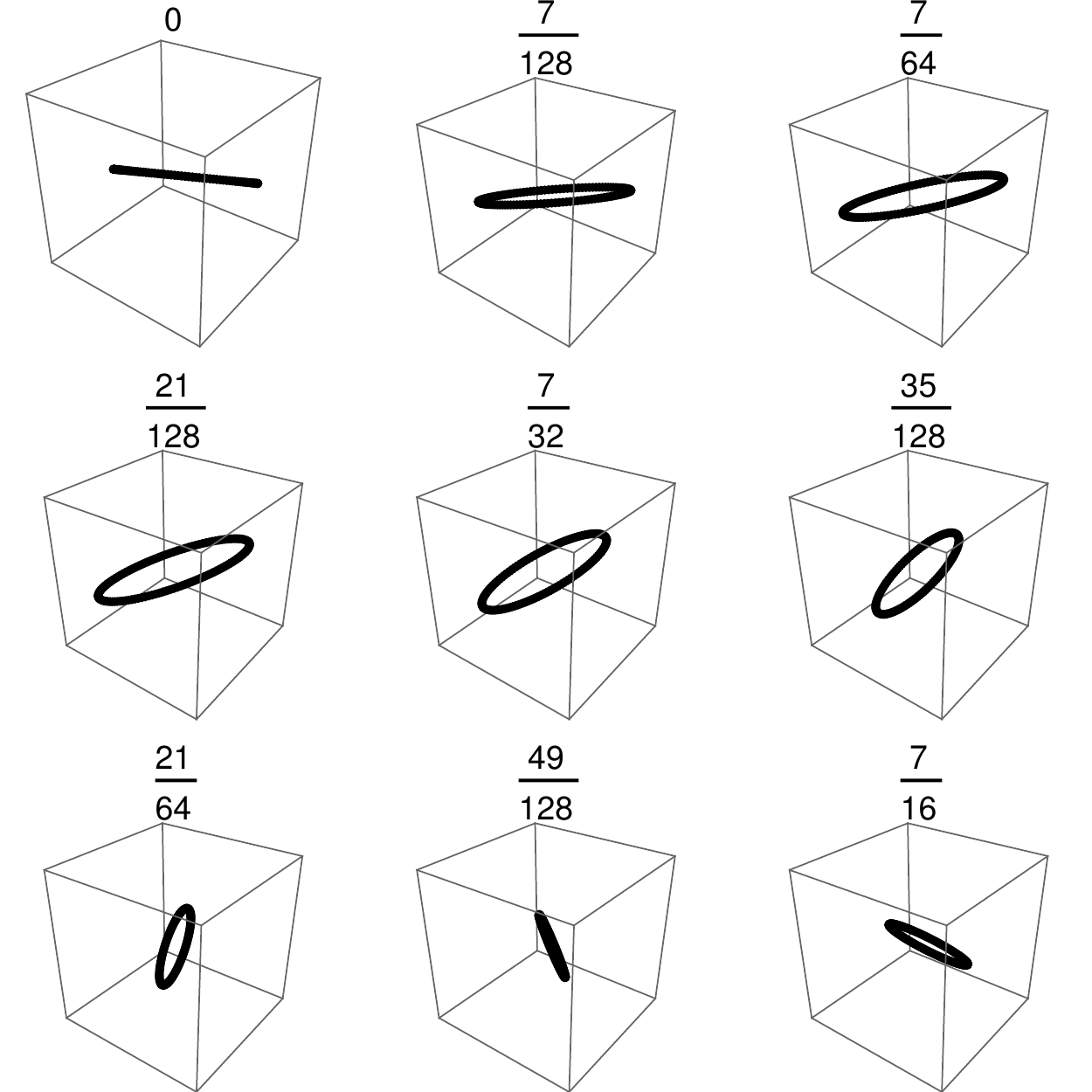}
\caption{\label{fig:VVloopspacetime} Snapshots of the KT string loop
  ($\alpha=0$ and $\phi=\pi/6$)
  configuration in spacetime, each labeled by time in units of $L$. All
  the boxes have the same size axes, $-1$ to $1$ for $L=2 \pi$, and
  fixed orientation. }
\end{figure}

We first focus on the case $\alpha=0$ and $\phi=\pi/6$
($N=M=1$ Burden loops \cite{Burden:1985md}).
Nine snapshots of the spacetime configuration of the loop
are shown in Fig.~\ref{fig:VVloopspacetime}.
The loop generally possesses an elliptical shape. It tumbles in space
while stretching and contracting. Twice per period it forms a
degenerate, line-like shape with a pair of cusps on opposite sides.
The tangent sphere representation is particularly simple:
there are two continuous great circles that cross
at $\tau + n \pi = \zeta + m \pi = 0$ for any integers $n$ and $m$.
Each crossing gives rise to a cusp and to a spacelike line
of string overlap in the center of momentum frame.
These two effects make the calculation of the self-force
particularly challenging.

We first compute the self-force at two points on the string (which we denote Case I and Case II):
\begin{align}
  (\tau, \zeta) &= (32 \pi/50,13 \pi/50)
  {\rm \ \ \ Case \ \ I} \\
  (\tau, \zeta) &\simeq (0.42,\pi/5)
  {\rm \ \ \ Case \ \ II}
\end{align}
For Case I, the field point is such that no cusp is present on the retarded image of the string.
%\noteme{Is the spacelike line ever visible?}
Since the left and right moving modes are continuous the loop stress-energy source is completely smooth except at
the field point itself.\footnote{In a patch of the world sheet that extends $\pm \pi$ about the
field point the cusps at $(\tau,\zeta)=(0,0)$ and $(0,\pi)$ are potentially visible for a
causal off-shell Green function.} The Case I results calculated by the [1D] method described in Sec.~\ref{sec:dh} are given in the first part of Table \ref{tab:VVanswers}. In this case, there are two important contributions to $F_1^\rho$:
the row labeled $\int$ is the integral
contribution arising from the $1D$ integral over the smooth worldsheet using Eq.~\eqref{eq:F1-convolution}; and
$\delta$ is the contribution from the field point obtained using Eq.~\eqref{eq:F-local-ST}. The total is $F_1 = \int + \delta$.
\begin{table}[H]
  \begin{center}
\begin{tabular}{|cc|cccc|}
  \hline
 Case \ \  & Force & \multicolumn{4}{c|}{contravariant spacetime components}\\
  & & $t$ & $x$ & $y$ & $z$ \\
  \hline
   I & $\int$     &  9.28612  & -4.96366 & 14.7739  & -1.68376 \\
     & $\delta$   & -0.680917 & 1.09474  & 1.16578  & -4.35077 \\
     & $F_1$      &  8.6052   & -3.86891 & 15.9397  & -6.03453 \\
     & $F_2$      &-12.181    & 4.56246  & -25.3768 & 14.1391 \\
%  I & $\int$     & $4.03438$ & $2.60165$ & $3.30902$ & $13.5473$\\
%    & $\delta$   & $11.5879$ & $0.779709$ & $12.1544$ & $17.4531$ \\
%    & $F_1$      & $15.6223$ & $3.38135$ & $15.4634$ & $31.0004$ \\
%    & $F_2$      & $-107.069$ & $2.24908$ & $-116.046$ & $-130.953$ \\
  \hline
   II & $\int$     & 44.5678 & 49.5374 & 22.8974 & -1.99924 \\
      & $\delta$   & 1.7937  &  1.5892 &  1.1546 & -4.30897 \\
      & $F_1$      & 46.3615 & 51.1266 & 24.052  & -6.30821 \\
      & $F_2$      & -75.6739 & -82.35   & -39.8936 & 21.8117  \\
%  II & $\int$     & $ -5.73906$ & $-8.29154$  & $-10.3183$ & $-3.53213$ \\
%     & $\delta$   & $-0.761867$ & $-0.217243$ & $ 2.42954$ & $-2.24653$ \\
%     & $F_1$      & $ -6.50093$ & $-8.50878$  & $-7.88876$ & $-5.77866$ \\
%     & $F_2$      & $   6.8884$ & $5.84377$   & $-5.28273$ & $12.5058$ \\
  \hline
\end{tabular}
\end{center}
  \caption{Self-force at two points on the KT string
($\alpha=0$ and $\phi=\pi/6$)
calculated by the $1$D method.}
\label{tab:VVanswers}
\end{table}

For Case II, we have carefully chosen a field point such that the cusp at $(\tau,\zeta)=(0,0)$
lies on the retarded string image. Numerical results for this case (which were again obtained using
the [1D] method) are given in the second part of Table \ref{tab:VVanswers}.

One notable feature of these numerical results is that the field point
contribution is comparable in magnitude to the contribution from the
integral. As such, this case provides a valuable and stringent test of
our derivation of the expression for the field point contribution. By comparing to a different
approach which doesn't rely on these terms we may distinguish between
$\int$ and $F_1$. The [2D] integration method (described in detail in
Appendix \ref{sec:2D}) provides just such a comparison. In Table
\ref{tab:summarytab} we tabulate the results of the [2D] integration
method and compare against the $1$D results for Case I in Table
\ref{tab:VVanswers}. This comparison unambiguously confirms that the
field point contribution is essential. The agreement provides a strong validation of our
formalism. Appendix \ref{sec:2D} includes analogous [2D] results for Case II. These are in equally good agreement so we omit additional discussion of
the comparison.
\begin{table}[H]
  \begin{center}
\begin{tabular}{|cccc|}
  \hline
  Force & Extrapolated Force & Extr. error & 2D-1D \\
  \hline
  $F_1^t$ &  $8.60882$ &  $0.0020$   & $-0.0036$ \\
  $F_1^x$ & $-3.87143$ & $-0.0016$  &  $0.0025$ \\
  $F_1^y$ & $15.9437$  &  $0.0013$  & $-0.0040$ \\
  $F_1^z$ & $-6.0318$  &  $0.0030$  & $-0.0027$ \\
  \hline
  $F_2^t$ & $-12.181$   &  $3.6 \times 10^{-5}$    &  $1.0 \times 10^{-5}$ \\
  $F_2^x$ & $4.56246$   & $-1.5 \times 10^{-6}$ & $-9.6 \times 10^{-7}$ \\
  $F_2^y$ & $-25.3768$  & $-1.6 \times 10^{-5}$ & $2.7 \times 10^{-5}$ \\
  $F_2^z$ & $14.1391$   & $-7.5 \times 10^{-6}$ & $1.5 \times 10^{-5}$ \\
\hline
\end{tabular}
\end{center}
\caption{Extrapolated self-force calculated by the [2D] method in Case I of the KT string ($\alpha=0$ and $\phi=\pi/6$).}
\label{tab:summarytab}
\end{table}

We now proceed to compute the self-force at \emph{all} points on the worldsheet. The results are shown in Figs.~\ref{fig:Force-components-VV} and \ref{fig:Force-VV}. Unlike the ACO case, the extra complexity in the KT solution means that there is no simple factorization of the force into a piece which only depends on $\zeta^+$ and another piece which depends on $\zeta^-$. As such, the self-force for the KT string is presented as a $2D$ surface plot, showing the force contributions to $F^\mu$ ($\log_{10}$ of the absolute value of a contribution, color coded by sign) at all\footnote{In all of our plots we show the segment of the worldsheet defined by $\tau \in [0, L/2]$, $\zeta \in [-L/2,L/2]$. This covers the entire set of unique points on the worldsheet; other values can be obtained by periodically extending in the $\tau$ and/or $\zeta$ direction.} points on the two-dimensional worldsheet. The green and red curves trace the advanced images of the cusps on the loop; each point on these curves has the cusp at $(\tau, \zeta) = (0,0)$ (red) or at
$(\tau, \zeta) = (0,L/2)$ (green) on its past light cone. The gross
variation of the self-force depends on the product of two factors
which have simple physical origins.
First, the loop's line-like structure, periodically
formed at $\tau=0$ and $L/2$, creates a ridge spanning all $\zeta$
at these particular times. Second, at any given time
the points along the
string loop which are least contracted and have the largest $\sqrt{-\gamma}$
occur at $\zeta = \pm L/4$. These produce a trough or minimum
in the force at $\zeta = \pm L/4$. The product of these two
factors yields the
egg-crate-like symmetry in the force with the cusps at the
corners.

\begin{widetext}
~
\begin{figure}[H]
\centering
\includegraphics[width=0.9\linewidth]{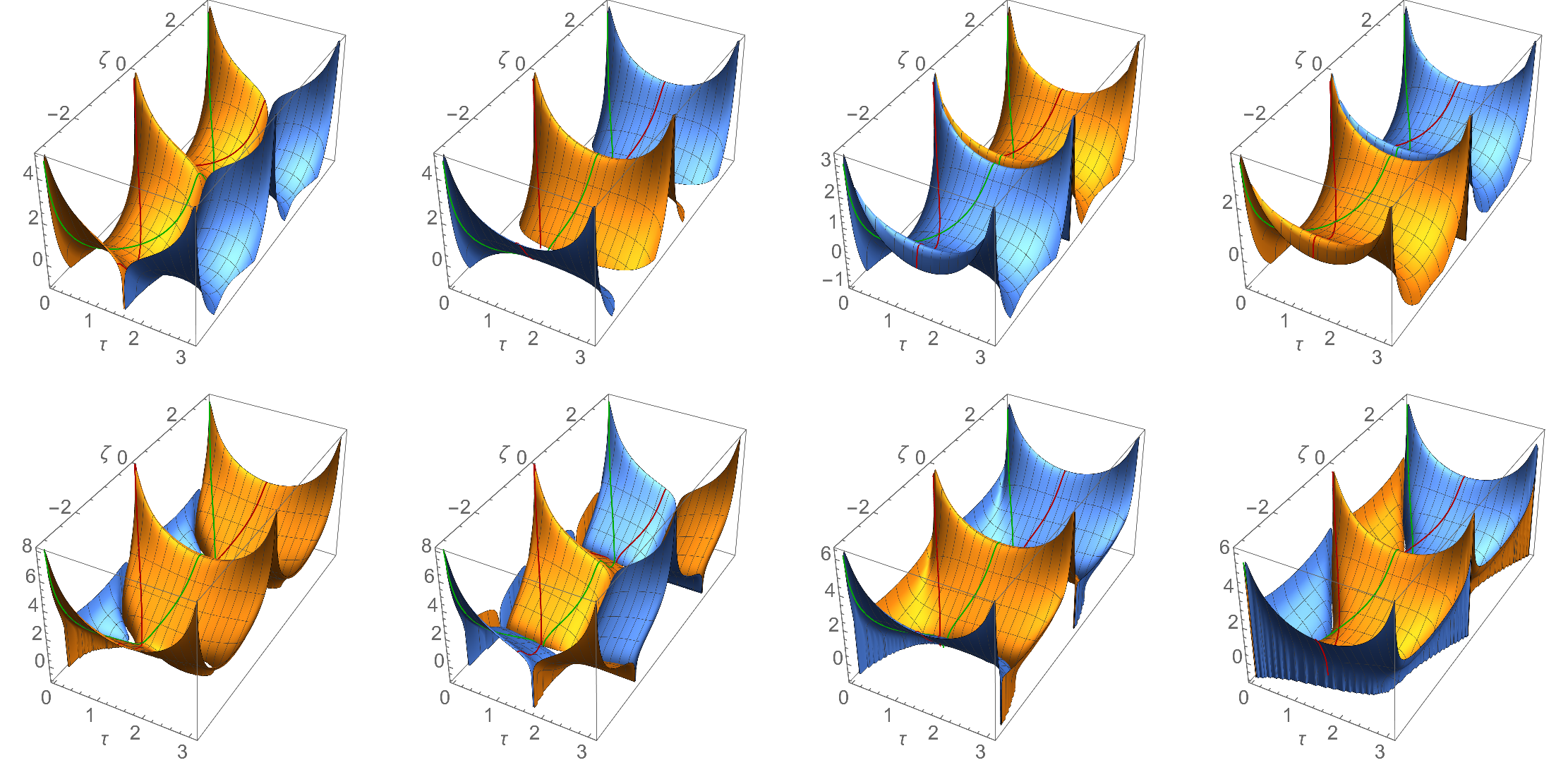}
\caption{\label{fig:Force-components-VV}
Contributions to $F_1^\mu$ for the KT string ($\alpha=0$ and $\phi=\pi/6$) when computed using the 1D integration method with integration with respect to $\zeta$. Each sub-figure shows the relevant contribution to the force at all points on the string in the region $\tau \in (0,L/2)$, $\zeta \in (-L/2,L/2)$; all other points can be obtained from the standard periodic extension of the string. Each column corresponds to a different component of the force: $F_1^t$, $F_1^x$, $F_1^y$, and $F_1^z$. The rows correspond to the contributions from: (i) the field point; and (ii) the integral over $\zeta$ (ignoring distributional contributions at the field point). For the purposes of the plots, we have set the string tension, $\mu$, and Newton's constant, $G$ equal to one; other values simply introduce an overall scaling. Note that we have used a logarithmic scale and denoted positive (negative) values by coloring the plot orange (blue).
}
\end{figure}
\begin{figure}[H]
\centering
\includegraphics[width=0.9\linewidth]{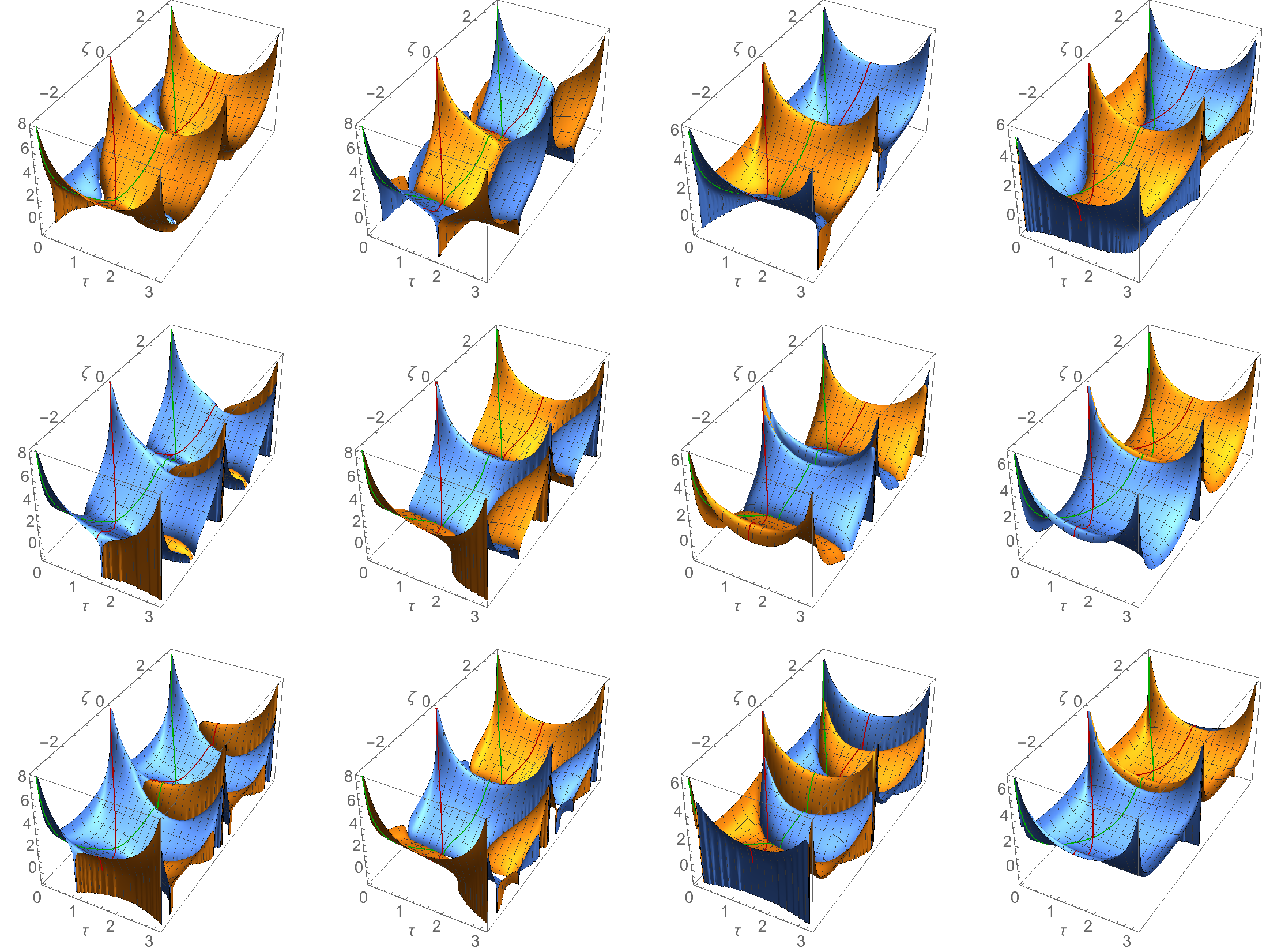}
\caption{\label{fig:Force-VV}
  The two pieces of the self-force, $F_1^\mu$ (row 1) and $F_2^\mu$ (row 2),
  and the total self-force $F^{\mu}$ (row 3) for the KT string ($\alpha=0$ and $\phi=\pi/6$) as a function of position on the string. The $F_1^\mu$ part can be obtained by summing the two rows in Fig.~\ref{fig:Force-components-VV}. For the purposes of the plots, we have set the string tension, $\mu$, and Newton's constant, $G$ equal to one; other values simply introduce an overall scaling. Note that we have used a logarithmic scale and denoted positive (negative) values by coloring the plot orange (blue).
}
\end{figure}
\end{widetext}

These plots show several interesting features:
\begin{enumerate}
\item The self-force is finite at almost all points on the worldsheet, the notable exceptions being the location of the two cusps, where it appears to diverge.
\item The two contributions to $F_1^\mu$ (coming from the integral over the smooth worldsheet and from the field point) are comparable in magnitude. It is therefore crucial that both contributions be included.
\item The contributions from $F_1^\mu$ and $F_2^\mu$ are both comparable in magnitude and both exhibit the same qualitative behavior in terms of divergence at the cusp and finiteness elsewhere.
\end{enumerate}

Although this case provides a good check of the general
methodology it involves special features
that can be traced to the self-intersections. In the next
section we modify the parameter choice to avoid self-intersections.

\subsection{KT strings with cusps without self-intersections}
\label{sec:VV-nonSI}

\begin{figure}[H]
\centering
\includegraphics[width=0.9\linewidth]{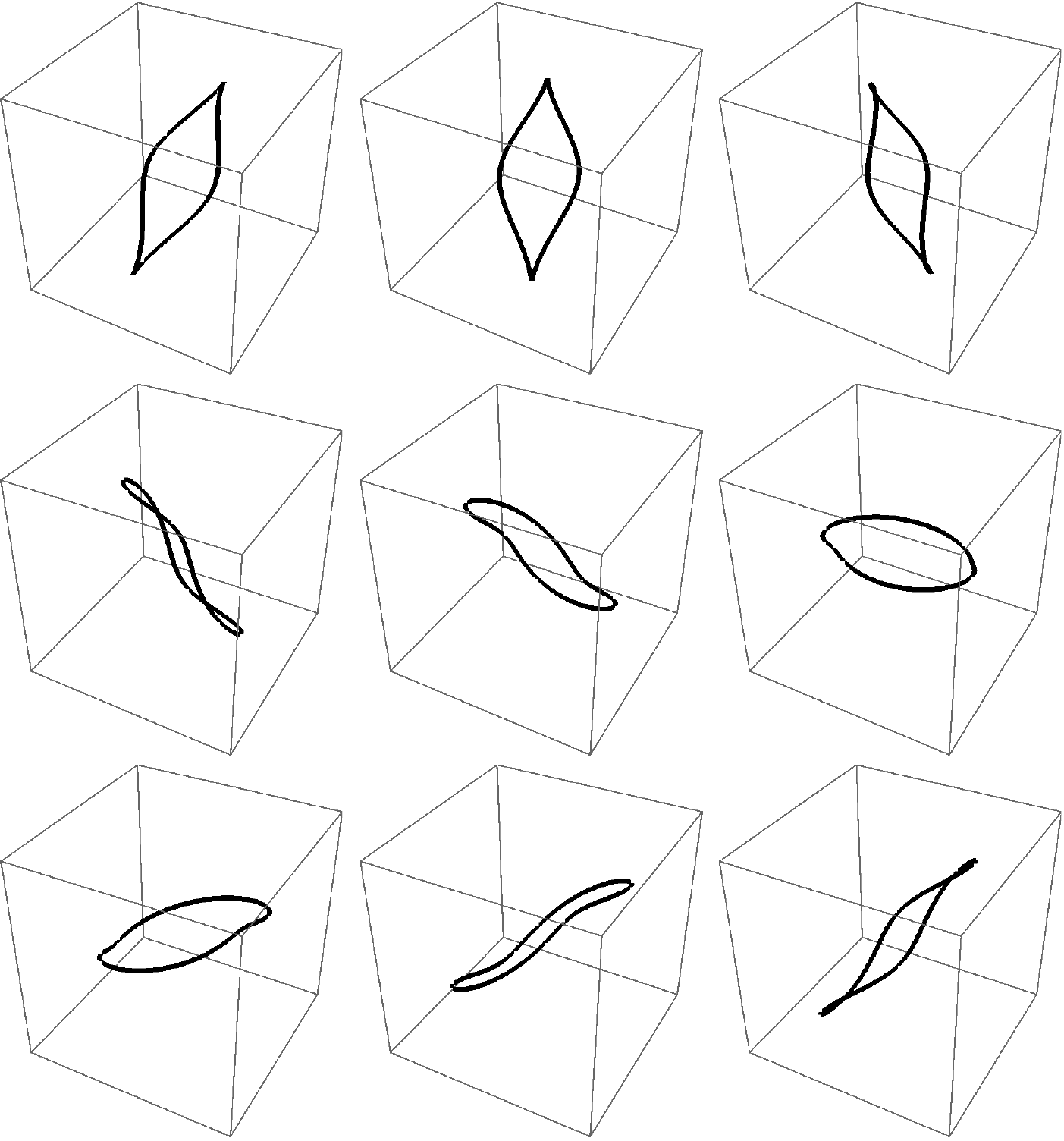}
\caption{\label{fig:VVloopNonSIspacetime}
  Snapshots of the KT string loop ($\alpha=1/2$ and $\phi=0$) configuration in spacetime, each
  labeled by time in units of $L$. All the boxes have the same size axes, $-1$ to $1$ for $L=2 \pi$,
  and fixed orientation.}
\end{figure}

Next we consider a KT string with parameter values $\alpha=1/2$ and $\phi=0$. Snapshots of this
loop are shown in Fig.~\ref{fig:VVloopNonSIspacetime}. The loop rotates about the z-axis and forms
cusps transiently at $(\tau,\zeta)=(0,0)$ and $(0,L/2)$. There are no self-intersections except
infinitesimally close to the cusp itself.

\begin{widetext}
~
\begin{figure}[H]
\centering
\includegraphics[width=0.9\linewidth]{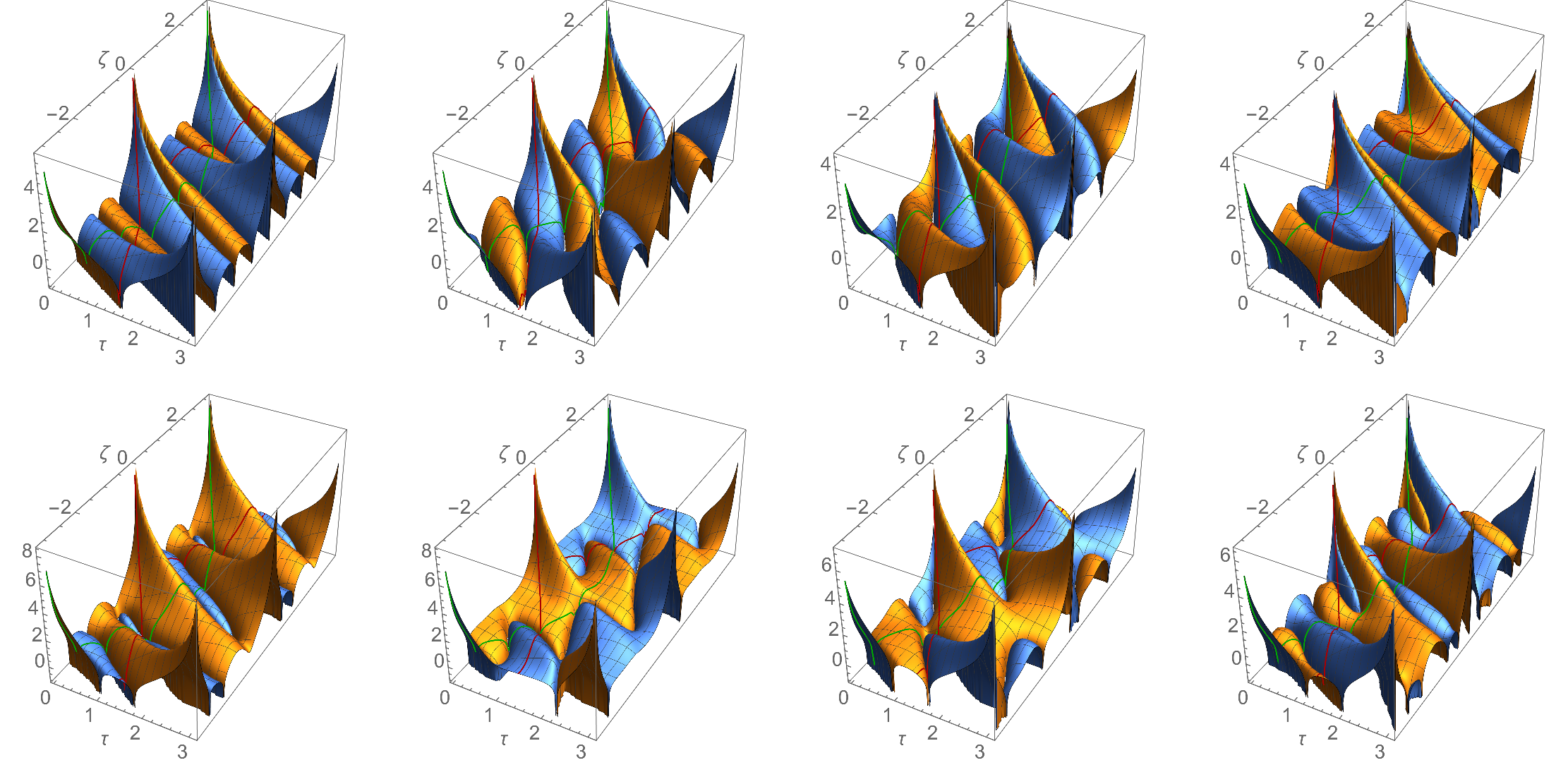}
\caption{\label{fig:Force-components-VVNonSIA}
Contributions to $F_1^\mu$ for the KT string ($\alpha=1/2$ and $\phi=0$) as otherwise described in Fig.~\ref{fig:Force-components-VV}.
}
\end{figure}
\begin{figure}[H]
\centering
\includegraphics[width=0.9\linewidth]{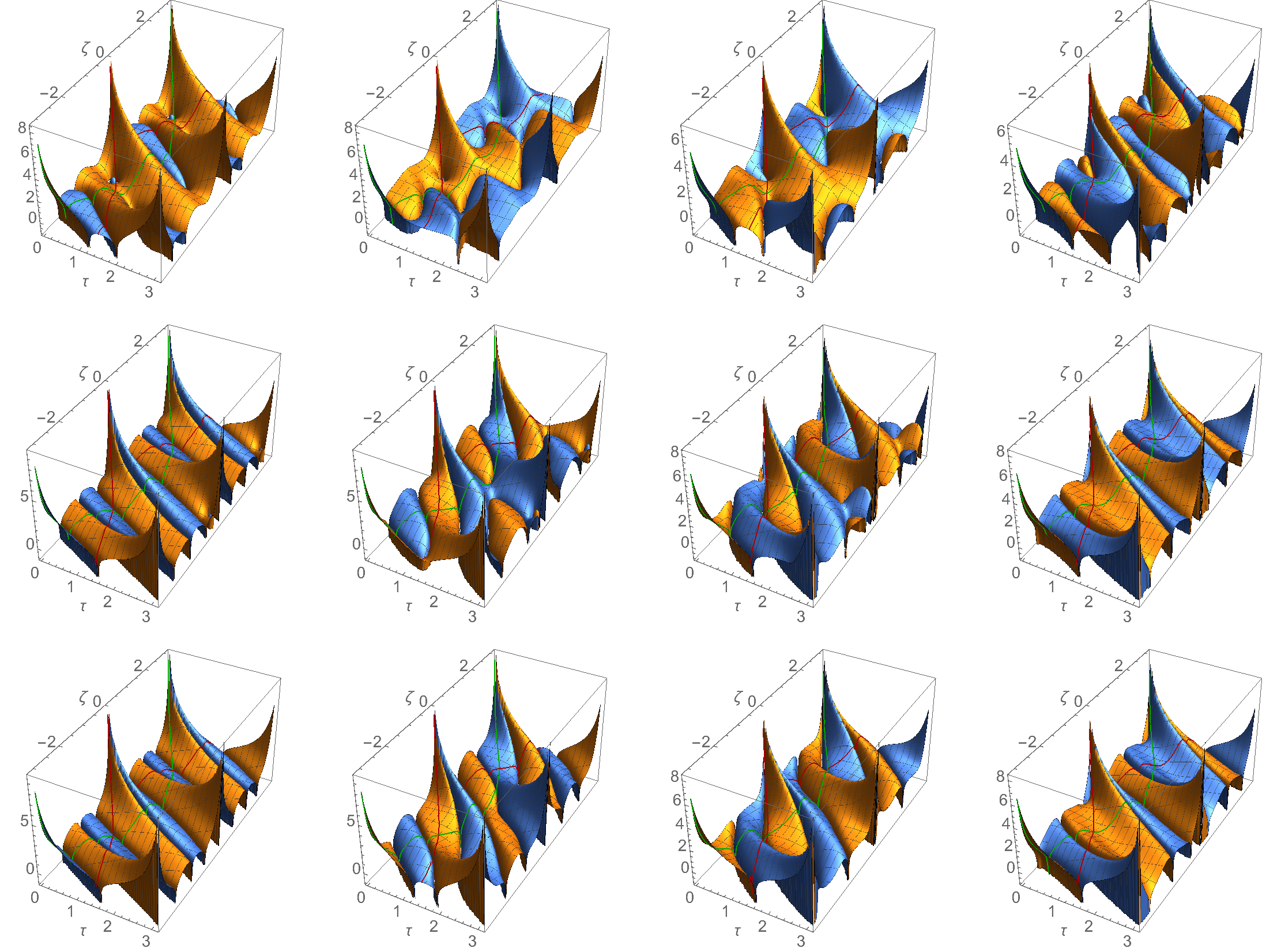}
\caption{\label{fig:Force-VVNonSIA}
  The two pieces of the self-force, $F_1^\mu$ (row 1) and $F_2^\mu$ (row 2), and the total
        self-force $F^{\mu}$ (row 3) for the KT string ($\alpha=1/2$ and $\phi=0$)
        as otherwise described in Fig.~\ref{fig:Force-VV}.
}
\end{figure}
\end{widetext}
Figs.~\ref{fig:Force-components-VVNonSIA} and \ref{fig:Force-VVNonSIA} show the self-force at all
points on the worldsheet of this KT string. These are analogous to the plots for the
self-intersecting KT string shown Figs.~\ref{fig:Force-components-VV} and \ref{fig:Force-VV}. The
peaks clearly show the cusp locations and the diagonal striping is related to the overall sense of
rotation of the loop. The spacelike line of overlap and the egg-crate symmetry seen in the previous
KT case are now absent.

This non-intersecting case allows for a detailed analysis of the behavior of the total backreaction
force in the vicinity of the cusp at $(\tau,\zeta)=(0,0)$. At times close to cusp formation the
tip's position (the string coordinate at fixed $\zeta=0$) is
\begin{eqnarray}
  {\mathbf{z}}^i & \sim & \{0,-0.83,-0.5\} + \{1,0,0\} \tau + \nonumber\\
  & &   \{0,1.5,0.5\} \frac{\tau^2}{2} + \{-3,0,0\} \frac{\tau^3}{6} + \nonumber\\
  & &   \{0,-7.5,-0.5\} \frac{\tau^4}{24} + \dots
\end{eqnarray}
The velocity lies in the x-direction and the acceleration in the y- and z-directions. Conversely,
the velocities in the y- and z-directions and the acceleration in the x-direction vanish. On
physical grounds we expect the y- and z-accelerations to source transverse gravitational waves and
the relativistic motion in the x-direction to lead to strong beaming.

The driving force $F^\alpha$ which enters the string loop's equation of motion, Eq.~\eqref{eq:final},
encodes the fully non-local, self-interacting gravitational dynamics. If we were to adopt the
conformal gauge at first order then ${\cal F}_{\rm conf}^\alpha$ would naturally appear as the driving
force in the equation of motion. We will not restrict ourselves to that choice for much of the
discussion in this section. We will show, however, that many of the features of the full worldsheet
variation of $F_1^\alpha$ can be understood based on the observed properties of the formally
defined quantity ${\cal F}_{\rm conf}^\alpha$ (which may be defined in any gauge; only its interpretation
as the driving force is restricted to conformal gauge). We will be explicit whenever our statements
demand the specification of the conformal gauge.

\begin{widetext}
        ~
\begin{figure}[H]
\centering
\includegraphics[width=0.45\linewidth]{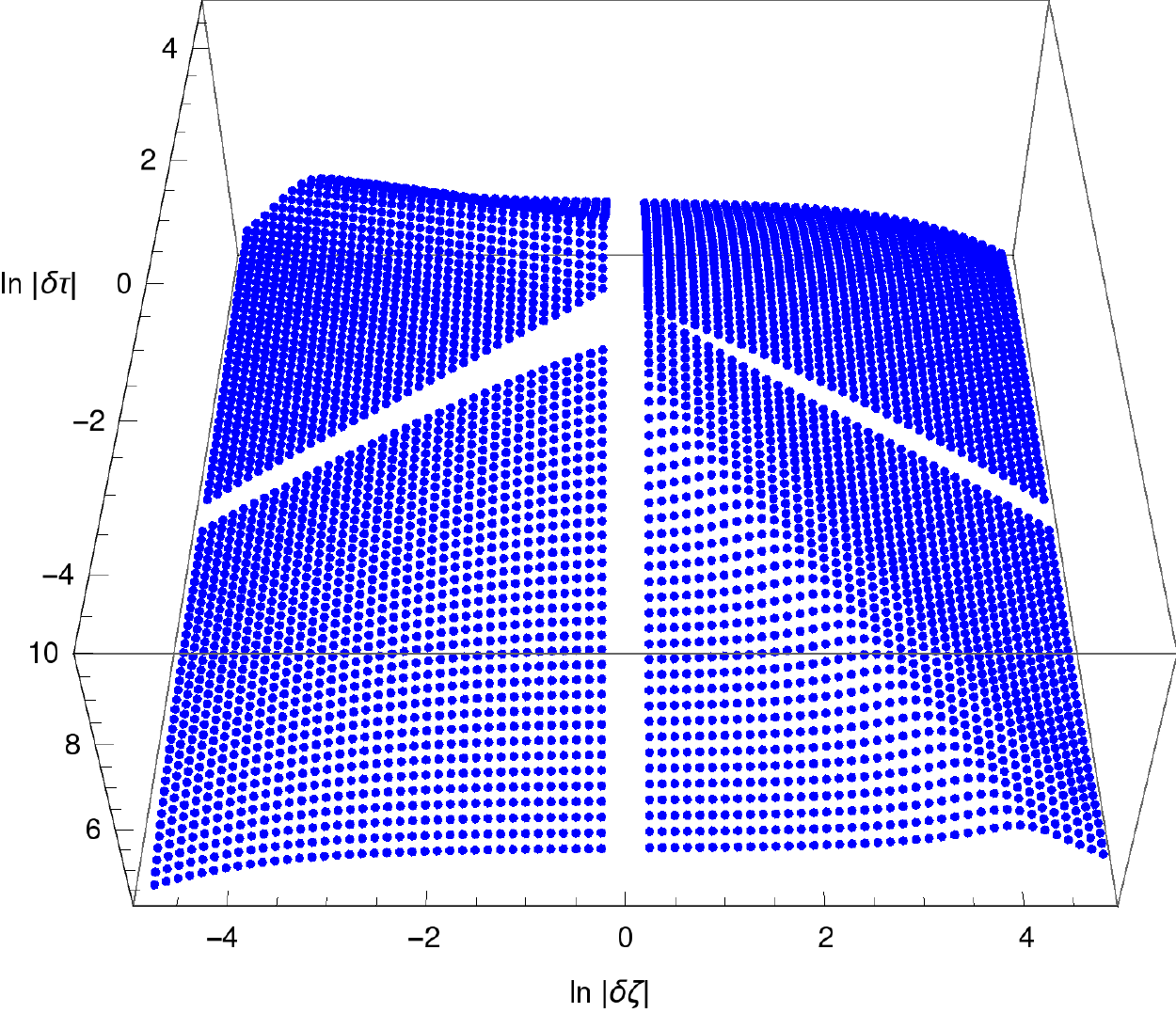}
\includegraphics[width=0.45\linewidth]{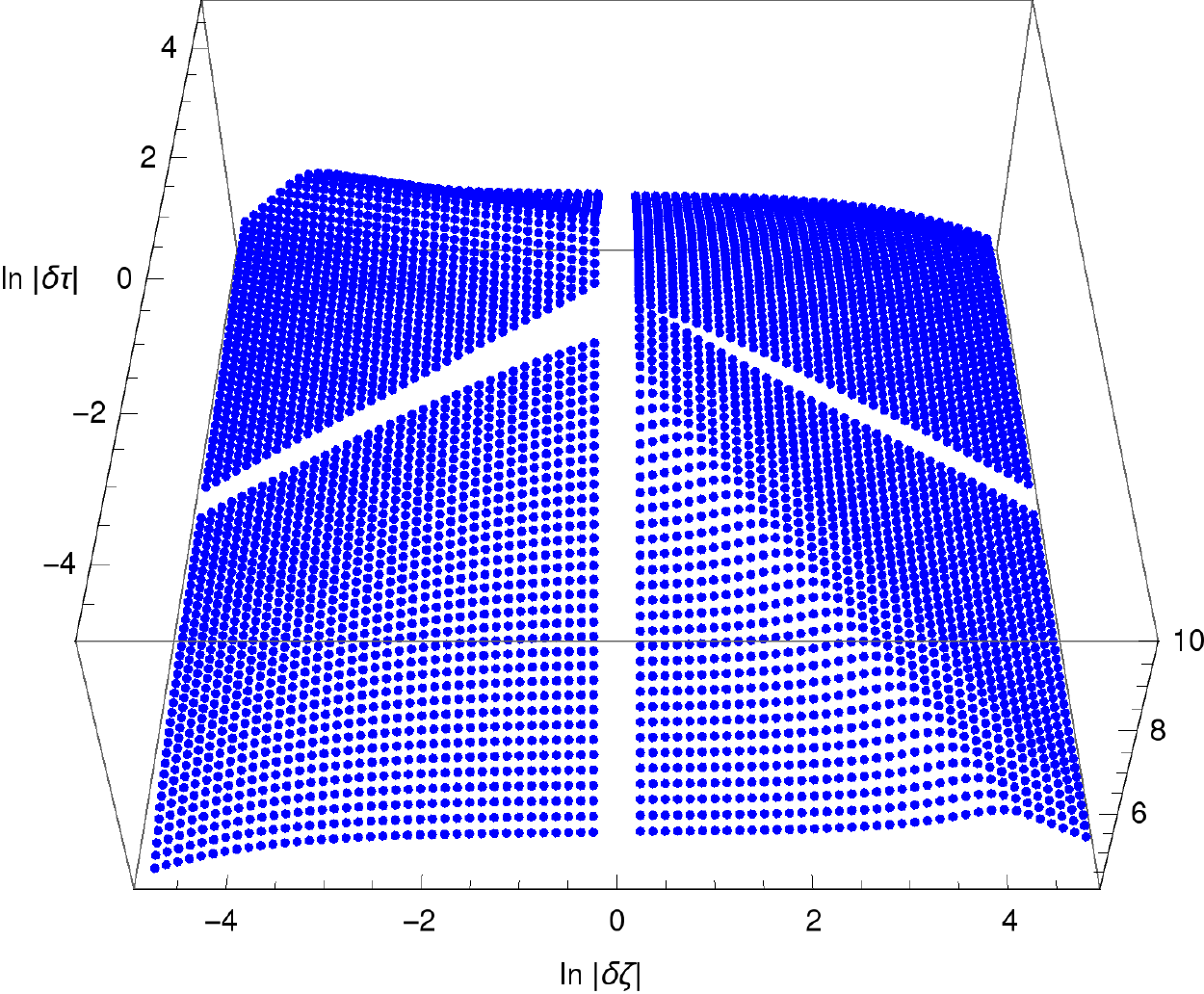}
\includegraphics[width=0.45\linewidth]{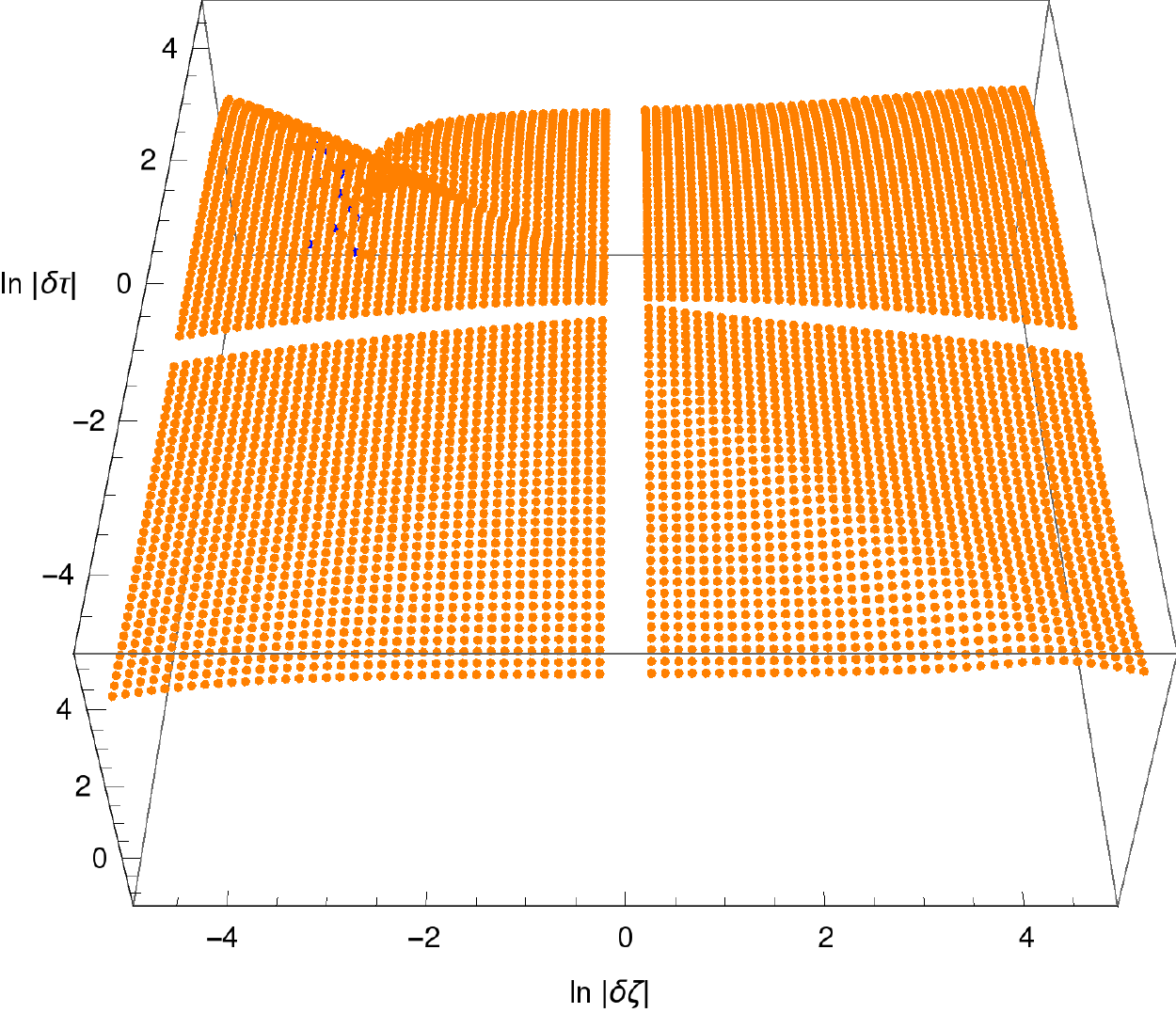}
\includegraphics[width=0.45\linewidth]{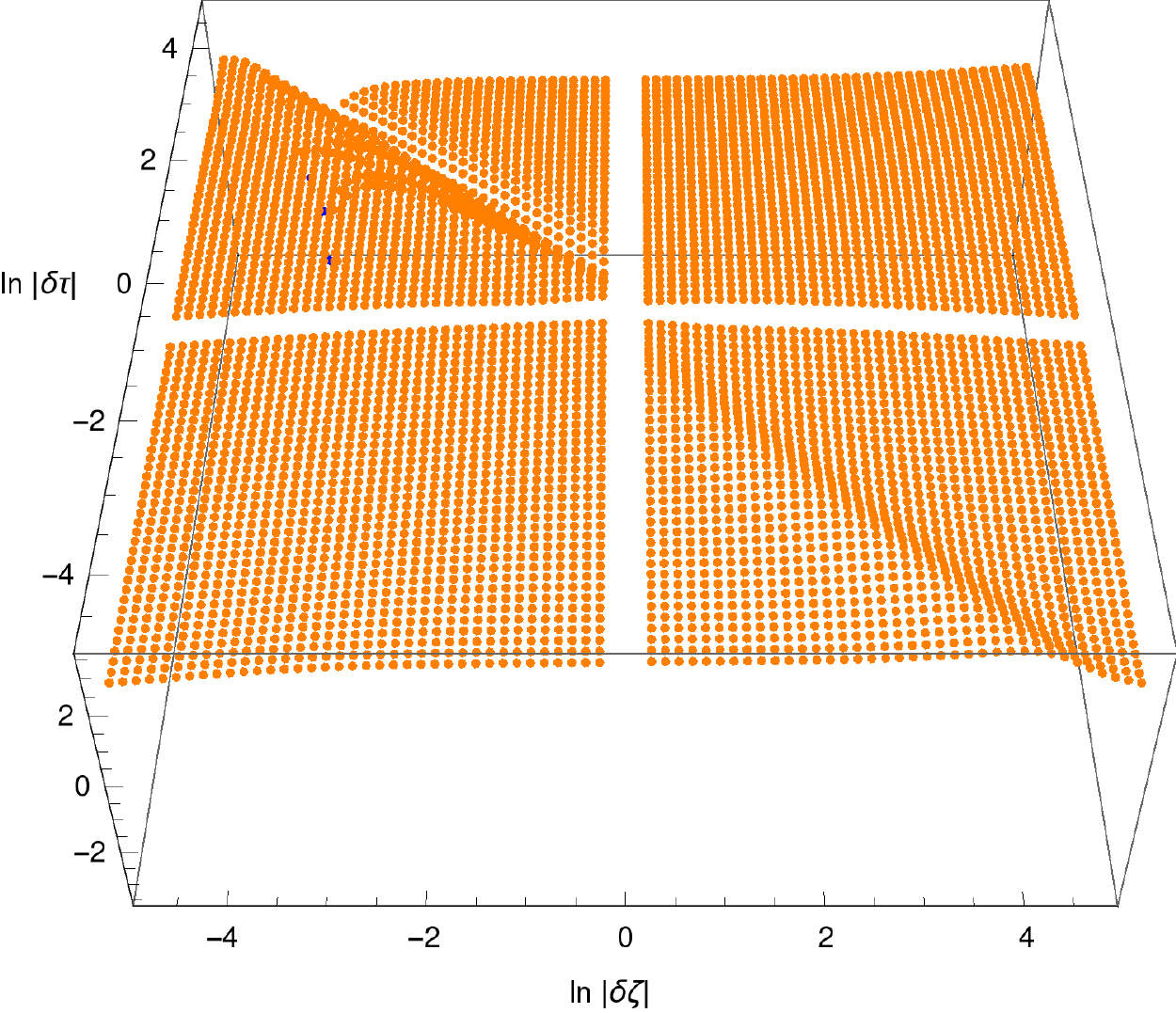}
\caption{\label{fig:trad-VVNonSI} The four components ($t$, $x$, $y$ and $z$) of
  $\ln |{\cal F}_{\rm conf}^\alpha|$ on a small patch of the worldsheet about the cusp.
  Four quadrants in $\{\log \tau, \log \zeta\}$ are displayed
  with the smallest $|\tau|$ and $|\zeta|$ at the center of
  the picture, oriented in the same way as the usual linear
  system about $\{0,0\}$. Orange (blue) represent positive (negative) values.
}
\end{figure}
\end{widetext}
The large dynamic range evident in Figs.~\ref{fig:Force-components-VVNonSIA} and
\ref{fig:Force-VVNonSIA} necessitates looking at small patches to examine special features like the
cusp. We begin by displaying ${\cal F}^\mu_{\rm conf}$ in Fig.~\ref{fig:trad-VVNonSI}. The special
coordinate system shows a small patch near the cusp which is located at $\zeta=0=\tau$. Results for
$\ln |{\cal F}^\mu_{\rm conf}|$ are displayed in these figures, color coded according to the sign of the
quantity: orange (blue) dots represent positive (negative) values. Each figure combines four plots
with axes $\{ {\rm sgn} (\zeta) \ln |\zeta|,\,{\rm sgn} (\tau) \ln |\tau|\}$, arranged and oriented
in the same way as a normal linear plot (plus a constant shift selected to bring small values close
to the center). The lower left hand quadrant has $\zeta < 0$ and $\tau < 0$. Smaller values of
$|\tau|$ and $|\zeta|$ lie near the center for all four quadrants. The gap encompasses all values
near the sign change of the independent coordinates.

We find that ${\cal F}^t_{\rm conf} < 0$ for the entire area of the patch. The magnitude of ${\cal F}^t_{\rm conf}$ is much less than $F^t$ and is less strongly divergent --- the two are related by a
projection factor and an overall factor of $1/\sqrt{-\gamma}$ (see Eq.~\eqref{eq:self-force}), both
of which diverge as the cusp is approached. Likewise, ${\cal F}^x_{\rm
  conf} < 0$, ${\cal F}^y_{\rm conf} > 0$
and ${\cal F}^z_{\rm conf} > 0$ have single, well-defined signs throughout most of the area of
corresponding patch.

In the conformal gauge the negative value for ${\cal F}_{\rm conf}^t$ implies (see
Eq.~\eqref{eqn:tradiationalforcesecondorderloss} in Appendix
\ref{sec:energymomentumlossdiscussion}) that the string is losing energy and decelerating in the
x-direction both before and after the cusp forms . This makes physical sense; the self-force saps
the mechanical energy during the period of large acceleration and the relativistic beaming ensures
that gravitational waves are emitted primarily in the x-direction, thus creating the largest
decelerating force in that direction. A small spatial segment of the string near where the cusp
forms behaves in a coherent fashion before and after the moment of cusp formation in terms of the
signs of ${\cal F}_{\rm conf}^\alpha$ for all components. ${\cal F}_{\rm conf}^\alpha$ shows a net positive
acceleration in y- and z-directions throughout most of the area of these figures.

As the figures of ${\cal F}^\mu_{\rm conf}$ make clear, the asymptotic behavior near the cusp varies
depending upon the direction of approach. A common diagonal feature is the locus in the worldsheet
where $\sqrt{-\gamma} \ge 0$ is small. Only at the cusp is $\gamma$ exactly equal to zero, but
along the visible fold its values are small.
\begin{widetext}
~
\begin{figure}[H]
\centering
\includegraphics[width=0.45\linewidth]{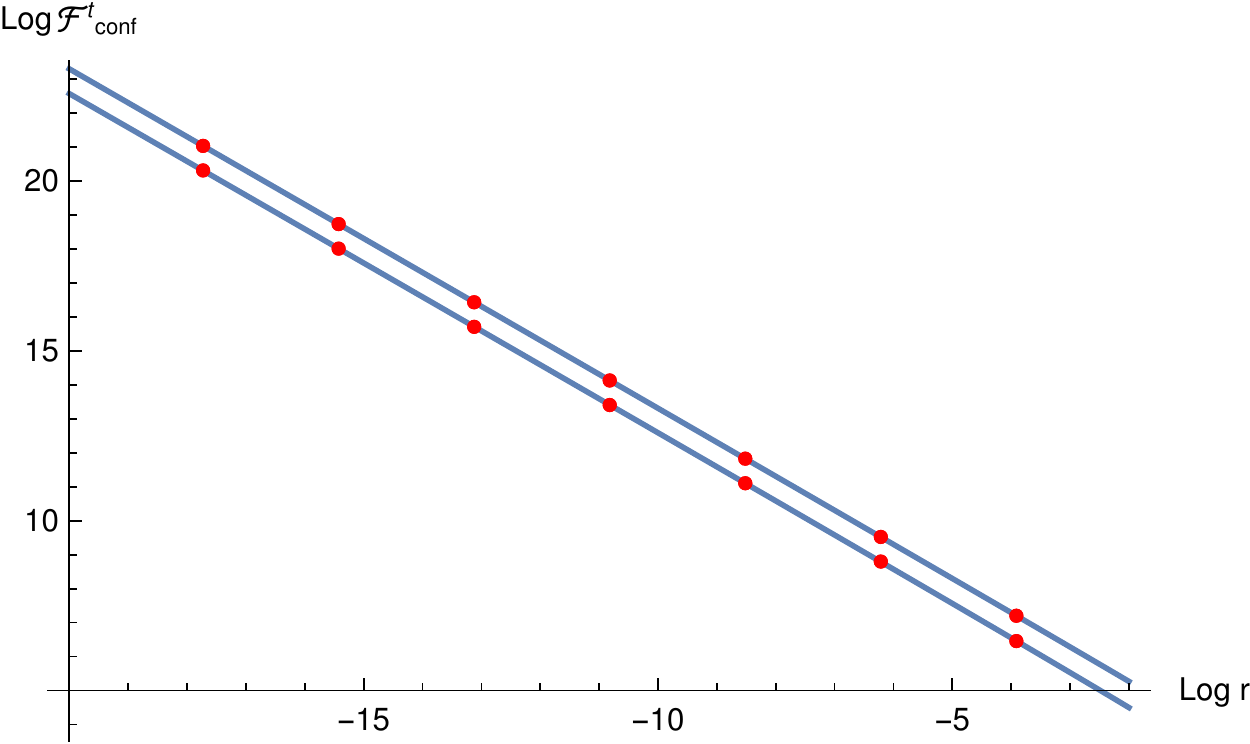}
\includegraphics[width=0.45\linewidth]{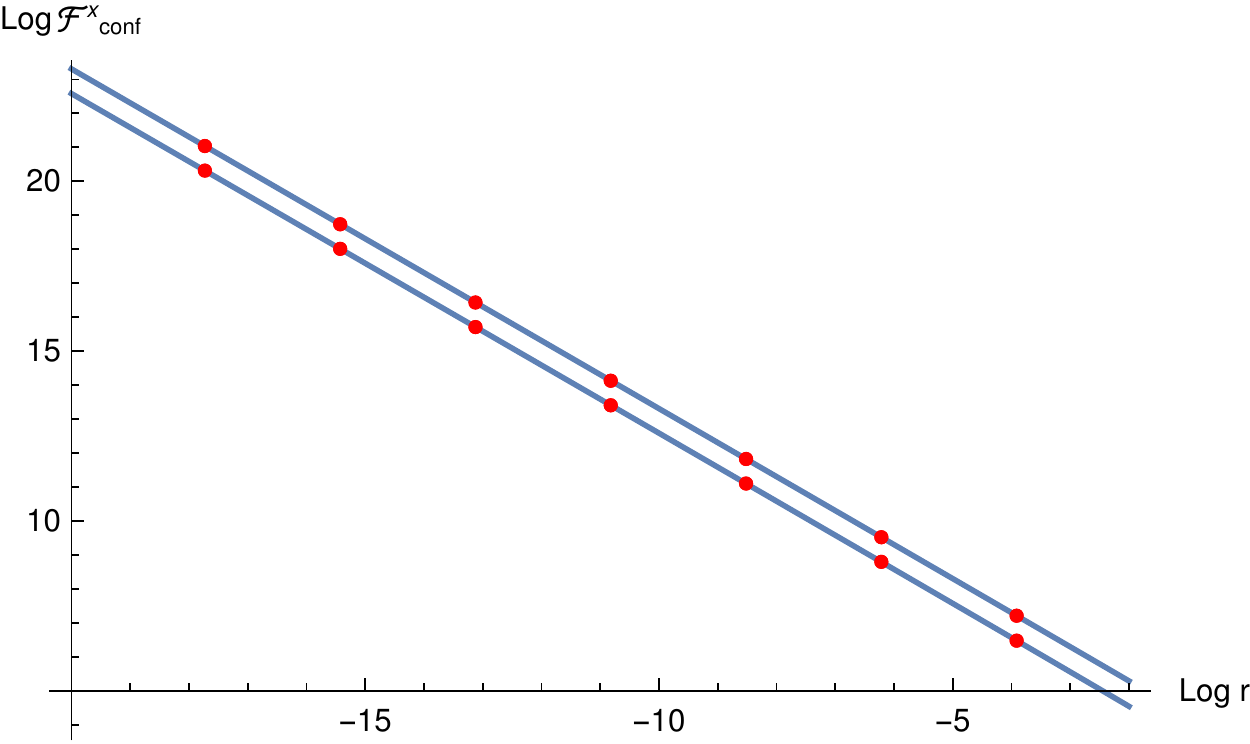}
\includegraphics[width=0.45\linewidth]{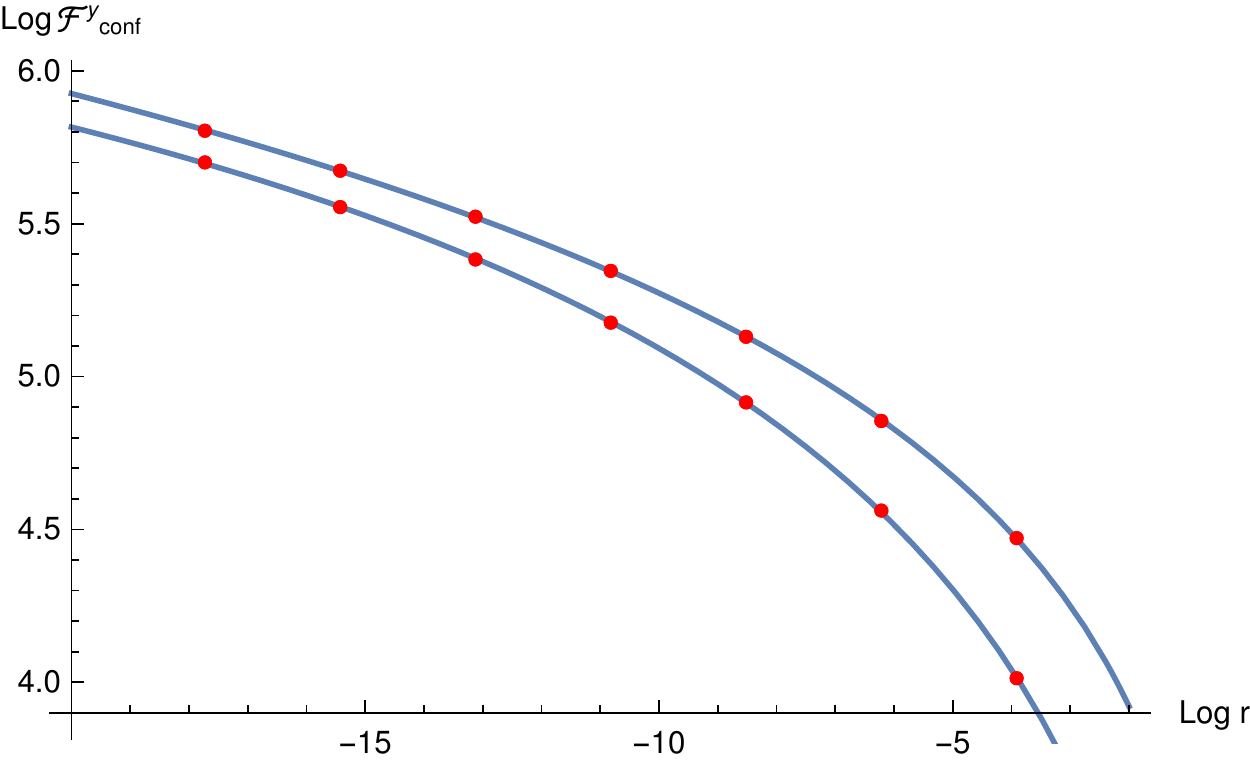}
\includegraphics[width=0.45\linewidth]{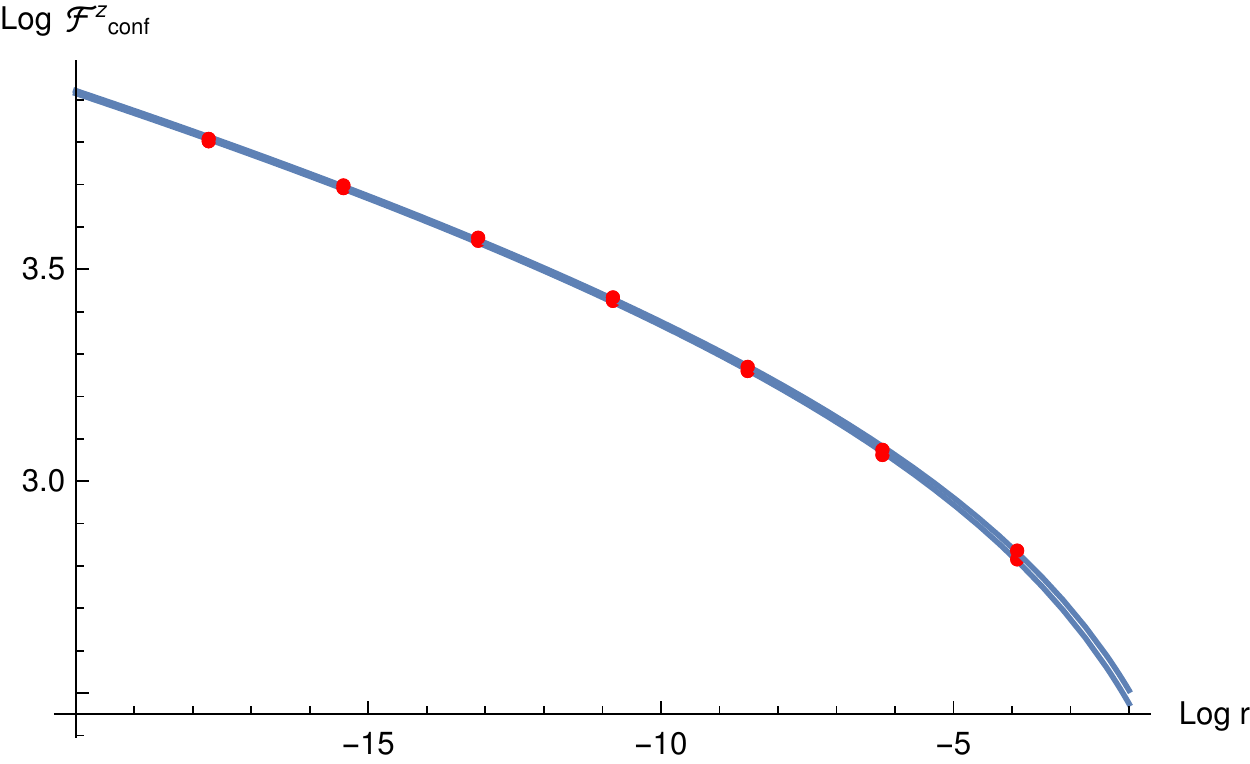}
\caption{\label{fig:VVloopNonSIcuspray} The components of the
  smooth integral contribution to ${\cal F}^\mu_{\rm conf}$ along two rays
  at angles $\theta=0$ and $\pi/2$ are shown. Red dots are numerical
  results and blue lines are fits of the form $\log |{\cal F}^\mu| = a \log |\log r| + b \log r +  c$.
  For $\mu=t$, $(a,b,c)=(0.098,-0.99,2.45)$ for $\theta=0$ and $(0.046,-1,3.24)$ for $\theta=\pi/2$;
  likewise, for $\mu=x$, $(a,b,c)=(0.044,-1,2.52)$ and $(0.018,-1,3.28)$.
  These fits show that the dominant behavior in the direction of motion
  of the cusp as $r \to 0$ is $1/r$. In the other directions, the behavior
  is consistent with a logarithmic divergence at leading order:
  $\mu=y$, $(a,b,c)=(0.79,-0.011,3.35)$ and $(1.23,0.013,2.39)$;
  $\mu=z$, $(a,b,c)=(0.43,-0.025,2.13)$ and $(0.41,-0.026,2.17)$.
}
\end{figure}
\begin{figure}[H]
\centering
\includegraphics[width=0.45\linewidth]{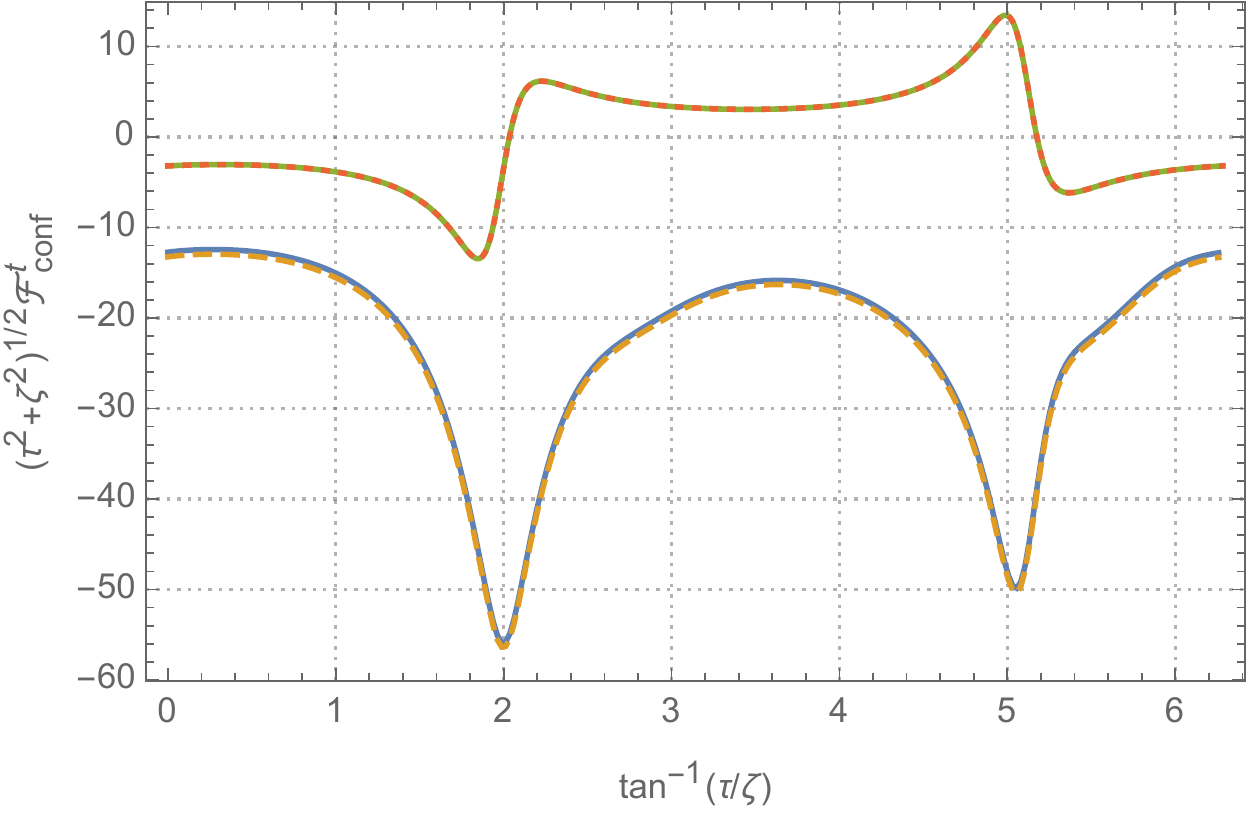}
\includegraphics[width=0.45\linewidth]{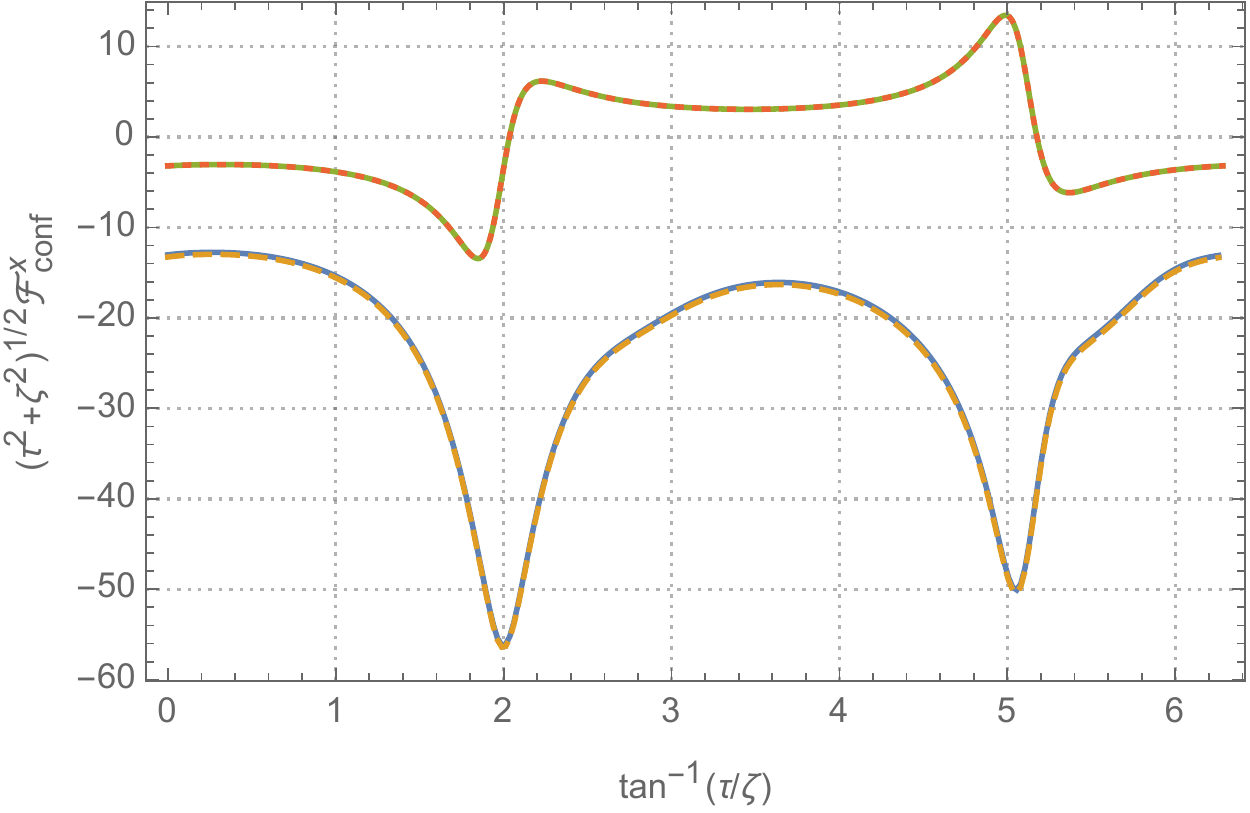}
\includegraphics[width=0.45\linewidth]{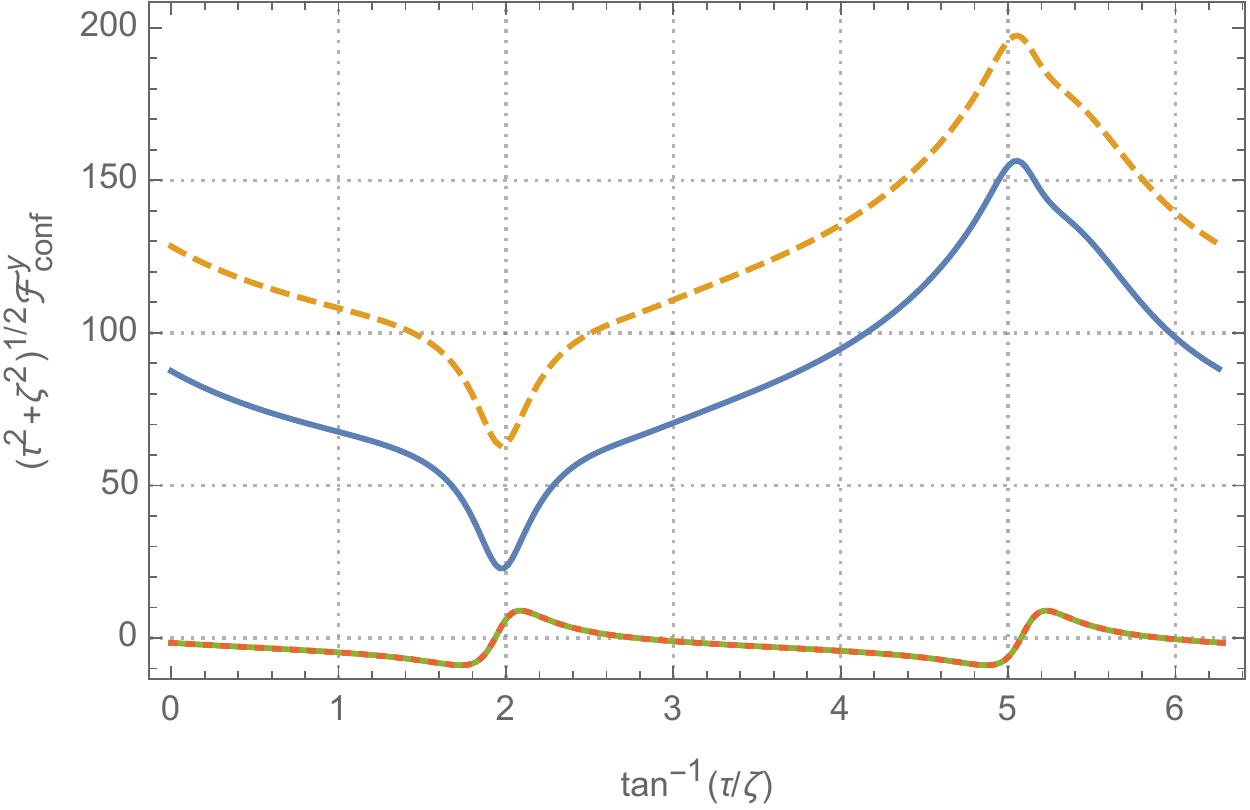}
\includegraphics[width=0.45\linewidth]{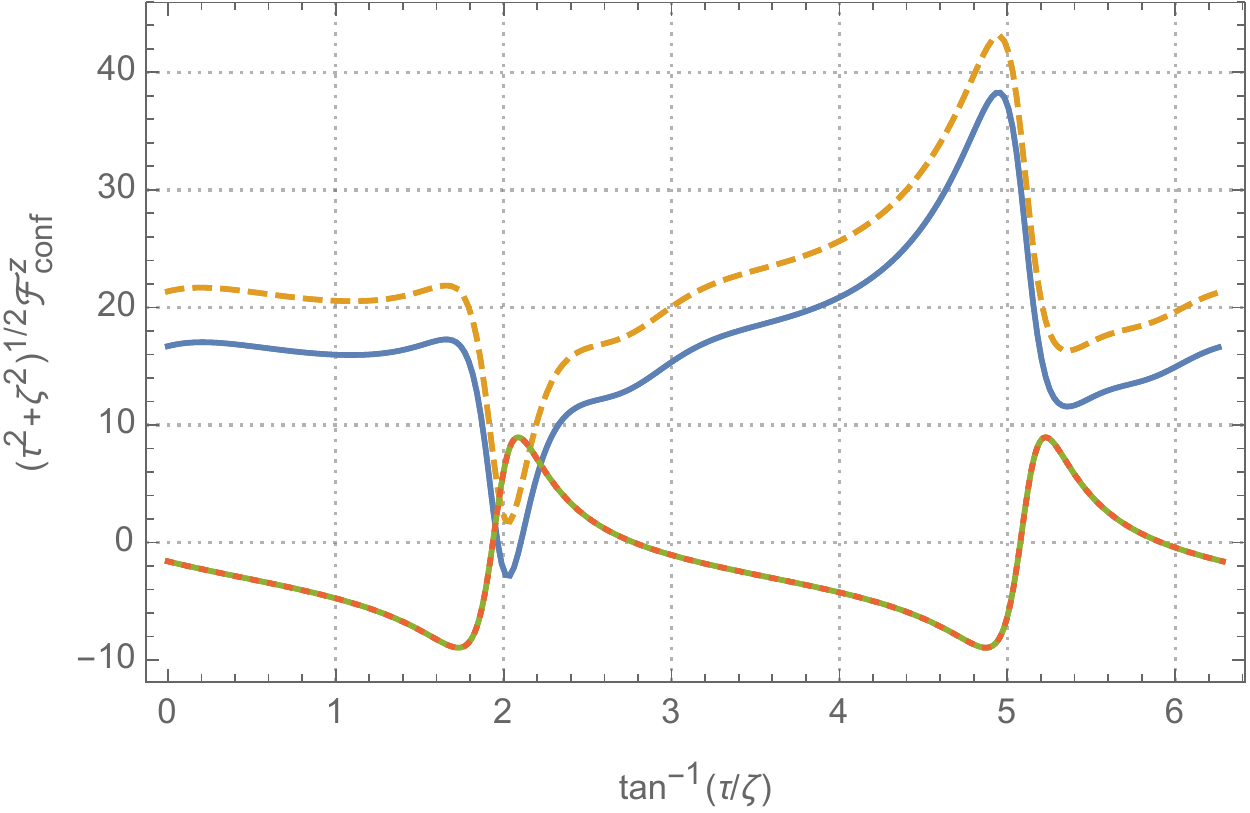}
\caption{\label{fig:VVloopNonSIcusp} ${\cal F}^\mu_{\rm conf}$ for the
  non-intersecting KT string ($\alpha=1/2$ and $\phi=0$) in the
  neighborhood of the cusp at $\tau = 0 = \zeta$. The $t$ and $x$
  components have been scaled by the radial (Euclidean) distance from
  the cusp, and the $y$ and $z$ components have not been scaled at
  all. Each panel displays two separate contributions to the total force. For
  $t$ and $x$ components the upper (green solid and red dotted) lines give the field point
  contribution. It is antisymmetric in angle and integrates over angle to give zero
        (in fact, when taken over the whole worldsheet the integral also vanishes exactly).
        The lower (blue solid and orange dashed) lines give the smooth
  integral contribution. It is single signed and large where
  $\sqrt{-\gamma}$ is small. We plot scaled results for $r=0.02$
  (solid lines) and $r=0.002$ (dashed/dotted lines) for each
  contribution to the total.  These overlap and show in a qualitative
  fashion the dominant $1/r$ scaling near the cusp for both contributions
  to the $t$ and $x$ components. In the lower panels we display the $y$ and $z$
        components. As before, the contribution from the field point is given by the green solid and
        red dotted curves. In this case it is independent of $r$ to lowest order (and
        integrates over angle to give $4\pi$ at this order). The solid blue and dashed
        orange lines show the integral contributions increase slowly as $r$ decreases, consistent 				with the $\log r$ type behavior.}
\end{figure}
\end{widetext}
Regardless of direction, however, the scaling with radial distance from the cusp is clear and
unambiguous in each of the components.
The smooth integral contribution to ${\cal F}^\mu_{\rm conf}$
is shown in Fig.~\ref{fig:VVloopNonSIcuspray}
for rays approaching the cusp with angle $\theta=0$ ($\delta \tau=0$) and $\theta=\pi/2$ ($\delta \zeta=0$).
The red dots are numerical results and the blue lines are fits of the form $\log |F^\mu| = a \log |\log r| + b \log r +  c$.\footnote{The occurence of both $\log|\log r|$ and $\log r$ are consistent with
  recently reported analytic results of Blanco-Pillado, Olum and Wachter [see acknowledgements].}
The angular variation and scaling of the delta-function term and the smooth integral contribution
are illustrated in Fig.~\ref{fig:VVloopNonSIcusp} and discussed in the caption.

Two-dimensional numerical fits for the integral part of ${\cal F}^\mu_{\rm conf}$ are
summarized in Appendix \ref{sec:VV-fits}. We find that
${\cal F}^t_{\rm conf}$ and ${\cal F}^x_{\rm conf}$ scale as the inverse distance from the cusp, and that
${\cal F}^y_{\rm conf}$ and ${\cal F}^z_{\rm conf}$ are at worst much less singular (consistent with a log divergence).
From this we conclude
that when one adopts the conformal gauge at first order the self-force near the cusp has a weak,
integrable divergence on the worldsheet and that any integrated quantities (such as the radiated
energy) are finite.

\begin{widetext}
        ~
  \begin{table}[H]
  \begin{center}
    \begin{tabular}{|c|cccccc|cc|}
      \hline
      Quantity     & $a$ & $b$ & $c$ & $d$ & $e$ & $f$ & $\log_{10} \epsilon$ & $\log_{10} Q$ \\
      \hline
$F^t-F^x$&$14.68$&$276.43$&$117.42$&$-41.51$&$12.47$&$-25.45$&$-3.08$&$1.06$ \\
$F^z(\tau+\zeta)+F^y(3\tau+\zeta)$&$0$&$521.81$&$175.92$&$-126.45$&$38.64$&$16.21$&$-2.67$&$0.91$ \\
      \hline
$H_{\tau\tau}$&$11.49$&$-6.53$&$-15.33$&$18.76$&$84.03$&$101.7$&$-7.22$&$3.9$ \\
$H_{\zeta\zeta}$&$0$&$0$&$0.01$&$64.4$&$196$&$237.13$&$-7.03$&$1.32$ \\
$H_{\tau\zeta}$&$0$&$-7.66$&$-17.94$&$22.71$&$69.29$&$56.75$&$-7.05$&$3.8$ \\
  \hline
\end{tabular}
  \end{center}
  \caption{First order fits
    for force combinations near the cusp with form
    $a + b \tau + c \zeta + (d \tau^2 + e \tau \zeta + f \zeta^2)/r$ with
    $r=\sqrt{\tau^2+\zeta^2}$ and
    second order fits for the worldsheet projected metric pertubations with
    form $a + b \tau + c \zeta + d \tau^2 + e \tau \zeta + f \zeta^2$
    over the radial range $2 \times 10^{-8} \le r \le 2 \times 10^{-4}$. The Table provides the two force
    combinations and 3 worldsheet projected metric perturbations that appear in the asymptotic
    forms for $F_1^\mu$ and $F_2^\mu$. The last two columns give the common log of $\epsilon$ (the root mean square
    error between the data and the fit), and $Q$ (the ratio of the variation in the data divided
    by $\epsilon$). 
  }\label{tab:fits}
  \end{table}
\end{widetext}

  \begin{table}[H]
  \begin{center}
    \begin{tabular}{|c|ccccc|}
      \hline
      Order & $-4$ & $-3$ & $-2$ & $-1$ & $0$ \\
      Component& & & & & \\      
      \hline      
$t$ &$ < $&$ < $&$-2.01$ &$ 1.28 $&$1.25$ \\
$x$ &$ < $&$ < $&$-2.01$ &$ 1.28 $&$1.25$ \\
$y$ &$ < $&$ < $&$ < $   &$ <$&$2.75$ \\
$z$ &$ < $&$ < $&$ < $   &$ < $&$2.03$ \\
      \hline
\end{tabular}
  \end{center}
  \caption{The numerical results for the
    expansion of $\sqrt{-\gamma}F^\mu$ in $r$ (averaged over angle)
    for the fit given in Table \ref{tab:fits}.
    For $\sqrt{-\gamma}(F_1 + F_2) \sim \sum_{n} c_n r^n$ the columns
    are the leading powers of
    the expansion ($n=-4$ to $0$), the rows are the spacetime
    components and the table values are the common log of the
    expansion coefficients ($d_n=\log_{10} | c_n | $).
    The symbol $<$ means a numerical result $|c_n| < 10^{-12}$.
    When the fitting range is narrowed about the cusp the $1/r^2$ contribution
    decreases $\propto r$ and the other pieces are fixed.
    From this we infer that the leading non-zero piece of $F^\mu$
    varies as $1/r$.
  }\label{tab:cancellations}
  \end{table}
Given the scalings for ${\cal F}^\mu_{\rm conf}$ nearby the cusp, it is straightforward to deduce the
corresponding scaling for $F^\mu_1 = -\tfrac{1}{\sqrt{-\gamma}} \perp^\mu{}_\nu {\cal F}^\nu_{\rm conf}$.
Working with the exact expression for the determinant of the induced metric in this case,
\begin{equation}
        \gamma = -\frac{1}{16} \left[2 - \cos (2 \zeta) - \cos (4\zeta+2\tau) \right]^2,
\end{equation}
and expanding the relationship between $F^\mu_1$ and ${\cal F}^\mu_{\rm conf}$ to next from leading order, we find
\begin{align}
        \label{eq:F1asym1}
  F^t_1 & \approx F^x_1 \approx - \frac{1}{\gamma}\bigg\{\left({\cal F}^t_{\rm conf}-{\cal F}^x_{\rm conf}\right) \\
         & - \frac12\left[\left({\cal F}^z_{\rm conf}+3 {\cal F}^y_{\rm conf}\right)\tau + \left({\cal F}^z_{\rm conf}+ {\cal F}^y_{\rm conf}\right)\zeta\right] + \cdots\bigg\}, \nonumber \\
        \label{eq:F1asym2}
        F^y_1 & \approx -\frac{1}{2\gamma} \left({\cal F}^t_{\rm conf}-{\cal F}^x_{\rm conf}\right) \left(\zeta + 3 \tau\right) + \cdots, \\
        \label{eq:F1asym3}
        F^z_1 & \approx -\frac{1}{2\gamma} \left({\cal F}^t_{\rm conf}-{\cal F}^x_{\rm conf}\right) \left(\zeta + \tau\right) + \cdots.
\end{align}
Since $\gamma$ scales as the fourth power of the distance from the cusp we infer that $F^\mu_1$ is
naively four orders more singular than ${\cal F}^\mu_{\rm conf}$. However, as can be seen in Table
\ref{tab:fits}, it turns out that ${\cal F}_{\rm conf}^t \approx {\cal F}_{\rm conf}^x$ near the cusp so the
leading-order divergence cancels and at worst $F^\mu_1$ diverges as the inverse fourth power of the
distance from the cusp. At next from leading order the asymptotic expression for $F^\mu_1$ is
antisymmetric about the cusp. (This behavior, combined with the mixing of components is what makes
the analysis of ${\cal F}^\mu_{\rm conf}$ clearer than working directly with $F^\mu_1$.)
With the worldsheet weighting we naively infer that $\sqrt{-\gamma} F^\mu_1$
diverges as the inverse quadratic power of the distance from the cusp,
one power worse than ${\cal F}^\mu_{\rm conf}$. Now we must consider the role of $F^\mu_2$.

To understand the behaviour of $F^\mu_2$ near the cusp we begin with the perturbed metric projected
along the worldsheet vectors $\partial_\tau z^\alpha$ and $\partial_\zeta z^\alpha$ according to
\begin{eqnarray}
  H_{\tau\tau}& =& \partial_\tau z^\alpha h_{\alpha\beta} \partial_\tau z^\beta \\
  H_{\tau\zeta}& =& \partial_\tau z^\alpha h_{\alpha\beta} \partial_\zeta z^\beta \\
  H_{\zeta\zeta}& =& \partial_\zeta z^\alpha h_{\alpha\beta} \partial_\tau z^\beta .
\end{eqnarray}
Evaluating the simple expression Eq.~\eqref{eq:F2-hproj} for the relationship between $F^\mu_2$ and
the worldsheet projections of the metric perturbation, we find
\begin{align}
        \label{eq:F2asym1}
        F_2^t &\approx F_2^x \approx \frac{1}{2(-\gamma)^{3/2}}\bigg[2\left(5H_{\tau \zeta} - H_{\tau \tau} - H_{\zeta \zeta}\right) \zeta  \nonumber \\
        & \qquad \qquad \qquad + \left(4H_{\tau \zeta} - H_{\tau \tau} - H_{\zeta \zeta}\right) \tau + \cdots \bigg], \\
        \label{eq:F2asym2}
        F_2^y &\approx \frac{2\left(6H_{\tau \zeta} - H_{\tau \tau} - H_{\zeta \zeta}\right) (\zeta^2+3 \tau \zeta + \tau^2) + \cdots}{2(-\gamma)^{3/2}}, \\
        \label{eq:F2asym3}
        F_2^z &\approx \frac{2\left(2H_{\tau \zeta} - H_{\tau \tau} - H_{\zeta \zeta}\right) (\zeta^2+3 \tau \zeta + \tau^2) + \cdots}{2(-\gamma)^{3/2}}.
\end{align}
With the finite behavior of the the worldsheet projections of the metric perturbation, it is
straightforward to deduce the corresponding scaling of the divergence in $F^\mu_2$. We find that at
worst $F_2^t$ and $F_2^x$ diverge as the inverse fifth power of distance from the cusp. With the
worldsheet weighting $\sqrt{-\gamma} F^\mu_2$ diverges as the
inverse cubic power of the distance from the cusp. However, as this leading-order divergence is
antisymmetric about the cusp its integral over a patch around the cusp cancels the leading-order
divergence to leave only subleading pieces.

We are now left with asymptotic forms for $\sqrt{-\gamma} F^\mu_1$ and $\sqrt{-\gamma} F^\mu_2$ each
scaling as the square of the inverse distance from the cusp. These
are individually non-integrable. However, once the asymptotic forms given in Table \ref{tab:fits} for the quantities
in Eqs.~\eqref{eq:F1asym1}-\eqref{eq:F1asym3} and \eqref{eq:F2asym1}-\eqref{eq:F2asym3} are taken
into account, we find that the leading order divergent behavior exactly cancels (see Table \ref{tab:cancellations}) to the level of
accuracy of the numerically fitted coefficients in the combination
$\sqrt{-\gamma} (F^\mu_1+ F^\mu_2)$ yielding the full force $\sqrt{-\gamma} F^\mu$ which
at worst diverges as the inverse distance from the cusp, and hence is integrable.
This divergence is no worse than ${\cal F}^\mu_{\rm conf}$ itself.

A detailed understanding of the behavior of these divergences near cusps
allows us to solve either the general covariant equation of motion
Eq.~\eqref{eq:final} or the corresponding Eq.~\eqref{eq:simple}
in which specific conformal gauge choices have been adopted.

\subsection{Garfinkle and Vachaspati string with kinks}

The third case we will explore is from a class of strings found by Garfinkle and Vachaspati (GV)
\cite{Garfinkle:1987yw}. These strings contain two kinks that travel in the same direction on an
oscillating and twisting string loop. We choose a particular representation from the general class
with the following right and left-moving modes
\begin{align}
  a^\mu(\zeta^+) & = \left[\zeta^+, 0, a^2(\zeta^+), a^3(\zeta^+) \right] \\
  b^\mu(\zeta^-) & = \left[\zeta^-,
  \frac{L}{2 \pi} \cos \frac{2 \pi \zeta^-}{L}, 0,
  \frac{L}{2 \pi} \sin \frac{2 \pi \zeta^-}{L} \right]
\end{align}
where
\begin{align}
  a^2(x) & =  \frac{L}{\pi} \sum_j \delta_{j, \lfloor \frac{2x}{L} \rfloor}
  \left( -1 \right)^{\lfloor \frac{j+1}{2} \rfloor}\times \nonumber \\&
  \cos \left( \frac{\pi}{4} + \left( - 1\right)^j \frac{\pi x}{L} \right) \nonumber \\
  a^3(x) & =  \frac{L}{\pi} \sum_j \delta_{j, \lfloor \frac{2x}{L} \rfloor}
  \left( -1 \right)^{\lfloor \frac{j}{2} \rfloor} \times \nonumber \\
  &\left[
  \sin \left( \frac{\pi}{4} + j \left( -1 \right)^j \frac{\pi^2}{L} \right)
  -
  \sin \left( \frac{\pi}{4} + \left( -1 \right)^j \frac{\pi x}{L} \right)
  \right],
\end{align}
and where the sums are over all integers $j$, $L$ is the invariant length, $\lfloor x \rfloor$ is
the floor function and $\delta_{j,k}$ is the Kronecker delta.

\begin{figure}[H]
\centering
\includegraphics[width=0.9\linewidth]{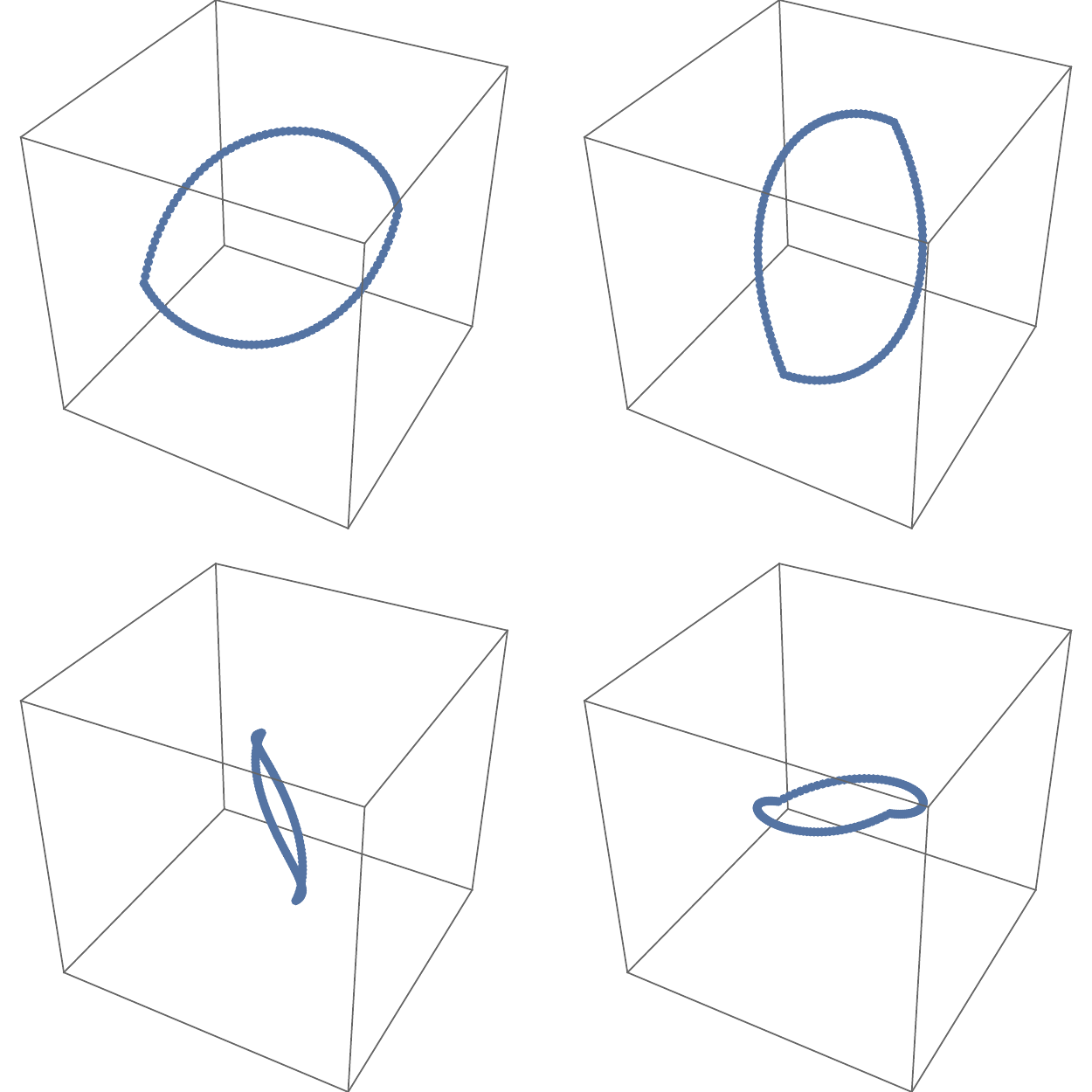}
\caption{\label{fig:GVloopspacetime} The GV string loop configuration in spacetime at four equally spaced moments in the basic loop oscillation cycle $\tau=0$, $L/8$, $L/4$ and $3L/4$. Each box is the same size with fixed axes $-1$ to $1$
  and fixed orientation. }
~\\
\includegraphics[width=0.9\linewidth]{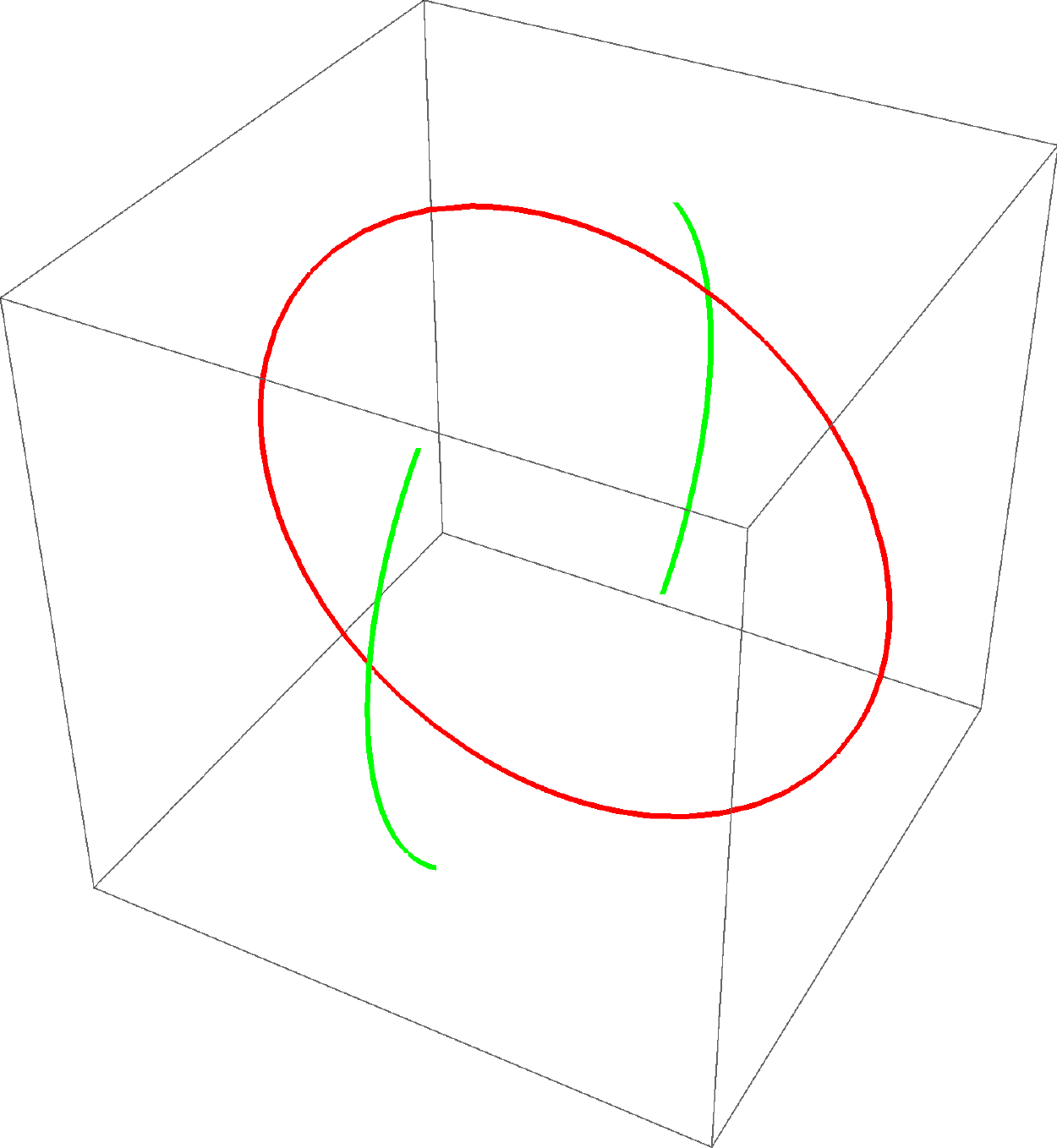}
\caption{\label{fig:GVlooptangentplot} The arcs on the tangent sphere
  traced by ${\mathbf b}'(x)$ (red line, a complete great circle passing
  through poles) and ${\mathbf a}'(x)$ (two green segments symmetrically
  cut from a great circle) for the GV string. The line is made by a series of points,
  equally spaced in argument $x$ for the left and right moving mode. A
  blow up of the line in the figure would show equal intervals between
  the points. When the kink is ``rounded off'' (non-zero $\Delta$ as
  described in the text) a few of these points will sit between the
  pictured green arcs and the green line is formally continuous.  The
  ${\mathbf a}'(x)$ tangent vector moves very rapidly from one side to the
  other.}
\end{figure}

Figure \ref{fig:GVloopspacetime} illustrates
the configuration in spacetime at equally spaced moments in the oscillation cycle. The kink
discontinuities are visible in all four snapshots. In the tangent sphere representation (shown in
Fig.~\ref{fig:GVlooptangentplot}), ${\mathbf b}'$ traverses a complete great circle through the
North and South poles at a steady rate; ${\mathbf a}'$ follows two disjoint segments of a great
circle (longitude offset by $\pi/2$ from the one traced by ${\mathbf b}'$) between latitudes
$\theta = \pm \pi/4$, also at a steady rate. The vector ${\mathbf a}'$ traces one segment and then
abruptly jumps from the point $(0,y,z)$ to $(0,-y,z)$ and
traces out the mirrored arc at a steady rate (and repeats).
Each jump from one segment to the other yields a kink discontinuity in the spacetime representation.

While the ACO string provided a useful analytic test case with kinks against which we could compare our numerical results, it turns out that the simplicity of the ACO string (in particular, that many quantities are a constant along the string) means that many important terms that appear in the general expression for the self-force are identically zero for the ACO string. Fortunately, the GV string is sufficiently general that this is not the case. Unfortunately, however, there is no known analytic solution for the self-force for the GV string. Instead, in order to use the GV string as a test of our method, we chose a particular point on the string ($\tau = 0.3L$, $\zeta = 0.4L$) and computed the self-force at that point using an extensive set of different and independent methods:
\begin{enumerate}
\item We used our exact $1$D method including a field point contribution and contributions from the two kinks (this is method [1D] discussed in Sec.~\ref{sec:1D}).
\item We repeated our $1$D calculation (again, method [1D]) using multiple choices of integration variable ($\zeta^+$, $\zeta^-$ and $\zeta$). In each case, the various contributions (from the integral, field point, and two kinks) were different. Indeed, in some cases there was no contribution picked up from the kinks.
\item We again repeated our $1$D calculation using method [1D], but using a mixed coordinate choice; we used $\zeta^+$ on one side of the field point and $\zeta^-$ on the other side. We then included a contribution at the point where these two segments meet up again, to account for the change in integration variable at that point. This contribution is exactly the one discussed in Secs.~\ref{sec:kinks}, and an explicit expression is the same as one obtains when breaking the integration at a kink, as discussed in Sec.~\ref{sec:coord-depend-integral}.
\item We repeated the previously mentioned $1$D calculations again, but instead of including the exact field point term, we considered an over-retarded image of the string (method [1DO]). In that case, we find that the over-retarded integrand picks up a $\delta$-function type feature nearby the field point (see Fig.~\ref{fig:over-retarded}). For finite over-retardation this manifests itself as a narrow Gaussian, and the Gaussian gets narrower and sharper as the over-retardation parameter is shrunk towards zero. Reassuringly, in the limit of the Gaussian shrinking down to zero size we recover a result which agrees with the previous calculations and can identify the $\delta$-function with the field point contribution.
\begin{figure}[H]
\centering
\includegraphics[width=0.9\linewidth]{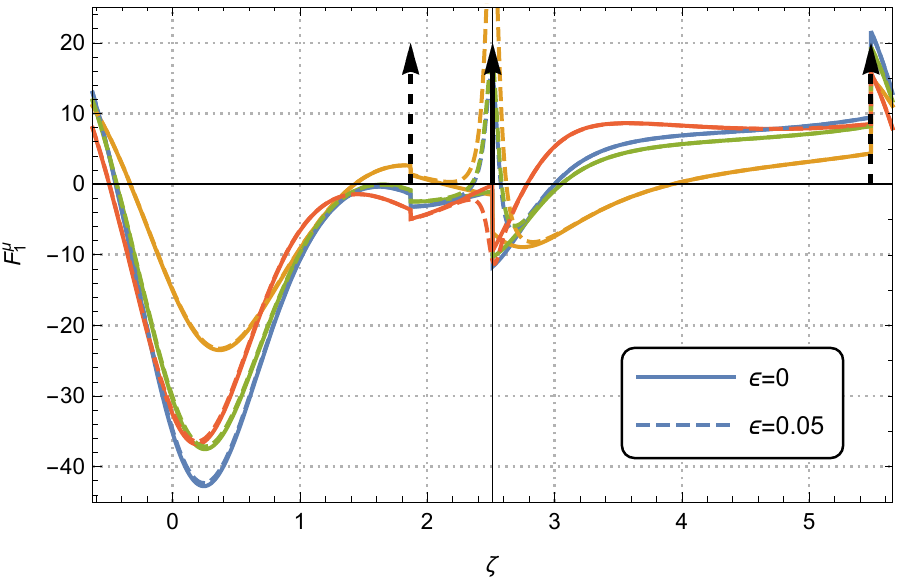}
\includegraphics[width=0.9\linewidth]{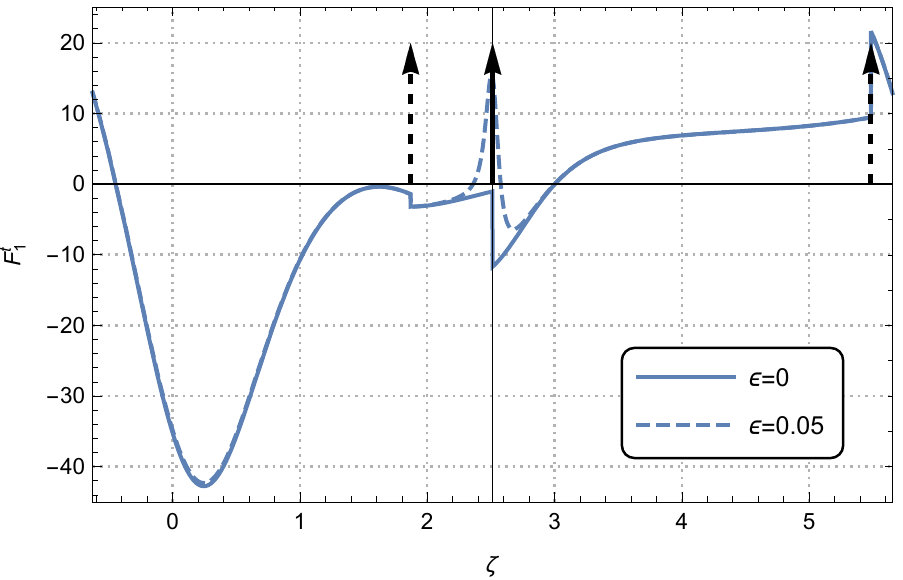}
\caption{\label{fig:over-retarded}
Integrand used to compute to the self-force for the Garfinkle-Vachaspati string.
These correspond to the values in the $\zeta$ column of Table \ref{tab:comparison}.
Distributional contributions from the kinks and field point are denoted by dashed and
solid arrows, respectively.
}
\end{figure}
\item Finally, we repeated the calculation in a completely independent way, directly evaluating the self-force from the $2$D integral (method [2D]) without the reduction to a $1$D integral. This was significantly less efficient, but provided an important check as there is no need to consider any split into field-point-plus-integral–plus-kink contributions. Instead, we smeared out the $\delta$ function in the Green function and also introduced a slight smoothing of the kinks, as discussed in detail in Appendix \ref{sec:2D}. Yet again, reassuringly, in the limit of our smearing and smoothing parameters going to zero we recovered a result which was in perfect agreement with all of the other methods.
\end{enumerate}

\begin{table*}
  \begin{center}
\begin{ruledtabular}
\begin{tabular}{c|c|cccccccc}
   & Contribution & $\zeta^-$ & $\zeta^+$ & $\zeta^-$/$\zeta^+$ & $\zeta$ & $2$D & $\zeta^-_\epsilon$ & $\zeta^+_\epsilon$ & $\zeta_\epsilon$ \\
  \hline \hline
  \multirow{6}{*}{$F_1^t$} & $\int$ & \multirow{2}{*}{$-330.558$} & \multirow{2}{*}{$-12.73(1)$} & $6.89229$ & $-25.6962$ & $-12.76(2)$ & $-330.60(4)$ & $-12.7488(1)$ & $-22.4707(2)$ \\
    & $\delta$ & $ $ & $ $ & - & $3.22536$ & - & - & - & - \\
    & Kink $1$ & $358.819$ & - & - & $16.7989$ & - & $358.819$ & - & $16.7989$ \\
    & Kink $2$ & $-41.0096$ & - & - & $-7.07683$ & - & $-41.0096$ & - & $-7.07683$ \\
    & $\zeta^-\leftrightarrow\zeta^+$ & - & - & $-19.6411$ & - & - & - & - & - \\
    \cline{2-10}
    & Total    & $-12.7488(1)$ & $-12.73(1)$ & $-12.7488$ & $-12.7488$ & $-12.76(2)$ & $-12.8(1)$ & $-12.7488(1)$  & $-12.7487(2)$ \\
  \hline \hline
  \multirow{6}{*}{$F_1^x$} & $\int$ & \multirow{2}{*}{$-300.675(1)$} & \multirow{2}{*}{$-2.64(7)$} & $8.62259$ & $-21.7023$ & $-2.58(3)$ & $-300.63(4)$ & $-2.56339(6)$ & $-14.8591(9)$ \\
    & $\delta$ & $ $ & $ $ & - & $6.84317$ & - & - & - &  - \\
    & Kink $1$ & $287.628$ & - & - & $14.7570$ & - & $287.628$ & - & $14.7570$ \\
    & Kink $2$ & $10.4832$ & - & - & $-2.46129$ & - & $10.4832$ & - & $-2.46129$ \\
    & $\zeta^-\leftrightarrow\zeta^+$ & - & - & $-11.186$ & - & - & - & - & - \\
    \cline{2-10}
    & Total    & $-2.5637(8)$ & $-2.64(7)$ & $-2.56342$ & $-2.56342$ & $-2.58(3)$ & $-2.52(8)$ & $-2.56339(6)$ & $-2.56333(9)$ \\
    \hline \hline
  \multirow{6}{*}{$F_1^y$} & $\int$ & \multirow{2}{*}{$-304.564$} & \multirow{2}{*}{$-10.7066(4)$} & $6.6211$ & $-23.4832$ & $-10.72(2)$ & $-304.59(3)$ & $-10.7068(1)$ & $-20.0615(2)$ \\
    & $\delta$ & $ $ & $ $ & - & $3.42159$ & - & - & - & - \\
    & Kink $1$ & $326.143$  & - & - & $15.4176$ & - & $326.143$ & - & $15.4176$ \\
    & Kink $2$ & $-32.2859$ & - & - & $-6.06281$ & - & $-32.2859$ & - & $-6.06281$ \\
    & $\zeta^-\leftrightarrow\zeta^+$ & - & - & $-17.3279$ & - & - & - & - & - \\
    \cline{2-10}
    & Total    & $-10.7069$ & $-10.7066(4)$ & $-10.7068$ & $-10.7068$ & $-10.72(2)$ & $-10.7(1)$ & $-10.7068(1)$ & $-10.7067(2)$ \\
    \hline \hline
  \multirow{6}{*}{$F_1^z$} & $\int$ & \multirow{2}{*}{$-190.140(1)$} & \multirow{2}{*}{$-13.82(7)$} & $2.25465$ & $-15.9946$ & $-13.90 (1)$ & $-190.22(8)$ & $-13.8960(1)$ & $-16.9795(2)$ \\
    & $\delta$ & $ $ & $ $ & - & $-0.985094$ & - & - & - & - \\
    & Kink $1$ & $234.551$ & - & - & $10.0429$ & - & $234.551$ & - & $10.0429$ \\
    & Kink $2$ & $-58.3072$ & - & - & $-6.95921$ & - & $-58.3072$ & - & $-6.95921$ \\
    & $\zeta^-\leftrightarrow\zeta^+$ & - & - & $-16.1507$ & - & - & - & - & - \\
    \cline{2-10}
    & Total    & $-13.8958(7)$ & $-13.82(7)$ & $-13.896$ & $-13.8960$ & $-13.90 (1)$ & $-13.97(9)$ & $-13.8960(1)$ & $-13.8958(2)$
\end{tabular}
\end{ruledtabular}
\caption{\label{tab:comparison}
Comparison of methods for computing the self-force at a generic point ($\tau=0.3L$, $\zeta=0.4L$) on the GV string.}
\end{center}
\end{table*}
The results of this extensive set of tests are given in Table~\ref{tab:comparison}. We see that all methods produce results which are consistent within their respective error bars. The [2D] method is least accurate, due the need for a $2$D rather than $1$D numerical integral. The [1DO] method also poses challenges for numerical accuracy due to the presence of sharp features (i.e. the Gaussian approximation to the delta function for the field point contribution), as does the [1DOS] method for portions of the integral nearby kinks.

The three exact [1D] methods all work reasonably well, however even in this case not all methods
are equally computationally efficient. In particular, calculations based on a single null
coordinate encounter a strong divergence in the integrand as the field point is approached from one
side (the particular side is dependent on whether one uses $\zeta^+$ or $\zeta^-$ as integration
variable). This diverging integral largely cancels against the field point contribution\footnote{In practice, we were only able to obtain finite results by evaluating the integral up to a short distance from the field point and evaluating the expression for the field point contribution at the point where the integral was cut off. We recovered a unique and consistent result as the cut-off point was pushed towards the actual field point.}, leaving a relatively small overall contribution from the field-point-plus-integral combination. We found that the remaining two approaches (integration with respect to $\zeta$; and half-$\zeta^+$ half-$\zeta^-$ plus coordinate change term) were comparable in terms of computational efficiency.

Importantly, other than accuracy concerns, all methods produced results which are unambiguous in agreeing with each other.

\begin{widetext}
~
\begin{figure}[H]
\centering
\includegraphics[width=0.9\linewidth]{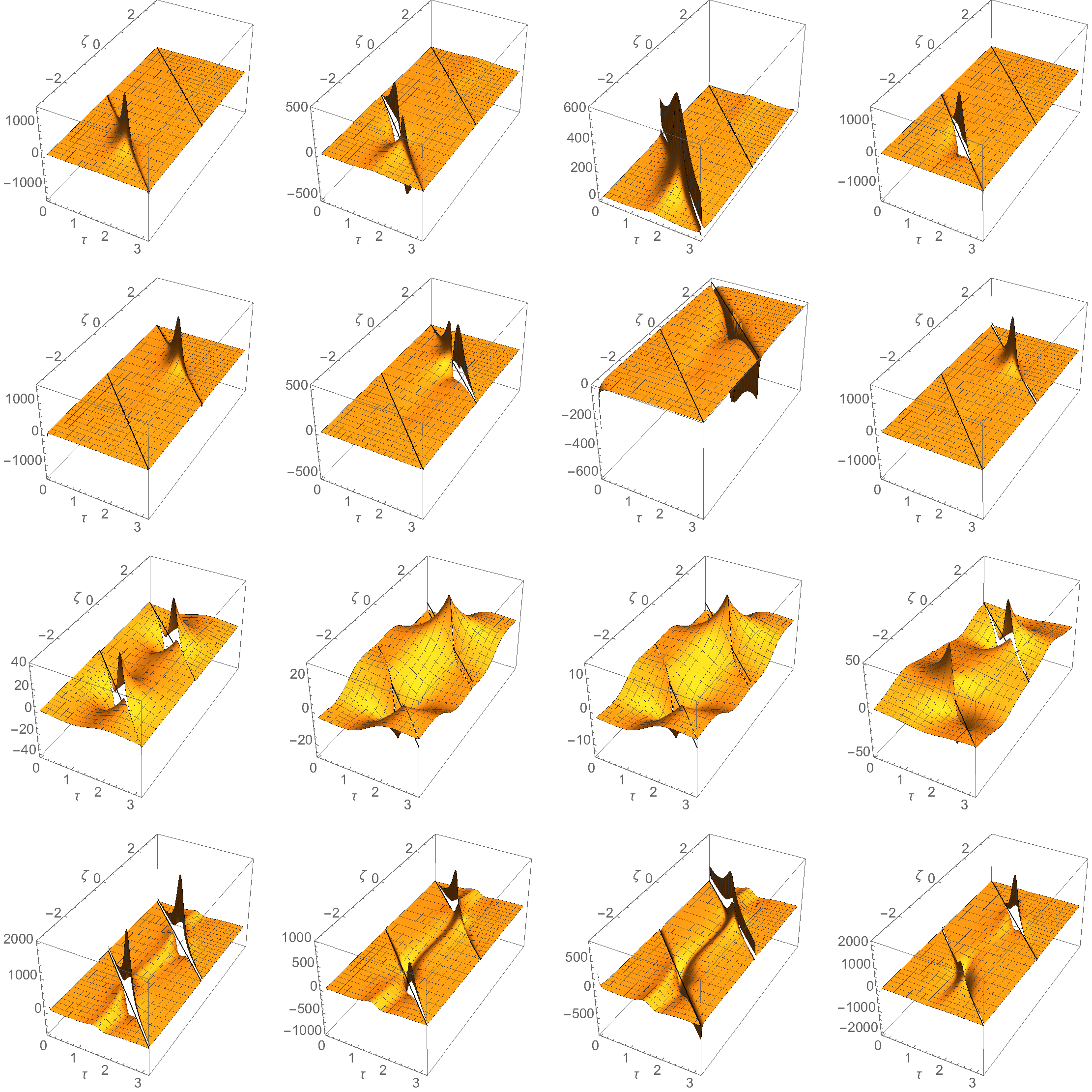}
\caption{\label{fig:Force-components-GV}
Contributions to $F_1^\mu$ for the Garfinkle and Vachaspati string when computed using the 1D integration method with integration with respect to $\zeta$. Each sub-figure shows the relevant contribution to the force at all points on the string in the region $\tau \in (0,L/2)$, $\zeta \in (-L/2,L/2)$; all other points can be obtained from the standard periodic extension of the string. Each column corresponds to a different component of the force: $F_1^t$, $F_1^x$, $F_1^y$, and $F_1^z$. The rows correspond to the contributions from: (i) the kink that passes through ($\tau = 0$, $\zeta = 0$); (ii) the kink that passes through ($\tau = 0$, $\zeta = \pi$); (iii) the field point; and (iv) the integral over $\zeta$ (ignoring distributional contributions at the kinks and field point). The two kinks are denoted by diagonal black lines. For the purposes of the plots, we have set the string tension, $\mu$, and Newton's constant, $G$ equal to one; other values simply introduce an overall scaling.
}
\end{figure}

\begin{figure}[H]
\centering
\includegraphics[width=0.9\linewidth]{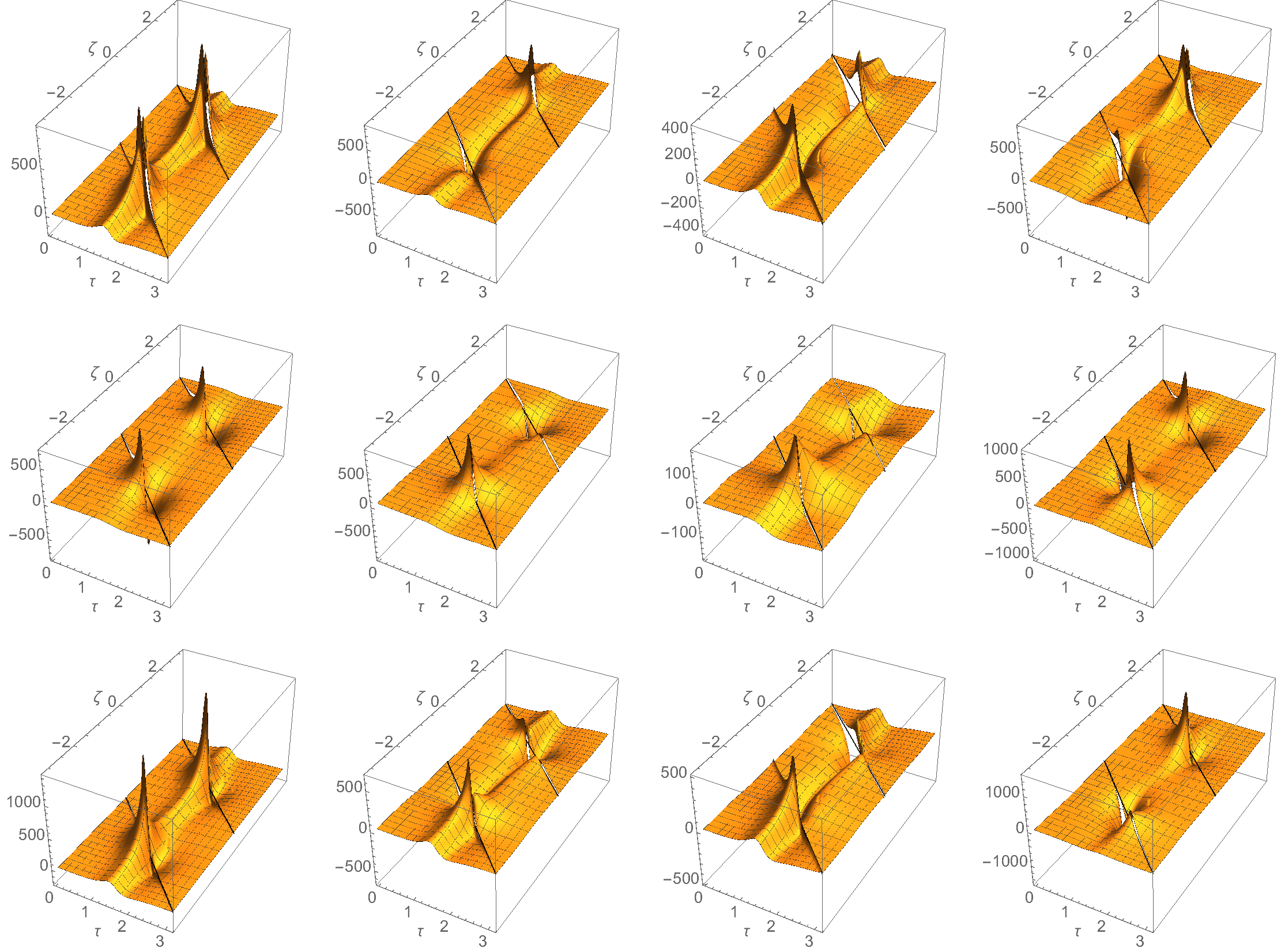}
\caption{\label{fig:Force-GV}
  The two pieces of the self-force, $F_1^\mu$ (row 1) and $F_2^\mu$ (row 2), and the total force $F^\mu = F^\mu_1+F^\mu_2$ (row 3), for the Garfinkle and Vachaspati string as a function of position on the string. The $F_1^\mu$ part can be obtained by summing the four rows in Fig.~\ref{fig:Force-components-GV}. For the purposes of the plots, we have set the string tension, $\mu$, and Newton's constant, $G$ equal to one; other values simply introduce an overall scaling.
}
\end{figure}
\end{widetext}

Finally, we used the [1D] method (specifically, integrating with respect to $\zeta$ and including field point and kink contributions) to evaluate the self-force at all points on the GV string. The results are shown in Figs.~\ref{fig:Force-components-GV} and \ref{fig:Force-GV}. Fig.~\ref{fig:Force-components-GV} shows how each of the contributions to $F^\mu_1$ contribute to the overall result, while Fig.~\ref{fig:Force-GV} shows $F^\mu_1$ and $F^\mu_2$ themselves, as well as their sum. As in the other string test cases, we find that the self-force is finite almost everywhere on the string, with the exception of exactly on the kinks, where it diverges.

We have analyzed the form of the divergence near the kink
by calculating the total backreaction force to high accuracy along a
set of worldsheet points for a line that runs perpendicular to the kink
with coordinates
$(\tau,\zeta)=(\pi/2 + \zeta^+/2, -\pi/2 + \zeta^+/2)$ for $-21 < \log |\zeta^+|  < -11$ for positive and negative $\zeta^+$.
Locally, the kink can be described in terms of the changes to the unit tangent
vector $e_t$, the velocity vector $e_v$ and
$e_\perp = e_v \times e_t/|e_v \times e_t|$ which form the
perpendicular coordinate system $\{e_t, e_v, e_\perp\}$.
We find $e_v$ and $e_t$ lie in the y-z plane and
$e_\perp$ along the x-direction. Letting $\Delta e = e_{+} - e_{-}$ stand for
the change in time of each unit vector,
\begin{eqnarray}
  \Delta e_\perp & = & \{2,0,0\} \\
  \Delta e_t & \simeq & \{0,-1.85,0\} \\
  \delta e_v & \simeq & \{0,-0.77,0\} .
\end{eqnarray}
The kink is a y-reflection of the
velocity and tangent vectors in the y-z plane.

On each side of the kink we fit each component of $F^\alpha$
with forms that include combinations of constant, linear
and ln terms in $|\zeta^+|$. We select the linear
or ln fit whichever is best; it turns out that this corresponds to
the term with coefficients that are of order unity.
We report the inferred scaling in Table \ref{tab:GVkinkasymptotics}.
\begin{table}[H]
  \begin{center}
\begin{tabular}{|c|cccc|}
  \hline
  sign & $F^t$ & $F^x$ & $F^y$ & $F^z$ \\
$\zeta^+ < 0$ & $|\zeta^+|^{-0.33}$ & $1$ & $|\zeta^+|^{-0.33}$ & $|\zeta^+|^{-0.33}$ \\
  $\zeta^+ > 0$ & $1$ & $1$ & $1$ & $1$ \\
    \hline
\end{tabular}
  \end{center}
  \caption{Asymptotic form for the force near the kink; $1$ means non-zero
    constant.}
\label{tab:GVkinkasymptotics}
  \end{table}
The results are similar to but not identical to the ACO
case. First, note that there is one redundancy $F^t=-F^z$
so we have 3 GV force components to compare to ACO.
The GV coordinate directions of the force
are not the same as the normal and longitudinal
directions in the ACO case and this complicates a
direct one-for-one comparison. Nonetheless, we see analogous
behavior. Most prominently the GV divergence for $\zeta^+<0$
of $F^t$ and $F^y$ scales close to $\propto (|\zeta^+|)^{-1/3}$
like ACO's $F^t$ and $F^N$. One difference is that the GV force for all
components with $\zeta^+>0$ approach non-zero constant values.
The ACO loop has no curvature
on one side of the kink, which is probably responsible for
the fact that two of its components approach zero. Curiously,
the ACO divergence for $F^L \propto \ln | \zeta^+ |$
on both sides of the kink is
absent for any components in the GV case. Likewise, the completely
finite GV result for $F^x$ on both sides of the kink
is absent in the ACO case. Despite these differences
the most important observation is that the GV divergent self-force
$\propto  (|\zeta^+|)^{-1/3}$ integrates to a finite value
so we expect the
physical displacement of the string to be finite.

\subsection{Kibble self-intersecting strings}
The ACO and GV string possess a pair of traveling kinks that circulate
around the loop throughout the period of oscillation while the KT
string forms two transient cusps each period.  In the tangent sphere
representation the kink discontinuities are jumps in ${\mathbf a}'$
and/or ${\mathbf b}'$ while the cusps form whenever ${\mathbf a}'$ and
${\mathbf b}'$ cross. The nature of the self-intersections of string
loops is not immediately apparent from the tangent sphere
representation. In the case of the KT string with $\alpha=0$ and
$\phi=\pi/6$ the string collapses to a line and the overlap is a
spacelike length of string.  Unless nature prefers special loop
configurations the generic type of self-intersection will be weaker
than in the above KT case. Here we investigate the Kibble string loop
\cite{Kibble:1976sj} which is simpler than any of the previous cases
in these respects: it has no discontinuities or crossings on the
tangent sphere, i.e. the loop is smooth and continuous everywhere,
{\it and} it self-intersects at a spacetime point not along a
spacelike line.

\begin{figure}[H]
\centering
\includegraphics[width=0.9\linewidth]{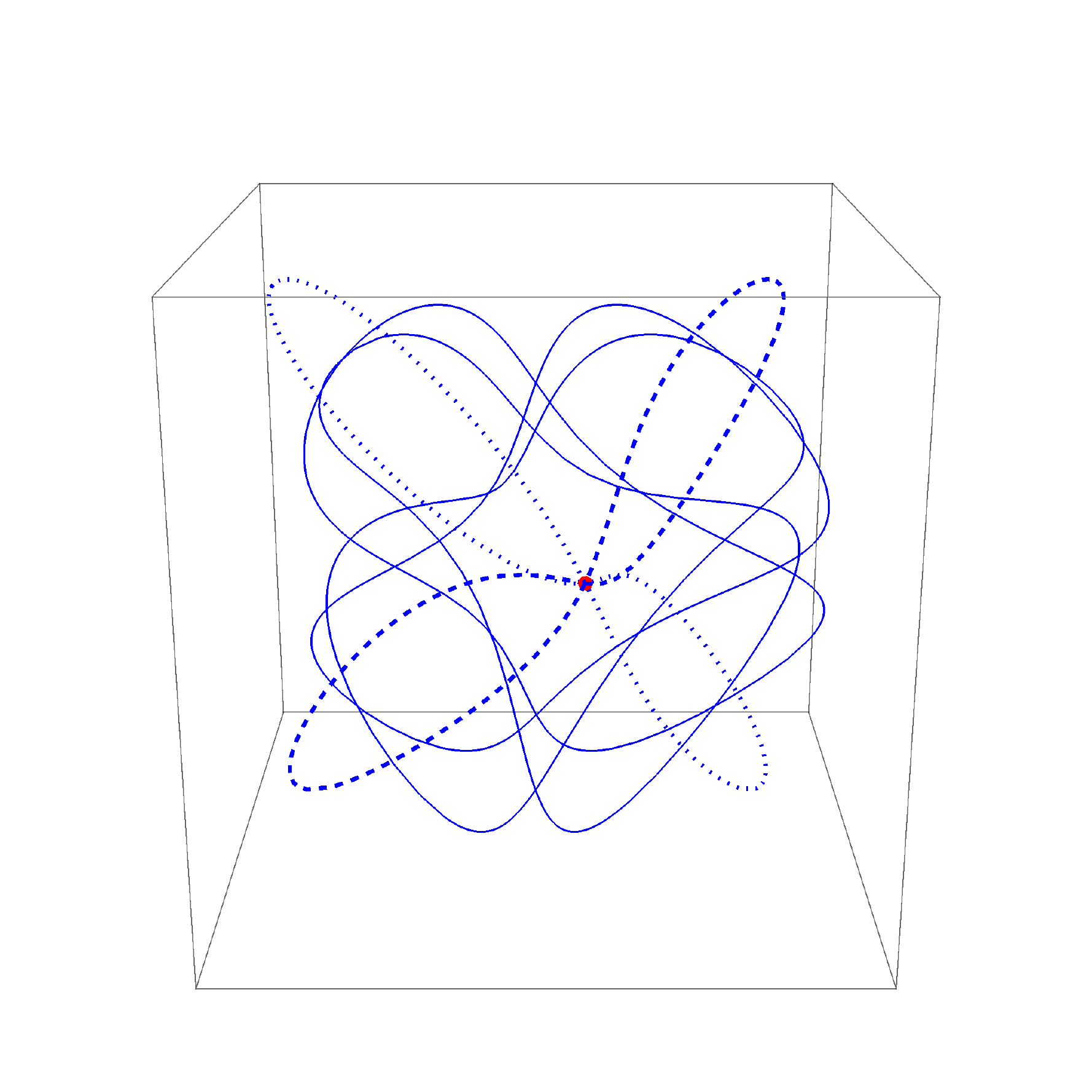}
\caption{\label{fig:Kibbleloopspacetime} The Kibble string loop
  configuration for $p=1/2$ in spacetime at six equally
  spaced moments $\tau = j \pi/6$ for $j=0$ to $5$ (invariant
  length $2 \pi$) in the basic loop oscillation cycle.
  The blue dashed and dotted loops self-intersect at
  the central red dot. The solid blue lines are non-intersecting configurations.}
~\\
\includegraphics[width=0.9\linewidth]{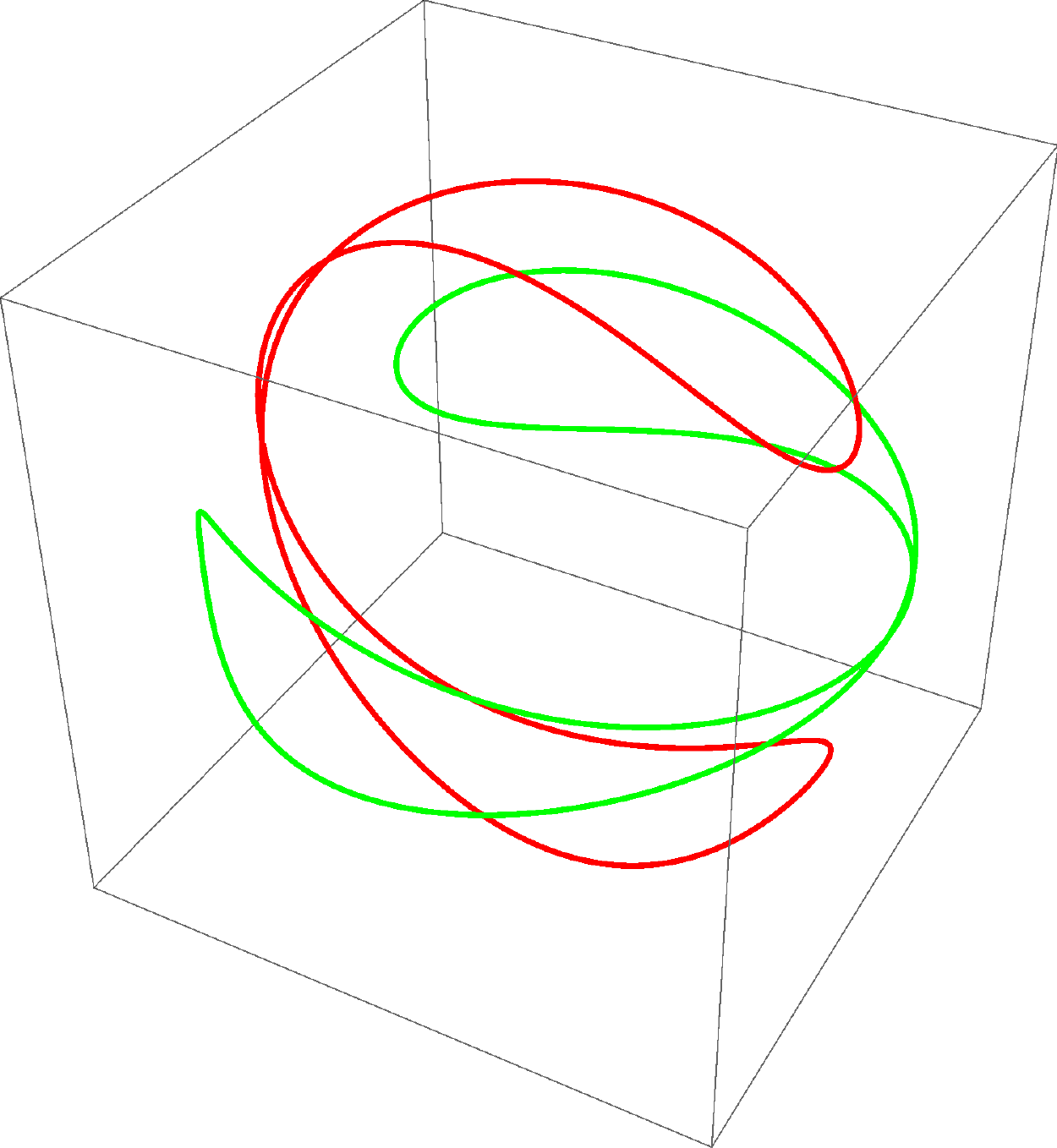}
\caption{\label{fig:Kibblelooptangentplot} The arcs on the tangent
  sphere traced by ${\mathbf a}'(x)$ and $-{\mathbf b}'(x)$ for the
  Kibble string resemble the seams on a baseball. The green and red
  lines are smooth and continuous and do not intersect each
  other. They satisfy an integral condition such that the loop has
  zero total momentum. }
\end{figure}

We integrate the tangent vectors \cite{Garfinkle:1987yw} to give
explicit forms for the right and left modes:
\begin{align}
  a^\mu(\zeta^+) & = \left[\zeta^+,
    f_1(\zeta^+), f_2(\zeta^+), f_3(\zeta^+) \right] \\
  b^\mu(-\zeta^-) & = \left[\zeta^-,
    -f_1(\zeta^-), -f_3(\zeta^-), -f_2(\zeta^-) \right]
\end{align}
where
\begin{align}
  f_1(x) & =  \frac{L}{2\pi} \left(
  \frac{
    (1+p^2)^2 \sin 2 y +
    (p^2/4) \sin 4 y}{2 + 5 p^2 + 2 p^4}
  \right)
       \\
  f_2(x) & = \frac{L}{2 \pi} \cos 2 y \times \nonumber \\
       & \quad \left(
  \frac{ -2 + 4 p^2 + 2 p^4 + p^2 \cos 2 y }
       {4 + 10 p^2 + 4 p^4}
       \right)   \\
  f_3(x) & = \frac{L}{2\pi} 2^{3/2} p \cos y \times \nonumber \\
  & \quad \left(
  \frac{ 5 + 3 p^2 + 2 \cos 2 y }
       {6 + 15 p^2 + 6 p^4}
       \right)
\\
       y & = \frac{2 \pi x}{L}
\end{align}
where $p$ is a constant. We choose for the numerical example $p=1/2$.
This is a more complicated loop in terms of harmonic
content than either the GV or KT loops.
Fig. \ref{fig:Kibbleloopspacetime} shows 6 equally spaced
snapshots of the loop during the fundamental oscillation period. The
dashed and dotted lines show the times when a self-intersection
occurs at the center (red dot). Figure \ref{fig:Kibblelooptangentplot}
gives the tangent sphere representation which resembles the seams
of a baseball.

\begin{widetext}
~
\begin{figure}[H]
\centering
\includegraphics[width=0.9\linewidth]{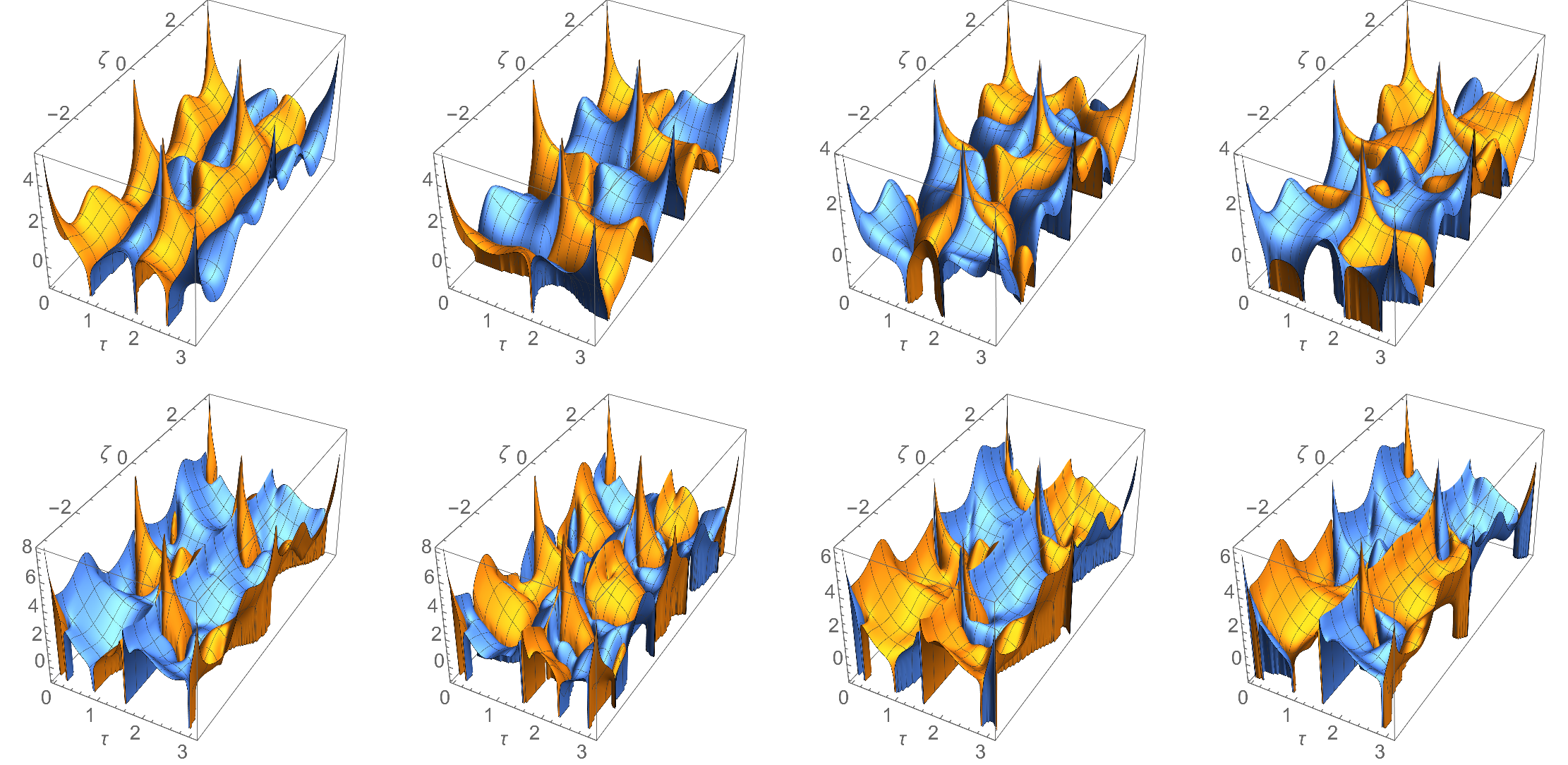}
\caption{\label{fig:Force-components-K}
Contributions to $F_1^\mu$ for the Kibble string when computed using the [1D] integration method with integration with respect to $\zeta$. Each sub-figure shows the relevant contribution to the force at all points on the string in the region $\tau \in (0,L/2)$, $\zeta \in (-L/2,L/2)$; all other points can be obtained from the standard periodic extension of the string. Each column corresponds to a different component of the force: $F_1^t$, $F_1^x$, $F_1^y$, and $F_1^z$. The rows correspond to the contributions from: (i) the field point; and (ii) the integral over $\zeta$ (ignoring distributional contributions at the field point). For the purposes of the plots, we have set the string tension, $\mu$, and Newton's constant, $G$ equal to one; other values simply introduce an overall scaling. Note that we have used a logarithmic scale and denoted positive (negative) values by coloring the plot orange (blue).
}
\end{figure}
\begin{figure}[H]
\centering
\includegraphics[width=0.9\linewidth]{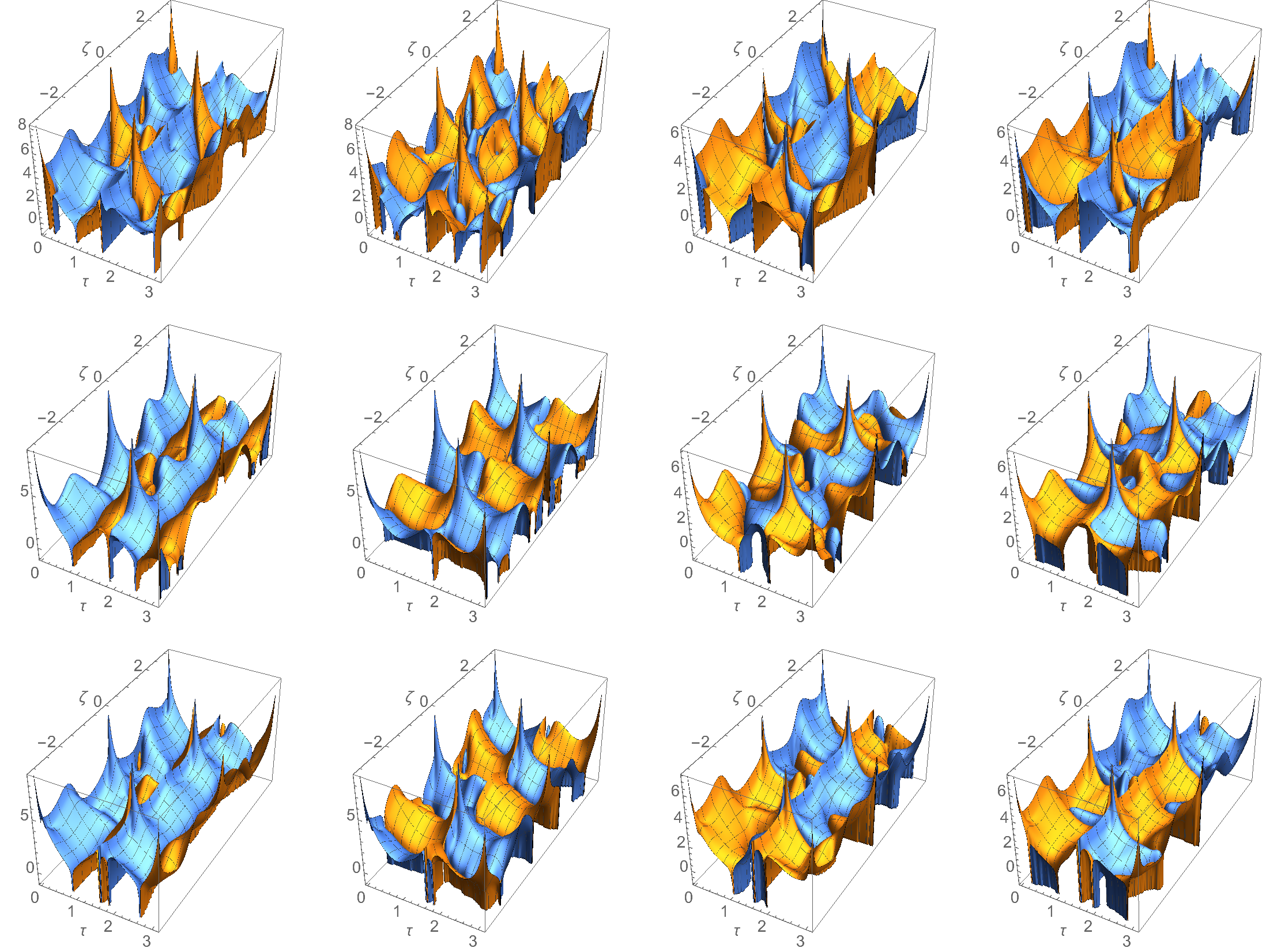}
\caption{\label{fig:Force-K}
The two pieces of the self-force, $F_1^\mu$ (row 1) and $F_2^\mu$ (row 2), for the Kibble string as a function of position on the string. The $F_1^\mu$ part can be obtained by summing the two rows in Fig.~\ref{fig:Force-components-K}. For the purposes of the plots, we have set the string tension, $\mu$, and Newton's constant, $G$ equal to one; other values simply introduce an overall scaling. Note that we have used a logarithmic scale and denoted positive (negative) values by coloring the plot orange (blue).
}
\end{figure}
\end{widetext}

This loop has collisions at worldsheet coordinates
$\{\tau,\zeta\}=\{0,\pm\pi/2\}$ and $\{\tau,\zeta\}=\{\pi/2,0\}$
and $\{\pi/2,\pi\}$.
We describe the limiting behavior near $\{\tau,\zeta\}=\{0,\pm\pi/2\}$.
The velocities of the two
bits of string are equal and opposite: ${\dot z}^i = \pm \{
0, -0.26, -0.26\}$. The tangent vectors are
$dz^i/d\zeta = \{-0.85, \pm 0.26, \mp 0.26 \}$ (an angle of
$\sim 0.94$ rad). The acceleration vectors are
${\ddot z}^i = \{0,-0.41, 0.41\}$. The gravitational radiation
emitted by each piece of string should be similar.

The net effect of the crossing is small. The bumps at the collision
points on the full scale worldsheet representations in Fig.
\ref{fig:Force-K}
are difficult to distinguish
at all. Here we look in more detail near those crossings.

Component $F^t$ is displayed in a small two-dimensional patch about
the crossing point in the top left plot of Fig. \ref{fig:Cross1-Kibble}. As $\tau \to 0$
at fixed $\zeta=\pi/2$ (the vertical line of small dots in
the picture) $F^t$ diverges $\propto \tau^{-1}$ with change
of sign as $\tau$ passes through zero. The results at $\pm \tau$ are nearly
equal and opposite. We find that the sum of the two components at $\pm \tau$ is
nearly constant as $|\tau| \to 0$, numerically approximately $\propto
|\tau|^{0.05}$. As $\zeta$ varies near $\pi/2$ (fixed
$\tau=0$, the horizontal line of small dots)
the results on each side of the crossing point are finite and the zero
value is not exactly at $\delta \zeta=0$. These results are quite
sensitive to the size of $\delta \tau$ since the surface changes
sign (from plus to minus infinity) near $\delta \tau=0$.
\begin{widetext}
	~
\begin{figure}[H]
\centering
\includegraphics[width=0.45\linewidth]{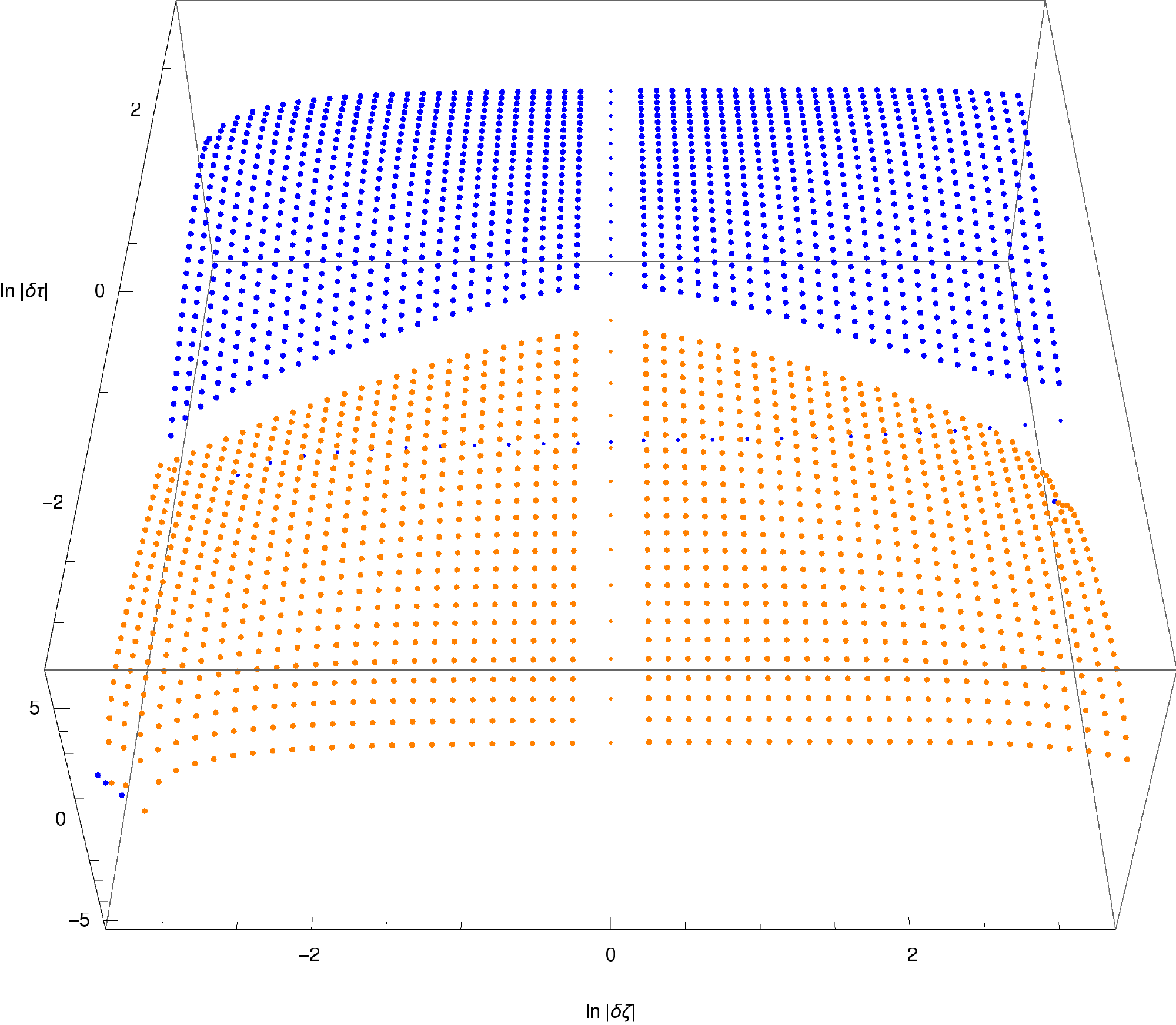}
\includegraphics[width=0.45\linewidth]{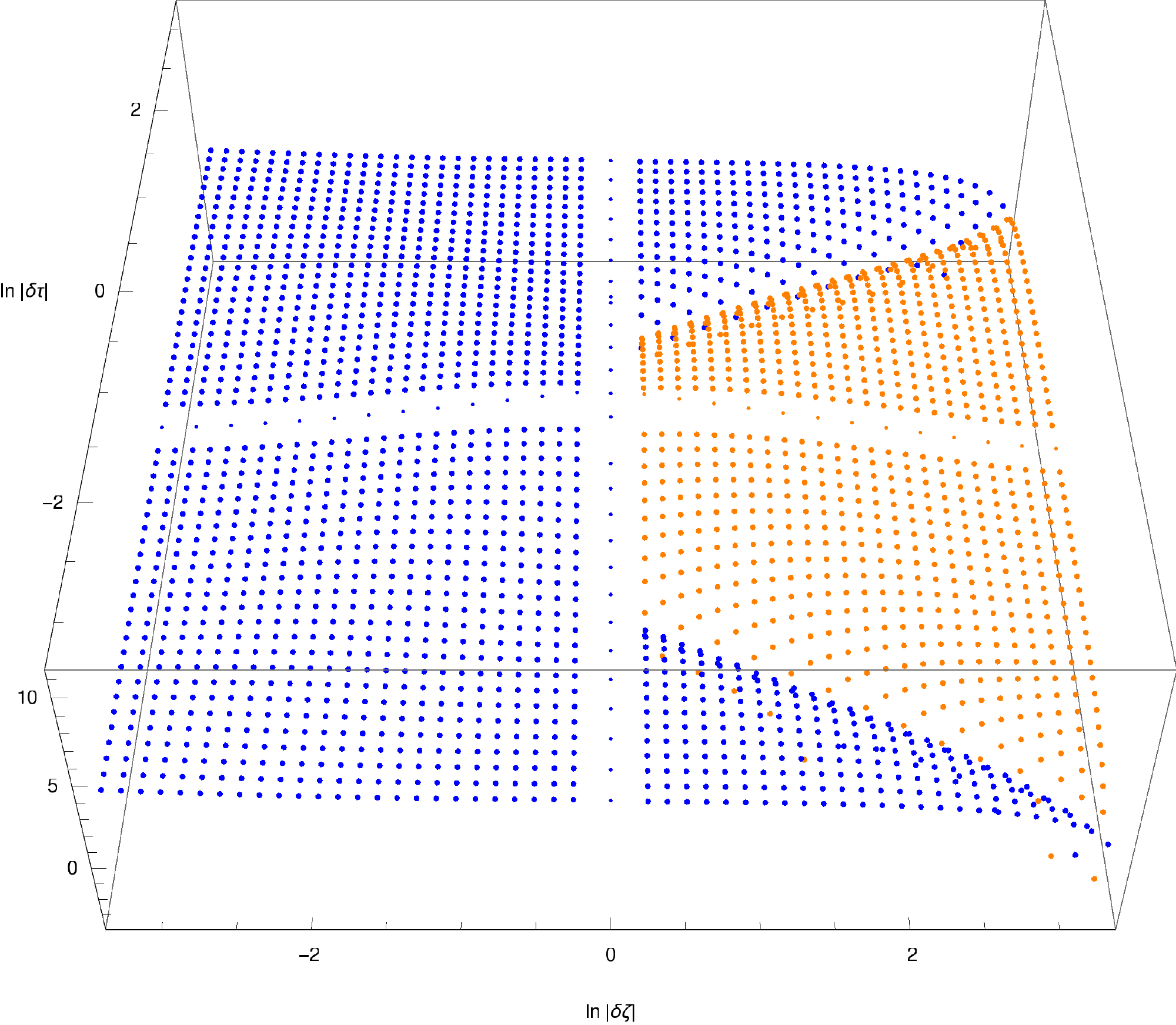}
\includegraphics[width=0.45\linewidth]{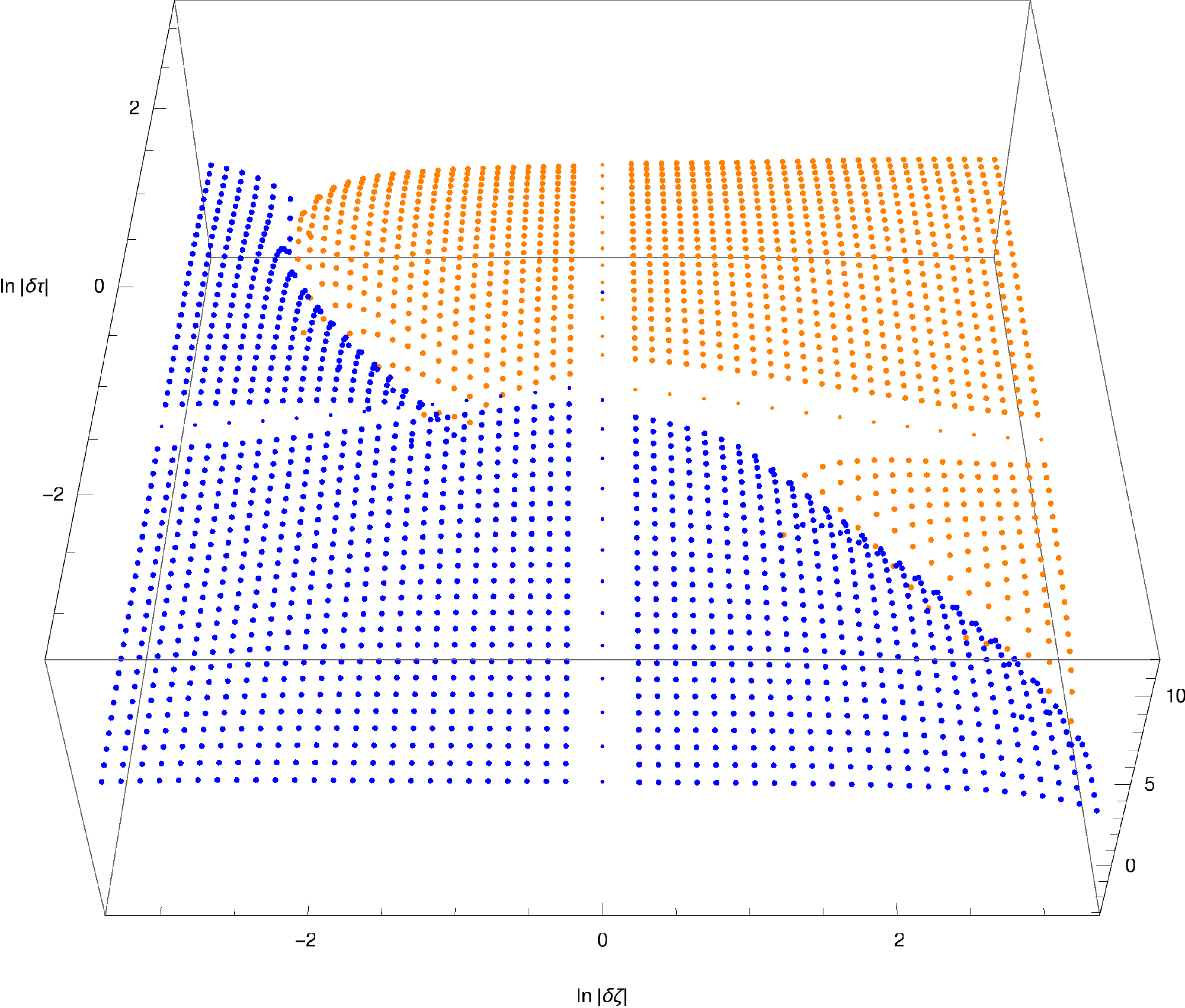}
\includegraphics[width=0.45\linewidth]{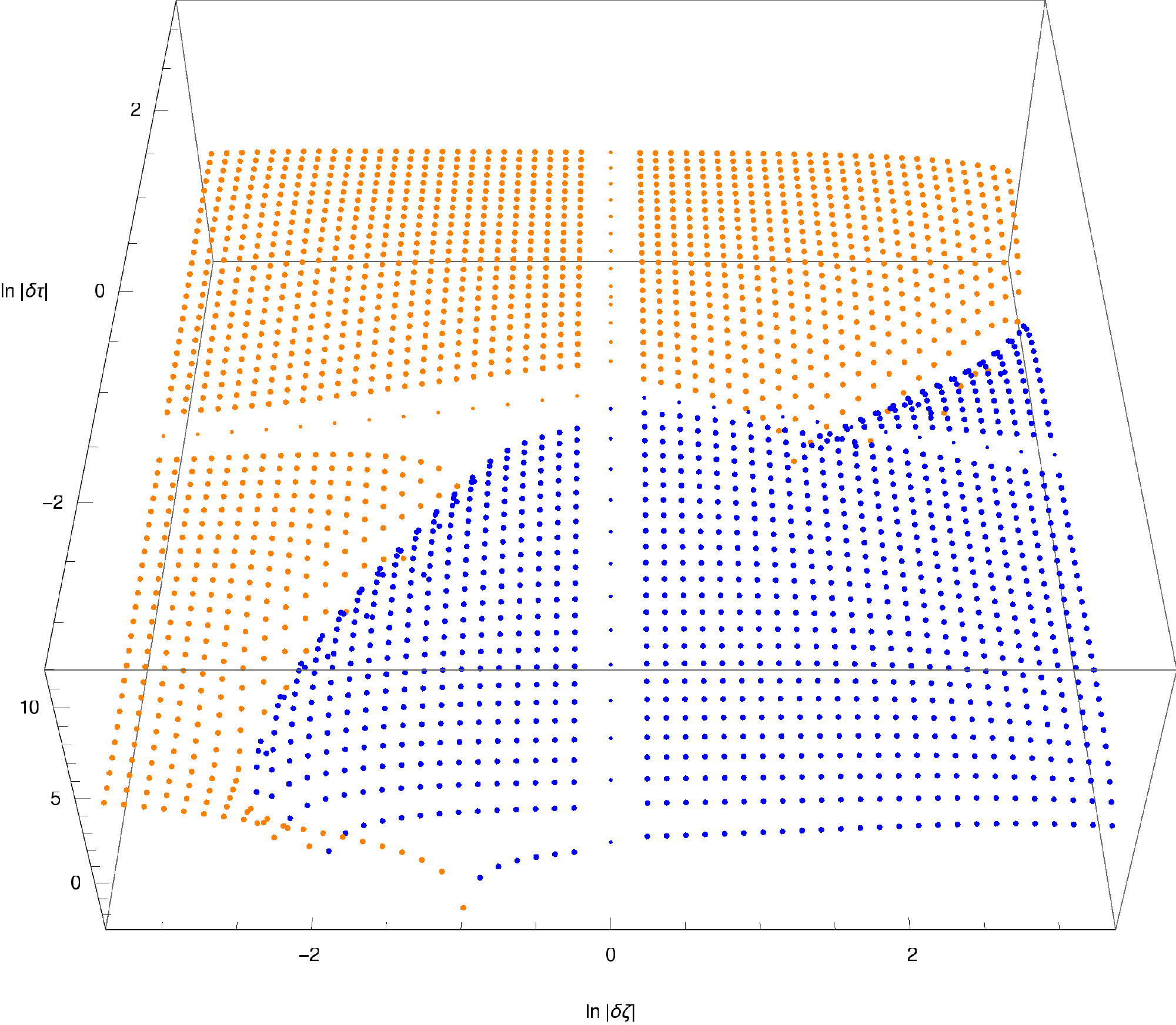}
\caption{\label{fig:Cross1-Kibble} The four components ($t$, $x$, $y$, and $z$) of
  $\ln |{\cal F}^t_{\rm conf}|$ for a small patch of the worldsheet about the crossing
  point.
  Four quadrants in $\{\delta \tau, \delta \zeta\}$ are displayed in
  log absolute value coordinates (oriented to match the usual linear
  system about $\{\tau,\zeta\}=\{0,\pi/2\}$).
  Blue (orange) are negative (positive) values. The small dots have
  been added for $\delta \tau=0$ and $\delta \zeta=0$.
}
\end{figure}
\end{widetext}

We have
formally fit the power law variation for $\delta \tau$ near $\tau=0$
and for $\delta \zeta$ near $\zeta = \pi/2$.
Table \ref{tab:Kibblescaling} summarizes the slopes extracted for
$F^\alpha$ along fixed $\tau$ and fixed $\zeta$ coordinates passing
exactly through the crossing point. Some components vary such that an
integral over just one side would yield a divergent quantity, however,
the symmetric sum is always integrable.
\begin{table}[H]
  \begin{center}
\begin{tabular}{|c|cc|cc|}
  \hline
  Component & \multicolumn{2}{c|}{$\zeta$ varies; $\tau=0$} & \multicolumn{2}{c|}{$\tau$ varies; $\zeta=\pi/2$} \\
            & One side & Net & One side & Net \\
  \hline
  $F^t$ & - & - & $-1.0$ & $0.04$ \\
  $F^x$ & $-1.1$ & $-0.04$ & $-0.04$ & $-0.03$ \\
  $F^y$ & $-1.1$ & $-0.06$ & $-1.0$ & $-0.05$ \\
  $F^z$ & $-1.1$ & $-0.05$ & $-0.96$ & $-0.04$ \\
  \hline
\end{tabular}
\end{center}
  \caption{Kibble loop divergent behavior at the crossing point
    $\tau=0$ and $\zeta=\pi/2$. The columns labeled $\tau=0$ give $\nu$
    for the scaling of the force component along the string near the
    crossing point $\propto |\zeta-\pi/2|^\nu$. Likewise,
    the ones labeled $\zeta=\pi/2$ describe the scaling
    $\propto |\tau|^\nu$ for times before and after the appearance of
    the crossing. ``One side'' means the scaling of the absolute value
    (approximately the same on each side);
    ``Net'' means the scaling of the symmetric sum of points on opposite sides
    of the crossing point. Small $\nu$ results are numerically close to
    finite limits but in any case are integrable.
    The ``-'' indicates values that are not
    well-defined because of a zero-crossing at $\tau=0$.}
\label{tab:Kibblescaling}
\end{table}

The plot of $F^t$ shows it to be
approximately a product of individual functions of
$\tau$ and $\zeta$. The other force components are more complicated.
Components $F^x$, $F^y$ and $F^z$ are shown in small two-dimensional
patches in the other plots of Fig.~\ref{fig:Cross1-Kibble}. The small dots show the
variation along the coordinate axes.

In summary, we find the effect of the string crossing leads to integrable forces
for all components in this example.

%%%%%%%%%%%%%%%%%%%%%%%
\subsection{Comparisons to radiated quantities evaluated in the far field}%
%%%%%%%%%%%%%%%%%%%%%%%

As an additional consistency check on our results, we compute the radiated energy and
compare it to the energy dissipated through the local self-force. The latter can be computed using the change in the 4-momentum of the string,
\begin{eqnarray}
  \Delta P^\mu & = & \mu \int {\cal F}^{\mu}_{\rm conf} d\zeta d\tau,
\end{eqnarray}
where the conformal-gauge force ${\cal F}^{\mu}_{\rm conf}$ is given by Eq.~\eqref{eq:trad-self-force}
and where the region of integration is given by the fundamental period of the worldsheet: $-L/2 \le
\zeta < L/2$ and $0 \le \tau < L/2$. In practice, we evaluate the integrand at
$N \sim 10^4$ equally spaced
points on a two dimensional surface and approximate the integral
as the sum of the function values at the points
times the worldsheet area per point. This is a low accuracy method that is
suited to the occurrence of steep spikes at various points on the worldsheet; we estimate the accuracy of the results to be within 1-5\%.

The work done on the string by the self-force lowers its energy, $\Delta P^0 < 0$, and should be
exactly balanced by the flux carried to infinity, which must be $-\Delta P^0 > 0$. We separately
compute this flux to infinity using the formalism of Allen and Ottewill
\cite{2001PhRvD..63f3507A}, in
which the stress energy tensor is a sum of individual Fourier components of the undamped string.
For each overtone $n$ we numerically integrate $dP^{(n)}/d\Omega$ over the sphere. We compute $N$
overtones and then fit and sum a power law extrapolation for $N \to \infty$. This yields a result
which is approximately 1\% accurate.

Table \ref{tab:energy} compares the results of the two calculations. We find good numerical
agreement within the expected accuracy of the result.
\begin{table}[H]
  \begin{center}
\begin{tabular}{|c|cc|cc|}
  \hline
   Case & far field & far field & direct & direct \\
    & (numerical) & (analytic) & (numerical) & (analytic) \\
  \hline
  ACO & 122.537 & 122.53 & 125.515 & 122.53 \\
  \hline
  KT & 349.677 & & 355.643 & \\
  $\alpha=0, \phi=\pi/6$& & & & \\
  \hline
  KT & 241.321 & & 238.259  & \\
  $\alpha=1/2, \phi=0$& & &  & \\
  \hline
  GV & 131.304 & & 132.486 & \\
  \hline
  Kibble & 137.6 & & 135.428 & \\
  \hline
\end{tabular}
\end{center}
\caption{The total energy loss integrated over one fundamental period of the loop oscillation in
  the center of mass frame of the loop. The far field is calculated with the formalism of Allen and Ottewill \cite{2001PhRvD..63f3507A}.
  The analytic result for the ACO loop in the far field is from
Ref.~\cite{Allen:1994bs}. The numerical results for the
direct energy loss integrate ${\cal F}^{\mu}_{\rm conf}$ over the world sheet according to the
description in this section. The analytic results for the direct energy loss for the ACO loop is
from Ref.~\cite{Anderson:2005qu}}
\label{tab:energy}
\end{table}

\section{Discussion}
We have developed a general method for calculating the self-force due to gravitational
perturbations of a lightly damped string loop. Our approach breaks up the calculation into smooth
integrals over the retarded image of the loop plus boundary terms. The latter are used to take
account of the special contributions when the source and field point coincide and when
discontinuities are visible on the past image of the loop. These may be from kinks or cusps or crossings (spacetime points where
intercommutation events might occur). Our methodology is quite general and can be used for arbitrary choices of
spacetime and worldsheet gauges.

There are some existing calculations of the gravitational self-force for cosmic strings
\cite{Quashnock:1990wv,Anderson:2005qu,Wachter:2016rwc}, however these results have all relied on
simplifications or approximations that do not hold in general. For example, although Quashnock and
Spergel \cite{Quashnock:1990wv} used a numerical approach not too different from ours, they do not
discuss any of the various distributional-type contributions (near kinks or the field point; they
do, however, discuss transitions between integration variables) that we have studied in detail
here. Our results\footnote{For example see the third column in Table \ref{tab:comparison} for the
GV case, but we also performed the same check for the other configurations discussed in this
paper.} suggest that their use of a pair of null coordinates sidesteps the issue of a contribution
from the field point. The issue of contributions from kinks, however, remains unaddressed.
Additionally, given the limited computational resources available at the time, their numerical
calculations were restricted to a low-resolution study in a restricted set of cases. In the case of
Refs.~\cite{Anderson:2005qu,Wachter:2016rwc}, approximations based on simple string configurations
were made which, while reasonable in some cases, do not fully capture the behavior for generic
string configurations.

Our numerical calculations have passed a number of validation checks including: comparisons with
existing analytical results; comparisons of the integrated power radiated over a fundamental period
against the flux of gravitational energy measured at large distances; and cross-comparisons of
several semi-independent methods for computing the self-force. From the perspective of
computational efficiency it is clear that the [1D] methods based on either integration with respect
to $\zeta$ or a Quashnock-Spergel type mixed integration with respect to $\zeta_+$ and $\zeta_-$
are the best choice. The other [1D] methods (using a single null coordinate or over-retardation)
inevitably encounter large numerical cancellations nearby the field point, making them significantly
more computationally demanding. The [2D] method is even worse, and is orders of magnitude more
demanding than any of the [1D] methods.

While the preferred [1D] methods work well in general, there are certain cases where they also run
into numerical challenges. Since the self-force diverges as one approaches kinks and cusps (in a
way such that the displacement of the worldsheeet is finite) it is unavoidable
that one would encounter numerically divergent quantities at one point or another. In this work, we
handled the issue of divergences in a brute force manner by simply evaluating quantities to a
sufficiently high accuracy that they can be canceled to leave a residual which is still accurately
determined. While this approach works reasonably well, the calculation could be made significantly
more efficient by developing an alternative approach to the problem. One promising possibility is
to borrow from results in the point particle case
\cite{Vega:Detweiler:2008,Barack:Golbourn:2007,Wardell:2015ada}, where it was found that the
separation of the full metric perturbation into a so-called ``puncture field'' that captures the
singular behavior plus a ``residual field'' that is more numerically well-behaved. In the point
particle case, by basing the puncture field on an approximation to the singular field proposed by
Detweiler and Whiting \cite{Detweiler-Whiting-2003}, one can work directly with the residual field
as it is entirely responsible for driving the motion. In the case of a cosmic string we do not yet
have an analogous Detweiler-Whiting type singular field. One could attempt to derive one following
the matched expansion methods of Ref.~\cite{Pound:2009}. Alternatively, even without such a
derivation a local analysis of the type done in Sec.~\ref{sec:local-expansion} may yield an
approximation to the singular behavior of the metric perturbation which leaves a numerically
well-behaved residual field, and which is sufficiently simple that its integrated contribution to
the motion can be determined analytically. Indeed, a preliminary analysis for the ACO string (where
the self-force is known analytically) suggests that exactly this approach will work well, and has
been found to significantly improve the accuracy with which the integrated motion can be
determined, even in the presence of a divergent self-force at the kinks.

The ultimate goal of our program is to evolve cosmic strings under the influence of the
self-force, and to study the consequences of backreaction on cusp formation, smoothing of kinks,
and other astrophysically relevant features of cosmic strings. This paper represents the first step
in such an endeavor. We can now compute the self-force for an arbitrary cosmic string with a
reasonable level of accuracy and with the freedom to arbitrarily choose coordinates and gauges which
are most suitable for evolution. The next step is to implement this into a numerical evolution
scheme. This will be presented in a future work.

\section*{Acknowledgments}
Concurrent to our
own work Blanco-Pillado, Olum and Wachter did related
work on cosmic string back-reaction; that paper and this one were
submitted at the same time. As far as we know, the results are in
agreement where they overlap.
We thank J.J. Blanco-Pillado, David Nichols, Ken Olum,
Adrian Ottewill, Joe Polchinski, Leo Stein, Peter Taylor, Henry Tye, Jeremy Wachter
and Yang Zhang for helpful conversations.
B.W. and D.C. gratefully acknowledge support from the John Templeton
Foundation New Frontiers Program under Grant No.~37426 (University of
Chicago) - FP050136-B (Cornell University). D.C. acknowledges that this
material is based upon work supported by the National Science
Foundation under Grant No. 1417132.  EF acknowledges the support of
the National Science Foundation under Grant Nos. 1404105 and 1707800.

\appendix

\section{Adapted Tetrads on the Worldsheet}

\subsection{Definition of adapted tetrad}

Suppose that we have a set of four linearly independent basis vectors ${\vec e}_{{\hat 0}}$, ${\vec
e}_{{\hat 1}}$, ${\vec e}_{{\hat 2}}$, ${\vec e}_{{\hat 3}}$, defined on the worldsheet, where
${\vec e}_{{\hat 0}}$ is timelike and the other vectors are spacelike. If ${\vec e}_{{\hat 0}}$ and
${\vec e}_{{\hat 1}}$ are tangent to the worldsheet, and if the other two vectors are orthogonal,
we'll say that the tetrad is adapted to the worldsheet. Such tetrads are convenient since the four
vectors can be used as a basis for spacetime tensors, while the first two vectors can be used as a
basis for worldsheet tensors. We will not require in the following that the basis be orthogonal or
orthonormal.

We now introduce the following index notations. Hatted lowercase Greek indices will run over ${\hat
0},{\hat 1},{\hat 2},{\hat 3}$, so for example an expansion of a vector ${\vec v}$ on the
orthonormal basis will be written as
\begin{equation}
{\vec v} = v^{\hat \alpha} {\vec e}_{\hat \alpha}.
\label{eq:ex}
\end{equation}
We will use hatted capital Roman indices ${\hat A}$, ${\hat B}$, $\ldots$
%from the start of the alphabet
to run over ${\hat 0}$, ${\hat 1}$, the directions along the worldsheet, and hatted capital Greek
indices ${\hat \Gamma}$, ${\hat \Sigma}$, $\ldots$ to run over ${\hat 2}$, ${\hat 3}$, the
directions orthogonal to the worldsheet. So the decomposition \eqref{eq:ex} of a general vector can
be rewritten as
\begin{equation}
{\vec v} = v^{{\hat A}} {\vec e}_{{\hat A}} + v^{{\hat \Sigma}} {\vec
  e}_{{\hat \Sigma}}.
\end{equation}

We define the dual basis of one forms $w^{{\hat \alpha}}_{\ \, \alpha}$ by
\begin{equation}
w^{{\hat \alpha}}_{\ \, \alpha} e_{{\hat \beta}}^{\ \, \alpha} =
\delta^{{\hat \alpha}}_{{\hat \beta}}.
\end{equation}
The basis vectors and dual basis vectors can be used to express tetrad basis components of tensors
in terms of coordinate basis components and vice versa in the usual way:
\begin{equation}
v^{\hat \alpha} = w^{{\hat \alpha}}_{\ \, \alpha} v^\alpha,
\ \ \ \ \ v_{\hat \alpha} = e_{{\hat \alpha}}^{\ \, \alpha} v_\alpha,
\end{equation}
etc.

\subsection{Examples of adapted tetrads}

Let us adopt the conventional parameterization of the background worldsheet \cite{Vilenkin:2000jqa}:
\begin{equation}
{\vec X}(\tau,\zeta) = \left( \tau, {1 \over 2} {\bf a}(\zeta -
  \tau) + {1 \over 2} {\bf b}(\zeta + \tau) \right),
\end{equation}
where ${{\bf a}'}^2 = {{\bf b}'}^2 = 1$. [Here the notation is that bold faced quantities are three
vectors, and quantities with arrows are four vectors.] We can then define an orthonormal tetrad of
basis vectors
\begin{subequations}
\begin{align}
{\vec e}_{\hat 0} = f_0 \bigg( 2, - {\bf a}' + {\bf b}'
\bigg), \\
{\vec e}_{\hat 1} = f_1 \bigg( 0,  {\bf a}' +  {\bf b}'
\bigg), \\
{\vec e}_{\hat 2} = f_2 \bigg( 1 - {\bf a}' \cdot {\bf b}',  -{\bf a}' +  {\bf b}'
\bigg), \\
{\vec e}_{\hat 3} = f_3 \bigg( 0,  {\bf a}' \times  {\bf b}'\bigg),
\end{align}
\label{eq:tetrad}
\end{subequations}
where $f_0 = f_1 = [2 (1 + {\bf a}' \cdot {\bf b}')]^{-1/2}$, $f_2 = \left[ 1 - ({\bf a}' \cdot
{\bf b}')^2 \right]^{-1/2}$ and $f_3 = 1 / | {\bf a}' \times {\bf b}'|$. This is an adapted tetrad
since the vectors ${\vec e}_{\hat 0}$ and ${\vec e}_{\hat 1}$ point along the worldsheet (they are
proportional to $\partial_\tau$ and $\partial_\zeta$), while ${\vec e}_{\hat 2}$ and ${\vec
e}_{\hat 3}$ are orthogonal to it. Another choice of adapted tetrad is given Eqs.~\eqref{eq:tetrad} but with the coefficients $f_{\hat \alpha}$ set to unity. This is an orthogonal
tetrad but not an orthonormal tetrad, and might be more convenient to use in computations.

\subsection{Geometric quantities in terms of the tetrad basis}

The spacetime metric on the tetrad basis is
\begin{equation}
g_{{\hat \alpha}{\hat \beta}} = {\vec e}_{\hat \alpha} \cdot {\vec
  e}_{{\hat \beta}}
\end{equation}
It follows from the definition of the adapted tetrad that this has block diagonal form with two
$2\times2$ subblocks, i.e. that $g_{{\hat A}{\hat \Gamma}} = 0$. Also the induced metric on the
tetrad basis is just one of the $2\times2$ subblocks:
\begin{equation}
\gamma_{{\hat A}{\hat B}} = {\vec e}_{{\hat A}} \cdot {\vec e}_{{\hat
    B}} = g_{{\hat A}{\hat B}}
\end{equation}
It also follows that
\begin{equation}
\gamma^{{\hat A}{\hat B}} = g^{{\hat A}{\hat B}}.
\end{equation}
Hatted indices are raised and lowered with $g_{{\hat \alpha}{\hat \beta}}$. For vectors $v^{\hat
\alpha}$ parallel to the worldsheet we have $v^{{\hat \Gamma}} = 0$ and $v_{{\hat \Gamma}} = 0$, so
indices can equivalently be raised and lowered just with $\gamma_{{\hat A}{\hat B}}$.

The projection tensor \eqref{eq:projection} can be expressed in terms of the tetrad vectors and
dual vectors as
\begin{equation}
P^\beta_{\ \,\gamma} = e_{{\hat A}}^{\ \,\beta} w^{{\hat A}}_{\ \,\gamma},
\end{equation}
and the orthogonal projection tensor is
\begin{equation}
\perp^\beta_{\ \,\gamma} = e_{{\hat \Gamma}}^{\ \,\beta} w^{{\hat \Gamma}}_{\
  \,\gamma}.
\end{equation}
Inserting these expressions into the definition \eqref{eq:Kdef} of the extrinsic curvature tensor
gives
\begin{align}
K_{\mu\nu\rho} &=& - P_\mu^{\ \,\alpha} P_{\nu\beta} \nabla_\alpha
\perp^\beta_{\ \,\rho} \nonumber \\
&=& - e_{\hat A}^{\ \,\alpha} w^{\hat A}_{\ \,\mu}
g_{\nu\lambda} e_{\hat B}^{\ \,\lambda} w^{\hat B}_{\ \, \beta}
\nabla_\alpha ( e_{{\hat \Gamma}}^{\ \,\beta} w^{\hat \Gamma}_{\ \,\rho})
\nonumber \\
&=& - e_{\hat A}^{\ \,\alpha} w^{\hat A}_{\ \,\mu}
g_{\nu\lambda} e_{\hat B}^{\ \,\lambda} w^{\hat B}_{\ \,\beta}
\nabla_\alpha ( e_{{\hat \Gamma}}^{\ \,\beta}) w^{\hat \Gamma}_{\ \,\rho}.
\end{align}
It follows that the nonzero components of the extrinsic curvature tensor on the tetrad basis are
\begin{equation}
K_{\hat A\ \  {\hat \Gamma}}^{\ \,{\hat B}} = - e_{\hat A}^{\ \,\alpha}
w^{\hat B}_{\ \,\beta} \nabla_\alpha e_{\hat \Gamma}^{\ \,\beta}.
\end{equation}
Note that this formula is valid for an arbitrary adapted tetrad, not just an orthonormal one.

\subsection{Explicit Form of equation of motion in orthogonal gauge using
adapted tetrad}

\label{app:gauge}
\subsubsection{Orthogonal Gauge}

An alternative worldsheet gauge choice is to impose that the displacement vector be orthogonal to
the worldsheet everywhere:
\begin{equation}
P^\alpha_\beta z_{(1)}^\beta = 0
\end{equation}
or
\begin{equation}
\perp^\alpha_\beta z_{(1)}^\beta = z_{(1)}^\alpha.
\end{equation}
From the general form \eqref{eq:gaugec} of linearized gauge transformations we see that is always
possible to achieve this gauge.

\subsubsection{Form of equation of motion}

In orthogonal gauge, the displacement vector is orthogonal to the worldsheet, so we can express it
as an expansion in terms of two components on an adapted tetrad
\begin{equation}
{\vec z_{(1)}} = z_{(1)}^{{\hat \Gamma}} {\vec e}_{{\hat \Gamma}}.
\label{eq:2c}
\end{equation}
We now write the general equation of motion \eqref{eq:final} in the schematic form
\begin{equation}
L[{\vec z_{(1)}}]^\alpha = {\cal F}^\alpha,
\end{equation}
where $L$ is the differential operator that appears on the left hand side, and ${\cal F}$ is the
forcing term linear in $h_{\alpha\beta}$ on the right hand side. Writing this equation on the
orthonormal basis gives
\begin{equation}
L[{\vec z_{(1)}}]^{\hat \alpha} = {\cal F}^{\hat \alpha}.
\end{equation}
Now the ${\hat \alpha} = {\hat 0}$ and ${\hat \alpha} = {\hat 1}$ components of this equation
vanish identically, since both sides are perpendicular to the worldsheet. So, we end up with an
equation with only two components:
\begin{equation}
L[{\vec z_{(1)}}]^{\hat \Gamma} = {\cal F}^{{\hat \Gamma}}.
\end{equation}
This will give two coupled equations for the two components $z_{(1)}^{{\hat 2}}$ and
$z_{(1)}^{{\hat 3}}$ of the displacement vector.

We now derive the following explicit form of the differential operator on the orthonormal basis:
\begin{equation}
L[{\vec z_{(1)}}]^{{\hat \Gamma}} = \triangle z_{(1)}^{{\hat \Gamma}} + {\cal
  K}^{A{\hat \Gamma}}_{\ \ \ {\hat \Sigma}} \partial_A z_{(1)}^{{\hat
    \Sigma}} + {\cal M}^{{\hat \Gamma}}_{\ {\hat \Sigma}} z_{(1)}^{{\hat \Sigma}}.
\label{eq:final2}
\end{equation}
Here $\partial_a$ denotes a derivative with respect to the worldsheet coordinates $\zeta^a$, and
$\triangle$ is the scalar differential operator
\begin{equation}
\triangle = {1 \over \sqrt{\gamma}} \eta^{ab} \partial_a \partial_b.
\end{equation}
The mass matrix ${\cal M}$ is given by
\begin{equation}
{\cal M}^{{\hat \Gamma}}_{\ {\hat \Sigma}} = w^{{\hat \Gamma}}_{\
 \,   \alpha} \triangle e^{\ \,\alpha}_{{\hat \Sigma}}
+ \lambda_{{\hat \Sigma}{\hat A}{\hat B}} K^{{\hat A}{\hat B}{\hat
    \Gamma}}.
\label{eq:massmatrix}
\end{equation}
Also
\begin{equation}
%\lambda_{{\hat \Sigma}{\hat A}{\hat B}} = -\partial_\alpha
%e^{\ \gamma}_{{\hat \Sigma}} ( e_{\hat A}^{\ \, \alpha} e_{{\hat
%    B}\,\gamma} + e_{\hat B}^{\ \, \alpha} e_{{\hat
%    A}\,\gamma} ).
\lambda_{{\hat \Sigma}{\hat A}{\hat B}} =
- \gamma_{{\hat B}{\hat C}}
w^{\hat C}_{\ \,\alpha} e_{\hat
  A}^{\ \,\mu} \nabla_\mu e_{\hat \Sigma}^{\ \,\alpha}
- \gamma_{{\hat A}{\hat C}}
w^{\hat C}_{\ \,\alpha} e_{\hat
  B}^{\ \,\mu} \nabla_\mu e_{\hat \Sigma}^{\ \,\alpha}.
\end{equation}
Note that all the derivatives in this expression are along the worldsheet, so the expression is
well defined (the basis vectors are not defined off the worldsheet, so orthogonal derivatives are
not well defined).
%The quantities $\lambda$ are related to the commutators
%  $c_{{\hat \alpha}{\hat \beta}{\hat \gamma}}$, which are defined by $[
%    {\vec e}_{{\hat \alpha}}, {\vec e}_{{\hat \beta}} ] = c_{{\hat
%        \alpha}{\hat \beta}}^{\ \ \ {\hat \gamma}} {\vec e}_{{\hat
%        \gamma}}$, by
%$\lambda_{{\hat \Sigma}{\hat A}{\hat B}} = c_{{\hat \Sigma}{\hat
%    A}{\hat B}} + c_{{\hat \Sigma}{\hat B}{\hat A}}$.  Not all of the
%commutators are well defined since the orthonormal basis is not
%defined off the worldsheet.
Finally the quantity ${\cal K}$ is given by
\begin{equation}
{\cal K}^{a{\hat \Gamma}}_{\ \ \ {\hat \Sigma}} = {2 \over
  \sqrt{\gamma}} w^{{\hat \Gamma}}_{\ \,\alpha} \eta^{ab} \partial_b
e_{{\hat \Sigma}}^{\ \,\alpha}.
\end{equation}

\subsubsection{Derivation}

From Eq.~\eqref{eq:final}), dropping the Riemann term since we are working in flat spacetime, and
replacing all indices with hatted indices, we get
\begin{equation}
L[{\vec z_{(1)}}]^{{\hat \Gamma}} = \perp^{{\hat \Gamma}}_{\ \,{\hat \chi}}
{\bar \nabla}_{{\hat \mu}} {\bar \nabla}^{{\hat \mu}} z_{(1)}^{{\hat
    \chi}}
- 2
{\bar \nabla}_{{\hat \mu}} z_{(1)}^{\hat \alpha} K^{{\hat \mu}\ \, {\hat
    \Gamma}}_{\ \,{\hat \alpha}}.
\label{eq:rr}
\end{equation}
Consider first the second term in Eq.~\eqref{eq:rr}). Since the extrinsic curvature tensor is
parallel to the worldsheet on its first two indices, we can drop the bar on the derivative
operator. Also we can replace the indices ${\hat \mu}$ and ${\hat \alpha}$ by worldsheet indices
${\hat A}$ and ${\hat B}$, giving
\begin{equation}
- 2
K^{{\hat A}\ \, {\hat
    \Gamma}}_{\ \,{\hat B}} {\nabla}_{{\hat A}} z_{(1)}^{\hat B} =
- 2
K^{{\hat A}\ \, {\hat
    \Gamma}}_{\ \,{\hat B}}
e_{\hat A}^{\ \,\alpha} w^{\hat B}_{\ \,\beta}
{\nabla}_\alpha z_{(1)}^\beta.
\end{equation}
Now inserting the expansion \eqref{eq:2c} of the displacement vector gives
\begin{align}
- 2
K^{{\hat A}\ \, {\hat
    \Gamma}}_{\ \,{\hat B}}
e_{\hat A}^{\ \,\alpha} w^{\hat B}_{\ \,\beta}
{\nabla}_\alpha (z_{(1)}^{\hat \Gamma} e_{\hat \Gamma}^{\ \,\beta}) =
- 2
K^{{\hat A}\ \, {\hat
    \Gamma}}_{\ \,{\hat B}}
e_{\hat A}^{\ \,\alpha} w^{\hat B}_{\ \,\beta}
z_{(1)}^{\hat \Gamma}
{\nabla}_\alpha
e_{\hat \Gamma}^{\ \,\beta}, \nonumber \\
\end{align}
where we have used the orthonormality of ${\hat A}$ and ${\hat \Gamma}$ directions. This gives the
second term in the mass matrix \eqref{eq:massmatrix}.

To evaluate the first term in Eq.~\eqref{eq:rr}), we temporarily return to the coordinate form of
this term, the first term on the left hand side of Eq.~\eqref{eq:final}). Then we use the general
result (valid in arbitrary gauges) for this term that
\begin{equation}
\perp^\rho_{\ \,\chi} {\bar \nabla}_\mu {\bar \nabla}^\mu z_{(1)}^\chi =
\perp^\rho_{\ \,\chi} {1 \over \sqrt{\gamma}} \partial_a \left(
  \sqrt{\gamma} \gamma^{ab} \partial_b z_{(1)}^\chi \right).
\end{equation}
We now use the fact that we have chosen conformal gauge to zeroth order, so that $\sqrt{\gamma}
\gamma^{ab} = \eta^{ab}$. This gives
\begin{equation}
\perp^\rho_{\ \,\chi} {\bar \nabla}_\mu {\bar \nabla}^\mu z_{(1)}^\chi =
\perp^\rho_{\ \,\chi} {1 \over \sqrt{\gamma}} \eta^{ab} \partial_a \partial_b z_{(1)}^\chi.
\end{equation}
We now convert the $\rho$ and $\chi$ indices in this equation to orthonormal indices, by
multiplying by appropriate factors of $e$ and $w$, and also by inserting the expansion
\eqref{eq:2c} of the displacement vector. We use the formula
\begin{equation}
\triangle ( z_{(1)}^{{\hat \Gamma}} e^{\ \,\chi}_{{\hat \Gamma}}) =
\triangle ( z_{(1)}^{{\hat \Gamma}}) e^{\ \,\chi}_{{\hat \Gamma}} +
z_{(1)}^{{\hat \Gamma}} \triangle e^{\ \,\chi}_{{\hat \Gamma}} +
{2  \over \sqrt{\gamma}} \eta^{ab} \partial_a z_{(1)}^{{\hat
    \Gamma}} \partial_b e^{\ \,\chi}_{{\hat \Gamma}}.
\end{equation}
The three terms in this expression generate, respectively, the first term in Eq.~\eqref{eq:final2}, the first term in the mass matrix \eqref{eq:massmatrix}, and the second term in
Eq.~\eqref{eq:final2}).

\section{Coordinate systems within the class of conformal gauges}

Within the conformal gauge, there is freedom in the particular choice of
worldsheet coordinates $\{\zeta^1, \zeta^2\}$. For example, they may be chosen
to be space-time ($\zeta^a = [\tau, \zeta]$), null ($\zeta^a = [\zeta^-,
\zeta^+] \equiv [\tau-\zeta, \tau + \zeta] = [2\tau - \zeta^+, \zeta^+]$) or
semi-null ($\zeta^a = [\tau, \zeta^+] \equiv [\tau, \tau + \zeta]$). The
relation between derivatives in the various coordinates is given by
\begin{equation}
 \diffp*{}{{{\zeta^{+}}}}{\tau} = \diffp*{}{\zeta}{\tau}
   = \diffp*{}{{{\zeta^{+}}}}{{{\zeta^{-}}}} - \diffp*{}{{{\zeta^{-}}}}{{{\zeta^{+}}}},
\end{equation}
\begin{equation}
 \diffp*{}{\tau}{{{\zeta^{+}}}} = \diffp*{}{\tau}{\zeta} - \diffp*{}{{\zeta}}{\tau} = 2\diffp*{}{{{\zeta^{-}}}}{{{\zeta^{+}}}}.
\end{equation}

As mentioned in Sec.~\ref{sec:local-expansion}, the conformal gauge condition imposes that
derivatives of the worldsheet are related. In space-time coordinates the $\tau$
and $\zeta$ derivatives are related by,
\begin{equation}
  \partial_\tau z^\alpha \partial_\tau z_\alpha + \partial_\zeta z^\alpha \partial_\zeta z_\alpha = 0,
  \quad \partial_\tau z^\alpha \partial_\zeta z_\alpha = 0.
\end{equation}
Equivalently, the null derivatives are related by
\begin{equation}
  \partial_{\zeta^+} z^\alpha \partial_{\zeta^+} z_\alpha = 0, \quad
  \partial_{\zeta^-} z^\alpha \partial_{\zeta^-} z_\alpha = 0,
\end{equation}
and the semi-null derivatives are related by
\begin{equation}
  \partial_\tau z^\alpha \partial_\tau z_\alpha = 0, \quad
  \partial_\tau z^\alpha \partial_{\zeta^+} z_\alpha + \partial_{\zeta^+} z^\alpha \partial_{\zeta^+} z_\alpha = 0.
\end{equation}
Additionally, the equation of motion \eqref{eq:conformal-eom} reduces to
\begin{equation}
\label{eq:EOM-coord-conformal}
  \phi^{-1} \big[\partial_{\zeta\zeta} z^\alpha - \partial_{\tau\tau} z^\alpha\big] = 0
\end{equation}
in space-time coordinates, to
\begin{equation}
  - 4 \phi^{-1} \partial_{\zeta^+ \zeta^-} z^\alpha = 0
\end{equation}
in null coordinates, and to
\begin{equation}
\label{eq:EOM-coord-conformal-semi-null}
  - \phi^{-1} \big[\partial_{\tau\tau} z^\alpha + 2 \partial_{\zeta^+ \tau} z^\alpha\big] = 0
\end{equation}
in semi-null coordinates. The solutions can be written in terms of left-moving and right-moving
waves, i.e. in terms of two functions $a^\alpha(\zeta^+)$ and $b^\alpha(\zeta^-)$ that satisfy the
tangent sphere condition $g_{\alpha \beta} \partial_{\zeta^+} a^\alpha \partial_{\zeta^+} a^\beta = 0 = g_{\alpha \beta} \partial_{\zeta^-} b^\alpha
\partial_{\zeta^-} b^\beta$.

\section{Energy-Momentum Loss Formulae}
\label{sec:energymomentumlossdiscussion}
We review the QS \cite{Quashnock:1990wv} result for energy-momentum
loss by self-forces in light cone coordinates and conformal gauge.  We
note that these results can be generalized to arbitrary gauge
choice. Finally, we rewrite the energy-momentum loss formulae directly
in terms of the conformal-gauge force using Eq.~\eqref{eq:simple}.
%Although the form appears covariant, the traditional force is valid
%only for particular gauge choices.
We utilize
the result to evaluate the dissipative effects on the string.

QS analyze a string with mass per length
$\mu$, spacetime position of the worldsheet $x^\mu=z^\mu(\zeta^1,\zeta^2)$
where the two coordinates covering the worldsheet are $\zeta^a = \{\zeta^1,\zeta^2\}$, $g_{\mu\nu}=\{-,+,+,+\}$ and
$\zeta^a=\{\zeta^1,\zeta^2\}=\{\tau,\zeta\}$.
The Nambu-Goto action is
\begin{eqnarray}
  S & = & -\mu \int d^2\zeta \sqrt{-\gamma} \\
  \gamma_{ab} & = & g_{\mu\nu}\frac{\partial z^\mu}{\partial \zeta^a}
  \frac{\partial z^\nu}{\partial \zeta^b} \\
  \gamma & = & \det \gamma_{ab}
\end{eqnarray}
and $\gamma<0$. We have
the stress energy tensor [Eq.~(12.2.2) of \cite{1972gcpa.book.....W}]
\begin{eqnarray}
  T^{\mu\nu}(x) & = & \frac{2}{\sqrt{-g(x)}} \left( \frac{\delta S}{\delta g_{\mu\nu}(x)} \right) \\
  g & = & \det g_{\mu\nu} .
\end{eqnarray}
Now using ${\dot z}^\mu = z^\mu_{,\tau}$, $z'^\mu = z^\mu_{,\zeta}$,
${\dot z}^2 = g_{\mu\nu}{\dot z}^\mu{\dot z}^\nu$,
$z'^2 = g_{\mu\nu}{z'}^\mu{z'}^\nu$,
$z' \cdot {\dot z} = g_{\mu\nu}{z'}^\mu{\dot z}^\nu$ and
$\delta g_{\alpha\beta}(x)/\delta g_{\mu\nu}(x') = \delta^4(x-x') \left(
\delta^\mu_\alpha \delta^\nu_\beta +\delta^\nu_\alpha \delta^\mu_\beta \right)/2$ we can write
\begin{eqnarray}
T^{\mu\nu}  & = & \frac{\mu}{\sqrt{-g}} \int \frac{d^2 \zeta}{\sqrt{-\gamma}}
\delta^4(x-z(\zeta)) C^{\mu\nu} \\
C^{\mu\nu} & = & D^{\mu\nu} + E^{\mu\nu}   \\
    D^{\mu\nu} & = & {\dot z}^\mu {\dot z^\nu} (z')^2
    + z'^\mu z'^\nu ({\dot z})^2 \\
    E^{\mu\nu} & = & \left( z' \cdot {\dot z} \right)
    \left( z'^\mu {\dot z}^\nu + {\dot z}^\mu z'^\nu \right)
\end{eqnarray}
This differs from QS Eq.~(3.2) in two details: the power of the determinant
of the induced metric is $-1/2$ not $1/2$ and there is an
explicit occurrence of the determinant of the
spacetime metric.

The Lagrangian $L = -\mu \sqrt{-\gamma}$ leads to the equations of
motion [QS Eq.~(3.3)]. With the gauge choices ${\dot x} \cdot {x'}=0$ and
$({\dot x})^2 + ({x'}^2)=0$ we have
\begin{equation}
  \sqrt{-\gamma} = -({\dot x})^2 = (x')^2
\end{equation}
and QS Eq.~(3.6)
\begin{equation}
  {\ddot x^\nu} - {x''}^\nu = -\Gamma^\nu_{\alpha\beta}
  \left( {\dot x}^\alpha {\dot x}^\beta - {x'}^\alpha {x'}^\beta \right)  .
\end{equation}

In light cone coordinates $u \equiv \tau + \zeta$ and
$v \equiv \tau - \zeta$
these are
\begin{eqnarray}
  \partial_u \partial_v x^\mu & = &
  - \Gamma^\mu_{\alpha\beta} \partial_u x^\alpha \partial_v x^\beta \\
  g_{\alpha\beta} \partial_u x^\alpha \partial_u x^\beta & = & 0 \\
  g_{\alpha\beta} \partial_v x^\alpha \partial_v x^\beta & = & 0 .
\end{eqnarray}

In flat background these reduce to
\begin{equation}
  x^\mu_{(0),uv}=0
\end{equation}
with gauge conditions
\begin{equation}
  \eta_{\mu\nu}x^\mu_{(0),u} x^\nu_{(0),u} =
  \eta_{\mu\nu}x^\mu_{(0),v} x^\nu_{(0),v} =  0 .
\end{equation}
The subscripts here and below label powers of $\mu$. Our
$\phi=-\gamma_{(0)} >  0$.

For general background
the stress energy tensor is $T^{\mu\nu} \equiv \mu I^{\mu\nu}$
\begin{eqnarray}
  I^{\mu\nu} & = & \frac{1}{\sqrt{-g}}
  \int du dv G^{\mu\nu} \delta^{(4)}(x-z) \\
  G^{\mu\nu} & = & z^\mu_{,u} z^\nu_{,v} + z^\mu_{,v} z^\nu_{,u} .
\end{eqnarray}
The form is exact and doesn't explicitly involve $\gamma$.
This is QS Eq.~(3.12) supplemented with an explicit determinant of
the spacetime metric.
We can regard $G$ (and $\gamma$) as functions of $u$ and $v$ via
$x=z(u,v)$.

Integrating the covariant derivative of the stress
energy tensor $\nabla_\nu T^{\mu\nu}$ over a large cylinder with
spatial extent beyond the source and asymptotically flat
and between two time
slices over weight $\sqrt{-g} d^4x$ gives the covariant
conservation law
\begin{eqnarray}
  \int d^4x \sqrt{-g} \nabla_\nu T^{\mu\nu}  & = &
  \int d^4x \left( \sqrt{-g} T^{\mu\nu} \right)_{,\nu} + \nonumber\\
  & & \int d^4x \sqrt{-g} \Gamma^\mu_{\alpha\beta} T^{\alpha\beta} .
  \label{eqn:covariantconservationlaw}
\end{eqnarray}
The first term on the left is rewritten
\begin{eqnarray}
  \int d^4x \left( \sqrt{-g} T^{\mu\nu} \right)_{,\nu} & = &
  \int d^4x \left( \sqrt{-g} T^{\mu 0} \right)_{,0} + \nonumber\\
  & & \int d^4x \left( \sqrt{-g} T^{\mu i} \right)_{,i} .
\end{eqnarray}
Defining
\begin{eqnarray}
  P^\mu & \equiv & \int d^3x \sqrt{-g} T^{\mu0} \\
  \Delta P^\mu & \equiv & \int dt P^\mu_{,0}
\end{eqnarray}
and applying Gauss' law to the integral at spatial infinity
\begin{equation}
  \int d^3x \left( \sqrt{-g} T^{\mu i} \right)_{,i} = 0
\end{equation}
we find the change in the source momentum
\begin{equation}
  \int d^4x \left( \sqrt{-g} T^{\mu\nu} \right)_{,\nu} = \Delta P^\mu .
  \label{eqn:changeinsourcemomentum}
\end{equation}
Inserting this result into Eq.~\eqref{eqn:covariantconservationlaw}
and rearranging
\begin{eqnarray}
  \Delta P^\mu & = &
  \int d^4x \sqrt{-g} \left(
  \nabla_\nu T^{\mu\nu} - \Gamma^\mu_{\alpha\beta} T^{\alpha\beta} \right) \nonumber\\
  & = & \int d^4x \left( \sqrt{-g} T^{\mu\nu} \right)_{,\nu}
\end{eqnarray}
It is not too surprising that the form is identical to Eq.~\eqref{eqn:changeinsourcemomentum}.

Next, use the explicit form for the stress energy tensor on the right hand
side. How well do we need to know the terms?
Since $P \sim \mu$, if we wish $\Delta P \sim \mu^2$ then we need
$\sqrt{-g} T \sim \mu \sqrt{-g} I$ accurate to order $\mu^2$. Since
there is one factor of $\mu$ which is explicit one
needs $I=I_{(0+1)}$; but, as mentioned earlier, $I$ is exact.
\footnote{In our formalism, this is not the case; we start with
Eq.~\eqref{exactstressenergy}
with $g_{\mu\nu} \to \eta_{\mu\nu}$ and all other quantities evaluated for
the background (see
footnote 1). This expression is $I_{(0)}$.}

Integrate over a fundamental period in time and a
large volume containing the string source
\begin{eqnarray}
  \int (\sqrt{-g} T^{\mu\nu})_{,\nu} d^4x & = &
  \mu \int d^4x \partial_\nu \int du dv G^{\mu\nu} \times \nonumber\\
  & & \delta^4(x-z) \\
  & = & \mu \int d^4x \int du dv G^{\mu\nu} \times \nonumber\\
  & & \partial_\nu \left( \delta^4(x-z) \right)  .
\end{eqnarray}

To handle the derivative of the delta function we
introduce two additional independent variables for spanning
the space perpendicular to the worldsheet
\begin{eqnarray}
  y^a & = & \left\{ \zeta, \sigma_\perp \right\} \\
  \zeta & = & \left\{ u, v \right\} \\
  \sigma_\perp & = & \left\{  \sigma_1 , \sigma_2 \right\} .
\end{eqnarray}
Define $Z$ which extends $z$ off the worldsheet by adding
a perpendicular component $h$ for $\sigma_\perp \ne 0$:
\begin{eqnarray}
  Z^\mu(y) & = & z^\mu(\zeta) + h^\mu(\zeta, \sigma_\perp) \\
  h^\mu |_{\sigma_\perp = 0} & = & 0 \\
  g_{\mu\nu} (\partial_\zeta z^\mu) (\partial_{\sigma_\perp} h^\nu) |_{\sigma_\perp = 0} & = & 0 .
\end{eqnarray}
The third conditions is the requirement that the extension
lie off the worldsheet. This extension is meant to be
exact to all orders in $\mu$.

Now we can formally extend the integration
\begin{equation}
  \int du dv \delta^4(x-z) \to \int d^4y \delta^2(\sigma_\perp) \delta^4(x-Z)
\end{equation}
and rewrite the derivative of the delta function (this is possible
because the extra variables allow 1-to-1 coord. transformations)
\begin{eqnarray}
  \frac{\partial}{\partial x^\nu} \delta^4(x-Z) & = & -
  \frac{\partial}{\partial Z^\nu} \delta^4(x-Z) \\
  & = &
  -\frac{\partial y^a}{\partial Z^\nu} \frac{\partial}{\partial y^a} \delta^4(x-Z) \\
  & = &
  -(\partial_\nu y^a) \frac{\partial}{\partial y^a} \delta^4(x-Z) ,
\end{eqnarray}
insert and integrate by parts
\begin{eqnarray}
  \int \left( \sqrt{-g} T^{\mu\nu}\right)_\nu d^4x & = &
  - \mu \int d^4x d^4y \delta^2 \left( \sigma_\perp \right) G^{\mu\nu} \times \nonumber\\
  & &
  \left( \partial_\nu y^a \right)
  \left( \partial_a \delta^4(x-Z) \right) \\
  & = &
  \mu \int d^4y \partial_a \left( \delta^2(\sigma_\perp) G^{\mu\nu} \times \right. \nonumber\\
  & & \left. \partial_\nu y^a \right)  .
\end{eqnarray}
Finally, note that
$\frac{\partial Z^\mu}{\partial y^a} \frac{\partial y^b}{\partial Z^\mu} = \delta^b_a$ so that we end up with
\begin{eqnarray}
  \Delta P^\mu & = &
  2 \mu \int du dv \frac{\partial^2 z^\mu}{\partial u \partial v} .
\end{eqnarray}
This is exact and identical to what QS derive in Eq.~(4.6).
Second order results in $\mu$ follow by
writing $z^\mu = z^\mu_{(0)} + z^\mu_{(1)}$,
noting the unperturbed
equations of motion are
$z^\mu_{(0),uv}=0$, and finding
\begin{eqnarray}
  \Delta P^\mu_{(0+1)} & = & 0 \\
  \Delta P^\mu_{(2)} & = &
  2 \mu \int du dv \frac{\partial^2 z_{(1)}^\mu}{\partial u \partial v} .
  \label{eqn:secondorderloss}
\end{eqnarray}
One neeeds to know the perturbed string position to apply this
formula directly to calculate energy-momentum loss in the sense
that $\Delta P_{(n+1)}$ requires $z_{(n)}$.

QS re-write this using their first order equations of motion
$z^\mu_{(1),uv}=-\Gamma^\mu_{\alpha\beta(1)} z^\alpha_{(0),u} z^\beta_{(0),v}$
giving the final result for energy-momentum change
to second order in $\mu$:
\begin{equation}
  \Delta P^\mu_{(2)} = - 2\mu \int du dv \Gamma^\mu_{\alpha\beta(1)} z^\alpha_{(0),u} z^\beta_{(0),v} .
  \label{eqn:QSenergychange}
\end{equation}
In this form the numerical evaluation of the energy-momentum
change requires the first order metric perturbations instead of the
first order string perturbations. The linearized equations for the
metric are
\begin{equation}
  \square g_{\mu\nu(1)} = -16 \pi G \left( T_{\mu\nu(1)} - (1/2)g_{\mu\nu(0)} T^\rho_{\rho(1)} \right)
\end{equation}
where $T_{(1)}=\mu I_{(0)}$ (i.e. $\sqrt{-g}_{(0)}=1$ and unperturbed
worldsheet $z=z_{(0)}$).

We can generalize this procedure by following the identical logic
without make the choice of the conformal gauge. In summary,
\begin{eqnarray}
  S & = & -\mu \int d^2 \zeta \sqrt{-\gamma} \\
  T^{\mu\nu}(x) & = & \frac{2}{\sqrt{-g}} \frac{\delta S}{\delta g_{\mu\nu}(x)} \\
  & = & \frac{\mu}{\sqrt{-g}} \int d^2 \zeta
  \sqrt{-\gamma} \gamma^{ab} z^\mu_{,a} z^\nu_{,b} \delta^4(x-z) \\
  \Delta P^{\mu} & = & \mu \int d^2\zeta \frac{\partial}{\partial \zeta^b}
  \left( \sqrt{-\gamma} \gamma^{ab} z^\mu_{,a} \right) .
\end{eqnarray}

If we count orders then we need $\gamma_{(0+1)}$ and $z_{(0+1)}$ to
give $\Delta P_{(2)}$. In the previous case there was no
explicit appearance of the induced metric, only $z$ was present
so it was sufficient to give $z_{(0+1)}$ to find $\Delta P_{(2)}$.
Now there is the possibility that $z_{(0)}$ couples to $\gamma_{(1)}$
in addition to $z_{(1)}$ coupling to $\gamma_{(0)}$.

Finally, the equation of motion, $K^\mu=0$ (our Eq.~\eqref{eq:coordK}) is explicitly
\begin{equation}
  K^\mu = \frac{1}{\sqrt{h}} \frac{\partial}{\partial \zeta^b}
  \left( \sqrt{h} h^{ab} z^\mu_{,a} \right) + P^{\alpha\beta}\Gamma^\mu_{\alpha\beta} = 0
\end{equation}
and so we can also write the energy-momentum change as
\begin{equation}
  \Delta P^{\mu} = -\mu \int d^2\zeta \sqrt{h} P^{\alpha\beta}\Gamma^\mu_{\alpha\beta} .
\end{equation}
The above pair of equations
is analogous to QS's Eq.~(3.14) and (4.6). They are valid
for any coordinate and gauge choice and do not presume flat background.

To calculate energy-momentum loss start from Eq.~\eqref{eqn:secondorderloss}
above, rewrite $u$ and $v$ in terms of $\zeta$ and $\tau$ to
give
\begin{eqnarray}
  du dv & = & 2 d\zeta d\tau \\
  \partial_u \partial_v & = & (1/4)
  \left( \partial_\tau^2 - \partial_\zeta^2 \right) \\
  & = & -(1/4) \eta^{ab} \partial_a \partial_b
\end{eqnarray}
for our definitions
\begin{eqnarray}
  \eta & = & \left(
  \begin{tabular}{cc}
  -1 & 0 \\
  0 & 1
  \end{tabular}\right) \\
  \gamma_{ab} & = & \phi \eta_{ab} .
\end{eqnarray}
Now Eq.~\eqref{eq:simple} (conformal gauge at zeroth and first
order) is
\begin{equation}
  \eta^{ab}\partial_a\partial_b z_{(1)}^\mu = -{\cal F}^\mu_{\rm conf}
\end{equation}
so \eqref{eqn:secondorderloss} implies that
the change in 4-momentum of the string is
\begin{eqnarray}
  \Delta P^\mu_{(2)} & = & \mu \int  {\cal F}^{\mu}_{\rm conf} d\zeta d\tau
  \label{eqn:tradiationalforcesecondorderloss}
\end{eqnarray}
with the region of integration given by the fundamental period of the
worldsheet: $-L/2 \le \zeta < L/2$ and $0 \le \tau < L/2$.
For ${\cal F}^0_{\rm conf}<0$ the work
done on the string lowers its energy, $\Delta P^0 < 0$, and the
flux carried to infinity is $-\Delta P^0 > 0$. This form is
analogous to QS Eq.~\eqref{eqn:QSenergychange}.

\section{Expansion of the retarded time in null coordinates}
\label{sec:local-expansion-null}
In Sec.~\ref{sec:local-expansion} we derived an expansion which is useful in the case where spacelike and timelike worldsheet
coordinates are used. It is also useful to consider the case where null coordinates are used (and,
in particular, where the variable of integration is a null worldsheet coordinate). We will denote
these null coordinates by $\zeta^+$ and $\zeta^-$ and assume they can be related to spacelike and
timelike coordinates in the standard way, $\zeta^+ = \tau + \zeta$ and $\zeta^- = \tau - \zeta$.
Just like with the spacelike and timelike coordinates, we can write down conformal gauge
orthogonality relations for these null coordinates\footnote{
In the coordinates $\{\zeta^+,\zeta^-\}$, the conformal
factor $\phi$ is half of its value in the $\{\tau,\zeta\}$ system
at the same worldsheet point.},
\begin{align}
  g_{\alpha \beta} \partial_{\zeta^+} z^\alpha \partial_{\zeta^+} z^\beta &= 0, \\
  g_{\alpha \beta} \partial_{\zeta^-} z^\alpha \partial_{\zeta^-} z^\beta &= 0, \\
  g_{\alpha \beta} \partial_{\zeta^+} z^\alpha \partial_{\zeta^-} z^\beta &= -\phi.
\end{align}
We also have null coordinate version of the conformal gauge equation of motion:
\begin{equation}
  \partial_{\zeta^- \zeta^+} z^\alpha = 0.
\end{equation}

Now, proceeding exactly as we did with spacelike and timelike coordinates, we can seek a local
expansion of $\Delta \zeta^-(\Delta\zeta^+)$ [or, equivalently, $\Delta \zeta^+(\Delta\zeta^-)$].
As before, we will achieve this using the fact that source and field points are null-separated,
$\sigma(z, z') = 0$. Unlike before, however, we can no longer make the assumption that $\Delta
\zeta^-$ has an expansion in \emph{integer} powers of our order-counting parameter $\epsilon$,
where $\Delta \zeta^+ = \mathcal{O}(\epsilon)$. We can see this by starting from our expansion of
$\Delta\tau(\Delta \zeta)$, rewriting it in terms of $\Delta \zeta^-$ and $\Delta\zeta^+$, and
rearranging to get the expansion of $\Delta \zeta^-(\Delta\zeta^+)$.

Whether taking the direct approach or going via the spacelike and timelike expansion, we obtain the
same result upon gathering terms, namely that the expansion takes a distinct form depending on the
sign of $\Delta\zeta^+$:
\begin{enumerate}
  \item For $\Delta\zeta^+>0$ we get a standard integer power series;
  \item For $\Delta\zeta^+<0$ we find that the $\mathcal{O}(\epsilon)$ pieces cancel and we are
left starting with (at least) a cubic term. The result is that our expansion now has the form
\begin{equation}
  \label{eq:tret-semi-null-expansion}
  \Delta\zeta^- = \Delta\zeta^-_1 \epsilon^{1/3} + \Delta\zeta^-_2 \epsilon^{2/3} + \cdots.
\end{equation}
This is generically true, provided $\Delta \tau_3$ in Eq.~\eqref{eq:tret-spatial-expansion}
is non-zero, which is equivalent to the question of how the retarded image
deviates from being a straight line.
\end{enumerate}
As before, the expressions for the coefficients $\Delta\zeta^-_1, \Delta\zeta^-_2, \cdots$ are somewhat cumbersome. The important point this time is that they depend on $\sigma$ (which is $\mathcal{O}(\epsilon^2)$); on $\partial_\tau \sigma$, $\partial_\zeta \sigma$, and $\Delta \zeta$ (all of which are $\mathcal{O}(\epsilon)$); and on $\bar{\phi}$ and $\bar{z}^\alpha$ and their worldsheet derivatives.
Two different expansions are needed based upon the sign of $\Delta\zeta^+$.

\section{2D Calculation of the Self-force}
\label{sec:2D}

As an independent check on the numerical results of our $1$D method, we also performed a direct 2D
integration to determine the self-force. The results of the $2$D calculation are given in
Sec.~\ref{sec:results} and are found to be consistent with those of the $1$D method. In this
Appendix we give the explicit details of the calculation.

\subsection{Regularization}

In order to devise a numerical scheme for evaluating the self-force as a $2$D integral, we will
allow two modifications of the delta function that enforces the exact null condition between source
and field points. First, we replace the delta function itself with a non-singular, smooth function
of the spacetime interval. Second, we modify the null condition to select source points marginally
time-like separated from the field point. There are two parameters, one for the characteristic
width of the delta function replacement and one for the amount of over-retardation. When both
jointly approach zero then the smooth function tends to the singular delta function and the
source-field separation tends to the exact null separation. Our prescription is that any physical
answer is the limit of a sequence of calculations with (jointly) vanishing parameters, assuming, of
course, that a unique limit exists.

We integrate over the $2$D surface of the source distribution using a Gaussian approximation to the
delta function. The Gaussian form picks up small source contributions away from exact null
separations (off shell). This $2$D method of integration avoids potential pitfalls associated with
the elimination of one world sheet coordinate in terms of another coordinate when solving
$\sigma(x,z(\zeta^1,\zeta^2))=0$. Subtleties can easily arise (and be missed) in
the $1$D reduction. The $2$D integration method therefore provides a valuable (if computationally
costly) independent check on other methods.

The $2$D integration produces manifestly coordinate invariant results. It is potentially
susceptible to the effect of any world sheet singularity that lies off the exact null image. For
example, if a kink or cusp exists anywhere on the world sheet then the associated divergence must be
integrable for the method to give a well-defined limit for the force. As a practical matter, all
finite integrable off shell singularities are multiplied by Gaussian wings and strongly suppressed.
We believe but have not proven that contributions of singularities lying off of the image do not
survive the limiting process.

\subsection{Mathematical forms}

As summarized in the main text, in Minkowski spacetime we write the modified Synge world-function for two spacetime coordinates as
\begin{eqnarray}
  \sigma(x,z) & = & \frac{1}{2} (x-z)^\alpha g_{\alpha\beta}(x-z)^\beta + \sigma_0
\end{eqnarray}
where $\sigma_0$ is the over-retardation parameter (and $\sigma_0=0$ gives the usual world
function). The retarded Green function for a source at $x_s$ and field at $x_f$ is
\begin{eqnarray}
  {\cal G}(x_f,x_s) & = & \Theta(x_s,x_f) \delta(\sigma)
\end{eqnarray}
where the $\Theta=1$ when the time of the source $t_s$ precedes the time of the field point $t_f$
and $0$ otherwise. We replace the delta function with the Gaussian
\begin{eqnarray}
  \delta(q) & \to & \delta_G(q) \equiv \frac{e^{-q^2/(2 w^2)}}{\sqrt{2 \pi} w}
\end{eqnarray}
which is finite for any source and field point choices. We allow two possible representations for
the causality condition. In the ``exact'' representation we use the discontinuous $\Theta$
function. For the parts of the integrand that are directly proportional to ${\cal G} = \Theta
\delta$ we expect that using the exact $\Theta$ representation suffices for all field-source
combinations (i.e. we take the limit $w \to 0$ of an integrand with ${\cal G} \sim \Theta
\delta_G$). However, parts of integrand include derivatives with respect to $x_f$ of ${\cal G}$.
Then the situation is more complicated. We can justify ignoring the functional dependence of
$\Theta$ on $x_f$ {\it if} we use the exact delta function $\delta(\sigma)$ -- the only point where
both $\Theta$ and $\delta(\sigma)$ are non-vanishing is $x_f=x_s$ and things are undefined there
anyway. But it is dubious to ignore $x_f$ dependence if we're replacing the delta function with the
Gaussian $\delta_G$ because there are contributions from points $x_s$ (away from $x_f$) such that
the derivative of $\Theta$ with respect to $x_f$ multiplies a non-zero off-shell $\delta_G$. It is
not obvious that it is safe to drop $\Theta$'s functional dependence on $x_f$ {\it before} taking
the limit of $\delta_G \to \delta$. To study the situation we replace the cut-off with a smoothed
discontinuity i.e. $\Theta \to \Theta_{tanh} \equiv (1-\tanh((t_s-t_f)/w))/2$ and take full
derivatives of $\Theta_{tanh} \delta_G$. As a practical matter we use the same small parameter in
the Gaussian and in the tanh, and will seek convergent results in the joint limit $w \to 0$ and
$\sigma_0 \to 0$.

Let $f_1$ and $f_2$ be the self-force integrands
\begin{eqnarray}
  f_1^\rho(x,z)
  & = & -\perp^{\rho\lambda}(x) \eta^{\mu\nu}(x) \Delta H_{\mu\nu\lambda} \\
  f_2^\rho(x,z)
  & = & K^{\mu\nu\rho}(x) \Delta h_{\mu\nu} .
\end{eqnarray}
These are integrated with uniform weight on the
worldsheet to give Eq.~\eqref{eq:self-force}
\begin{eqnarray}
  F_1^\rho(x) & = & \iint f_1^\rho(x,z) d\zeta^1 d\zeta^2 \\
  F_2^\rho(x) & = & \iint f_2^\rho(x,z) d\zeta^1 d\zeta^2 .
\end{eqnarray}
The tangential and perpendicular projection operators and the extrinsic curvature are evaluated at
the field point, $x$. The metric perturbation quantities, $\Delta H$ and $\Delta h$, depend upon
both source and field points. Replacing covariant derivatives with partial derivatives (recall we
are working in Minkowski spacetime, in which case the two derivatives are equal), the integrands
for the trace-reversed metric perturbation and metric perturbation derivative are
\begin{eqnarray}
  \Delta{\bar h}_{\alpha\beta}
  & = & -4 G\mu \sqrt{-\gamma(z)} \eta_{\alpha\beta}(z) {\cal G}(x,z) \\
  \Delta{\bar h}_{\alpha\beta,\gamma}
  & = & -4 G\mu \sqrt{-\gamma(z)} \eta_{\alpha\beta}(z) \\
  & & \frac{\partial {\cal G}(x,z)}{\partial x_f^\gamma},
\end{eqnarray}
where source and field point dependence is explicitly indicated.

We now proceed with the calculation for a given choice of $w$ and $\sigma_0$ for the smoothed
Green function. We repeat the calculation for a sequence with $w \to 0$ and $\sigma_0 \to 0$. The
area of the world sheet includes all non-zero contributions to the integral but as a practical
matter we limit it to regions where the magnitude of the integrand exceeds some minimum threshold.

We have used several different techniques for estimating the integrals of interest. The first
approach is a simple quadrature along lines in which we evaluate $f_1^\rho$ at a lattice of points
$z$ and estimate
\begin{eqnarray}
  F_1^\rho(x) & \sim & \frac{A_{WS}}{N} \sum_{i=1}^N f_1^\rho(x,_{,i}) \\
\end{eqnarray}
where $A_{WS} = \int d \zeta^1 d \zeta^2$ is the world sheet area (and likewise for $F_2$). This
approach makes essentially no assumptions about the integrand's support but the achievable accuracy
is limited in practice by the inefficiencies of working with a low order scheme and a global grid.

Only a small part of the worldsheet is important in the limiting results. The second approach is
cubature, an adaptive algorithm for numerical integration based on the algorithm of Genz and
implemented in routine Cuhre in the Cuba package
\cite{Hahn:2004fe}. The basic integration
rule for the two dimensional application is degree 13. We begin by identifying a part of the world
sheet in which the integrand is above some threshold (say $> 10^{-12}$), apply an integration rule,
repeatedly subdivide the region to estimate the global integral and errors and stop when errors are
sufficiently small. This method, being higher order and more selective about the choice of points,
achieves higher accuracy for a given computational cost but sometimes terminates too early if
errors are mis-estimated.

\subsection{KT Cases I and II}

The main text
compares the Case I and II results for 1D to those found by direct integration over the 2D world
sheet. We used the rectangular area centered on the field point and of size $\pm L/2$ in both
$\zeta^1$ and $\zeta^2$ which encompasses the entire string loop image. We parameterized the
Gaussian width and the $\Theta$ function with $n$ and $m$:
\begin{eqnarray}
  d_{char} & = & \frac{L}{2^{n}} \\
  w & = & d_{char}^4 \\
  \sigma_0 & = & (m d_{char})^2
\end{eqnarray}
and let $n$ range from $4$ to $7$ and set $m \ge 0$.\footnote{Note
  $\delta(\sigma) \sim e^{-\sigma^2/2 \chi^2}/(\sqrt{2 \pi} \chi)$ and
  $\sigma \sim ds^2 \propto L^2$ implies that $\chi \propto d_{char}^2
  \propto \sigma_0$. However, we use Gaussian width $w \propto \chi^2
  \propto d_{char}^4$ and $\sigma_0 \propto d_{char}^2$. Of course,
  both the width and the over-retardation parameter decrease as $w \to
  0$ or $n \to \infty$.}. We accumulated the contributions to the
integrals only for points with ${\cal G} > 10^{-4}$ for each choice of
$n$ and $m$. This cutoff effectively removed any effect of cusps
lying off the string image. We checked sensitivity to the cutoff by
recalculating with ${\cal G} > 10^{-3}$ and ${\cal G} > 10^{-2}$,
finding no differences and will drop further mentions of this parameter.

Case I is intrinsically the simplest calculation since the image of
the string loop is smooth, without cusps or kinks. Several series of
integrations using 2D uniform grids were performed. We studied the
convergence for the following sequences:
\begin{itemize}
\item $m=0$ (no over-retardation), $\Theta$ (exactly causal),
  $n=4-6.25$
\item $m=0$ (no over-retardation), $\Theta_{tanh}$ (smoothed causality),
  $n=4-5.5$
\item $m=0.1$ (over-retardation),  $\Theta$ (exactly causal),
  $n=4-5.75$ .
\end{itemize}
For the integration for a specific $m$ and $n$ we estimated the
quadrature errors by halving the grid spacing and looking at
successive differences in the answers (hereafter, Cauchy errors). We
repeated the process until the Cauchy errors were small. We then
generated sequences for varying $n$. In these we observed smooth,
steady convergence to the [1D] results.

We extrapolated numerical results to $n = \infty$ by fitting each
component $F$ with the form $F(n) = F(\infty) + a e^{-b
  n}$. $F(\infty)$, $a$ and $b$ are found from numerical results at 3
specific values of $n$. Table \ref{tab:summarytab} summarizes the
results for one sequence with fixed $m=0$ and exact $\Theta$
function. The second column gives the extrapolated force solution for
each component, the third is an estimate of the size of the systematic
errors in extrapolation (by fitting different grid based calculations)
and the fourth is the difference of $F(\infty)$ and the [1D]
results. The [1D] results include the hidden delta function at the
field point and the line integral with no
over-retardation and exact causality. Note that the error with respect
to the [1D] results is $\lta 4 \times 10^{-3}$ for the individual
components of $F_1^\rho$ and $\lta 3 \times 10^{-5}$ for
$F_2^\rho$. The size of the 1D-2D differences are comparable,
component by component, to the Cauchy errors for the quadrature
itself. These are much smaller than the contribution of the delta
function at field point.

We found Case I sequences with or without over-retardation, with or
without smooth causal condition all smoothly converged to the [1D] answer.
It is perhaps important to emphasize that this consistency is
strong evidence that a single, well-defined limit exists and
is correctly recovered with the 1D analysis for smooth loops.

\begin{figure}[H]
\centering
\includegraphics[width=0.9\linewidth]{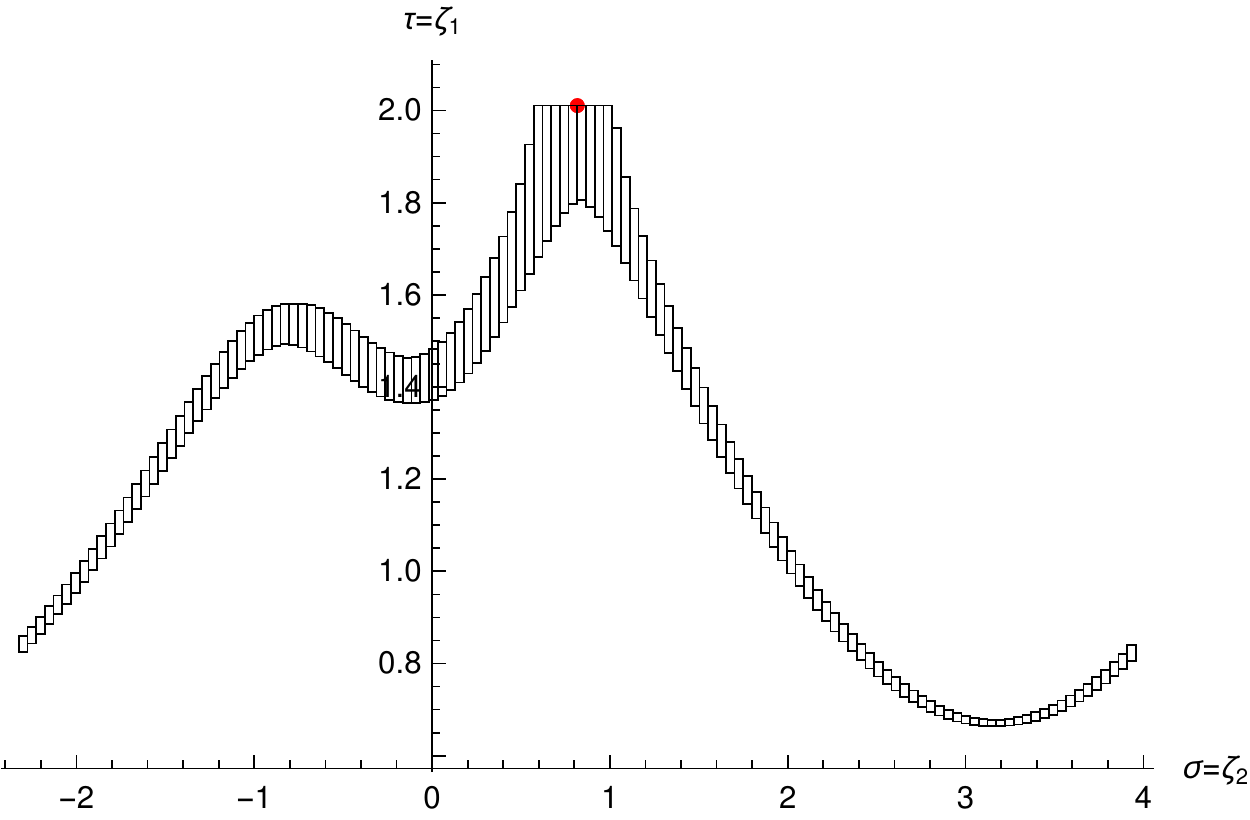}
\caption{\label{example64-5}The red dot shows the field point
  for Case I for the KT string loop. The boxes
  enclose the regions with RMS integrands $>10^{-8}$ for
$m=0$ and $n=5$.}
\end{figure}

We next repeated the Case I sequence for $m=0$ (no
over-retardation), exactly causal Green function, for $n=4-7$ using
the cubature method. Cubature will be used for most of the remaining
force evaluations because it is more efficient and has higher
accuracy. We began by selecting
rectangular subregions of the world sheet that can plausibly contribute
to the quadratures of interest. These closely trace the
retarded string image. For example, figure
\ref{example64-5} illustrates the field point (red dot) and
outlines 128 subregions of interest for $m=0$ and $n=5$.
The union of all the subregions encloses the area
containing non-trivial integrands in the sense
$\sqrt{\sum_\rho \left( (f_1^\rho)^2 + (f_2^\rho)^2 \right)} > 10^{-8}$. The
subregion bounds are found numerically. The string loop
image lies within the subregions as $n$ grows large.

The function is integrated by cubature over each separate subregion
and the full answer is the sum over the set. In the cubature method
the subregion is divided repeatedly until an estimated error
tolerance (absolute or relative errors $\lta 10^{-6}$)
is reached. Sometimes the error estimate is inaccurate
so we also systematically increased the number of rectangles
from $128$ up to $2048$ yielding finer and finer starting conditions.
This allows a check that the error estimates are robust.
Once all the subregions are accurately accounted for the total is calculated.

\begin{figure}[H]
\centering
\includegraphics[width=0.9\linewidth]{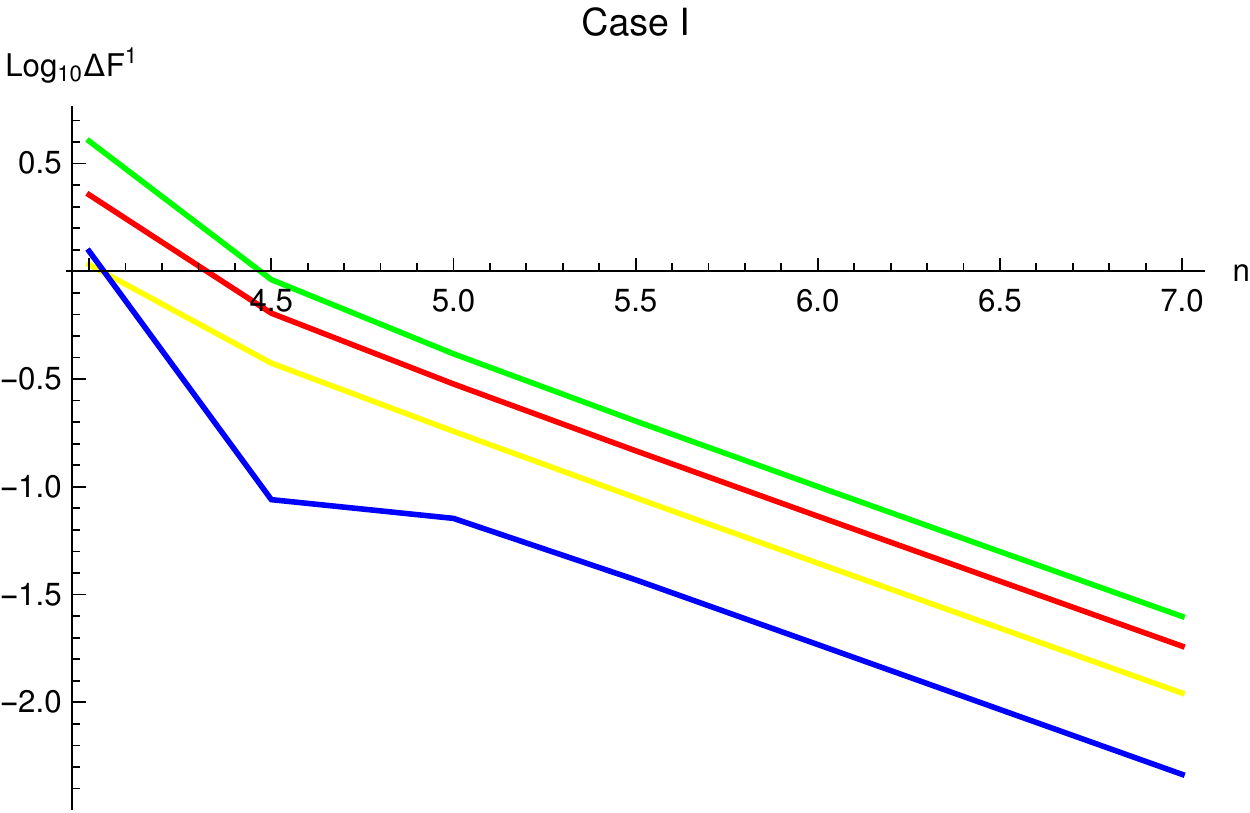}
\caption{\label{CaseIf1convergence} The difference between smoothed
  Green function based, 2D grid evaluation of $F_1^\rho$ and the
  exact delta function, [1D] evaluation, as a function of $n$ for loop
  I. The [2D] results should approach the [1D] results as $n \to \infty$
  when the Gaussian tends to the delta function.  These results have
  no over-retardation ($m=0$) and use the exact $\Theta$ function for
  causality.  The colors label results for the four spacetime
  components of $F_1$; red, yellow, green and blue correspond to
  components $0$, $1$, $2$ and $3$, respectively. The ordinate,
  $\log_{10} | F^\rho_{1,2D} - F^\rho_{1,1D} |$, quantifies the
  difference, component by component, as a function of $n$. The
  delta function contribution at the field point is of order
  unity and far larger than the 1D-2D differences. Table \ref{tab:summarytab}
  gives detailed information on individual contributions.}
\end{figure}

\begin{figure}[H]
\centering
\includegraphics[width=0.9\linewidth]{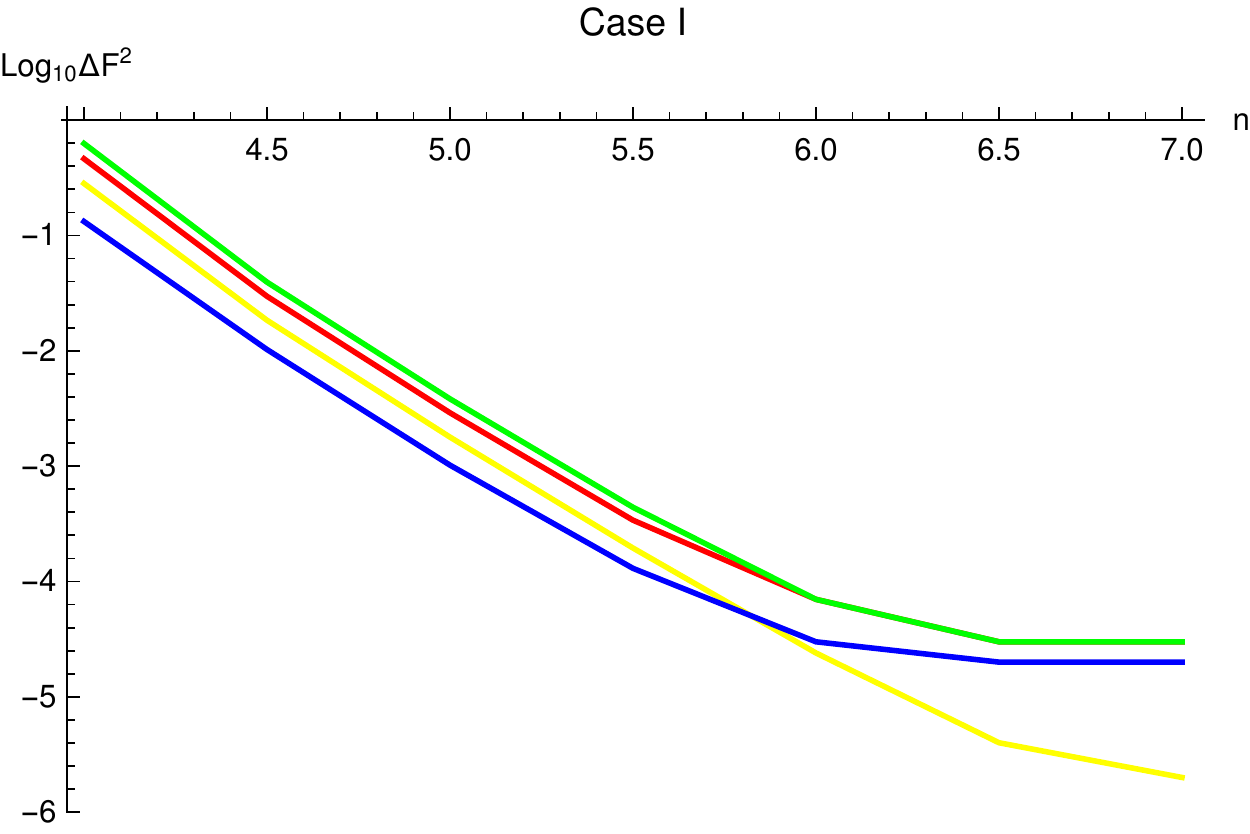}
\caption{\label{CaseIf2convergence} Same as Fig. \ref{CaseIf1convergence} except for $F_2^\rho$.}
\end{figure}

Figures \ref{CaseIf1convergence} and \ref{CaseIf2convergence} plot the
$\log_{10}$ absolute differences between the [2D] and [1D] calculations as
a function of $n$. The solid lines are for [1D] calculations with
the delta function contribution at the
field point. The scale for the y-axis on the figures is quite
different and, as expected, the error is much larger for $F_1$ than
for $F_2$. The errors cease to decrease exponentially with $n$ in
figure \ref{CaseIf2convergence} for 1D-2D differences of size $3\times
10^{-5}$, likely a consequence of the intrinsic accuracy of the
cubature integrations.

In summary, the Case I [2D] cubature calculations agree with the [1D]
calculations with delta function contribution at the field point.

\begin{figure}[H]
\centering
\includegraphics[width=0.9\linewidth]{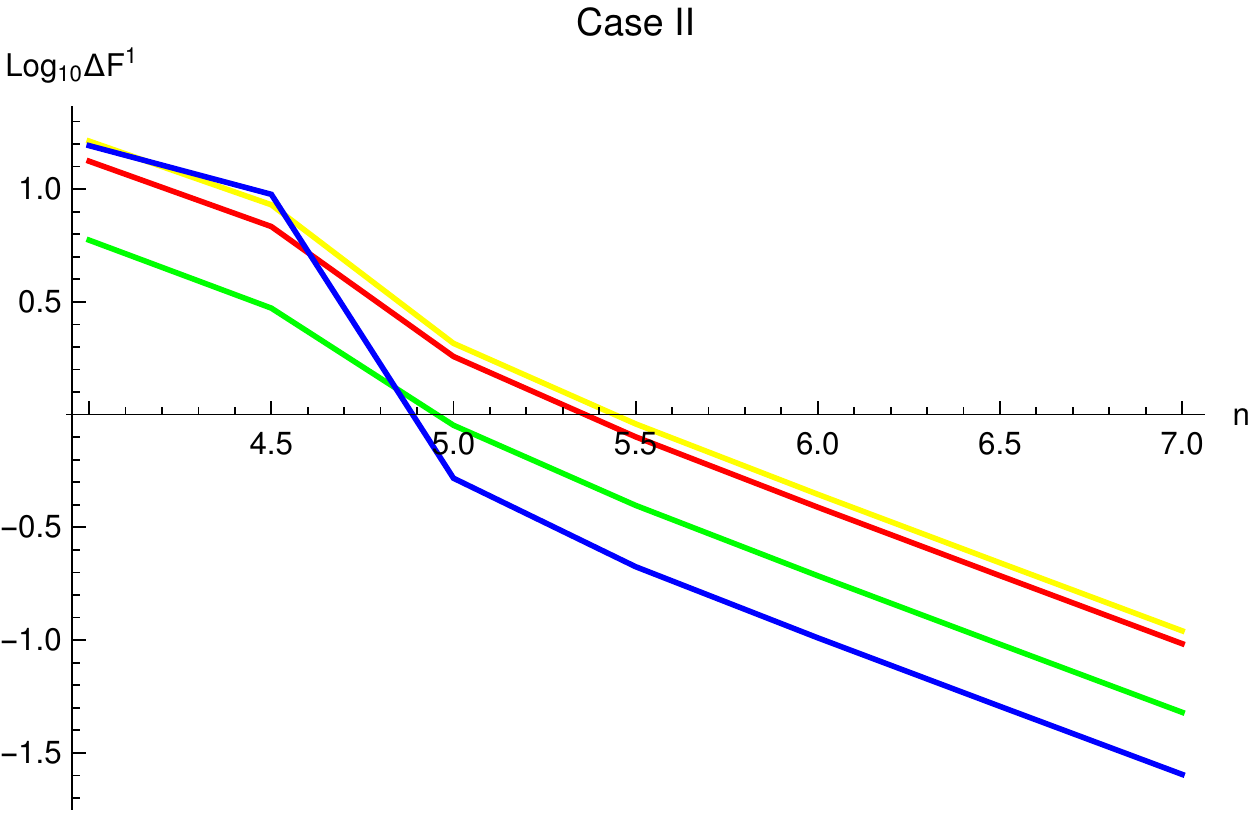}
\caption{\label{CaseIIf1convergence} Same as Fig. \ref{CaseIf1convergence} except for loop II.}
\end{figure}

\begin{figure}[H]
\centering
\includegraphics[width=0.9\linewidth]{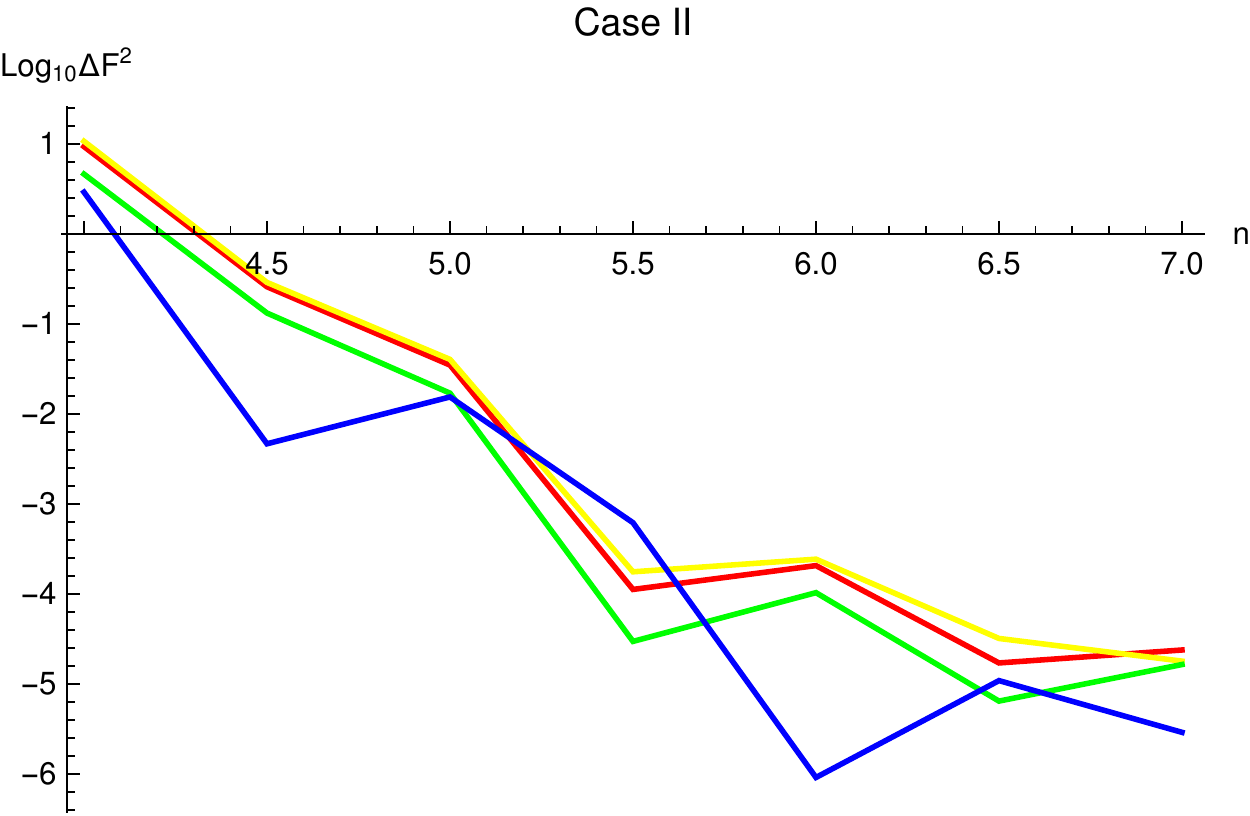}
\caption{\label{CaseIIf2convergence} Same as Fig. \ref{CaseIf1convergence} except for $F_2^\rho$ and loop II.}
\end{figure}

Case II is similar to Case I except that a cusp occurs on the image of
the string loop. We proceeded in the same manner, breaking the world
sheet up into small subregions with non-trivial integrands.  The main
difference from Case I is that there is a significant loss in
precision for source points near the cusp. We had to perform those
calculations with quadruple precision arithmetic. We avoided
evaluating quantities at the cusp itself but sampled points as closely
as needed to estimate the integrated quantities. The integrands are
extremely complicated and we did not perform an analytic expansion in
the vicinity of the cusp.  We found (1) no numerical evidence for singular
behavior and (2) well-behaved quadratures. Fig. \ref{CaseIIf1convergence} and
\ref{CaseIIf2convergence} show the convergence as $n$ increases.  The
basic conclusion is the same as for Case I: the [2D] numerical results
tend to the [1D] results when the latter include the delta function
contribution at the field point.

\subsection{Garfinkle and Vachaspati string}

The assumed field point is $(\zeta^1,\zeta^2) = (\tau,\zeta) = (0.3,0.4)L$. The kinks are located
at $\zeta^2$ approximately $-0.79$ and $1.87$. The [1D] results are given in Table
\ref{tab:comparison}. These are calculated from the line integrals, boundary terms at each kink and
the delta function at the field point as described in the main text.

We introduce a small parameter to ``round off'' the kinks as follows\footnote{To recap, in the various 1D methods
  we took care to accommodate the string discontinuities,
  the issue of field equal source point and patching different
  coordinate systems together.
  The [2D] calculations is not immune to the first of these issues.}: replace
$\delta_{j,\lfloor\frac{2x}{L} \rfloor}$ with a smoothly varying function of characteristic width
$\Delta$:
\begin{eqnarray}
  \delta_j(x) & = & \frac{1}{2} \left(
  \tanh \left( \frac{x-x_{lo}}{\Delta} \right) - \right. \nonumber\\
  & & \left. \tanh \left( \frac{x_{hi}-x}{\Delta} \right) \right) \\
  x_{lo} & = & \frac{j L}{2} \\
  x_{hi} & = & x_{lo}+\frac{L}{2}
\end{eqnarray}
For $L=2 \pi$ and $\Delta=10^{-2}$ changes in the spacetime configuration of the loop are largely
confined to the vicinity of the kink. They remain invisible at the resolution of Fig.~\ref{fig:GVloopspacetime}. In the tangent sphere representation, non-zero $\Delta$ connects the green
arcs. The linking curves do not lie on the unit sphere and the position ${\vec b}'$ along the curve
changes very rapidly with its argument, i.e. the tangent vector quickly passes through the gap.

\begin{figure}[H]
\centering
\includegraphics[width=0.9\linewidth]{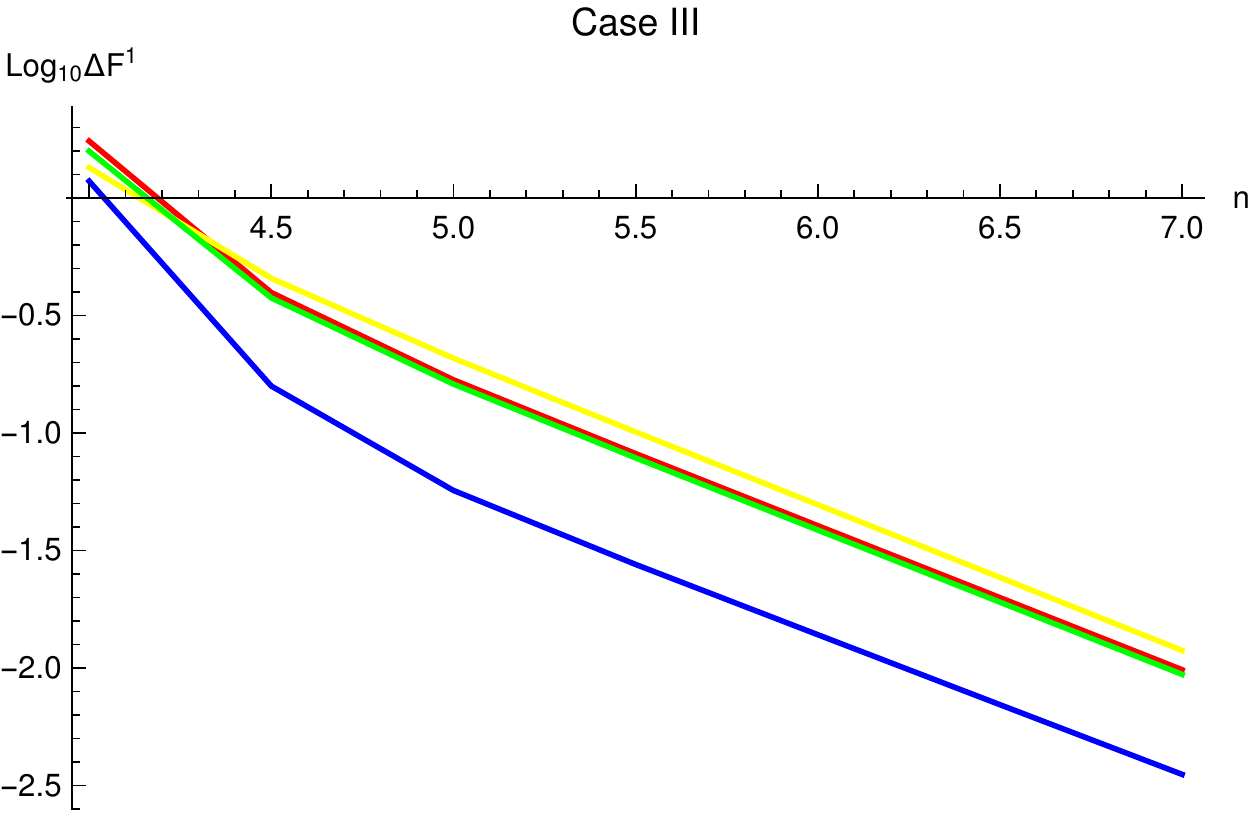}
\caption{\label{CaseIIIf1convergence} Same as Fig. \ref{CaseIf1convergence} except for loop III.}
\end{figure}

\begin{figure}[H]
\centering
\includegraphics[width=0.9\linewidth]{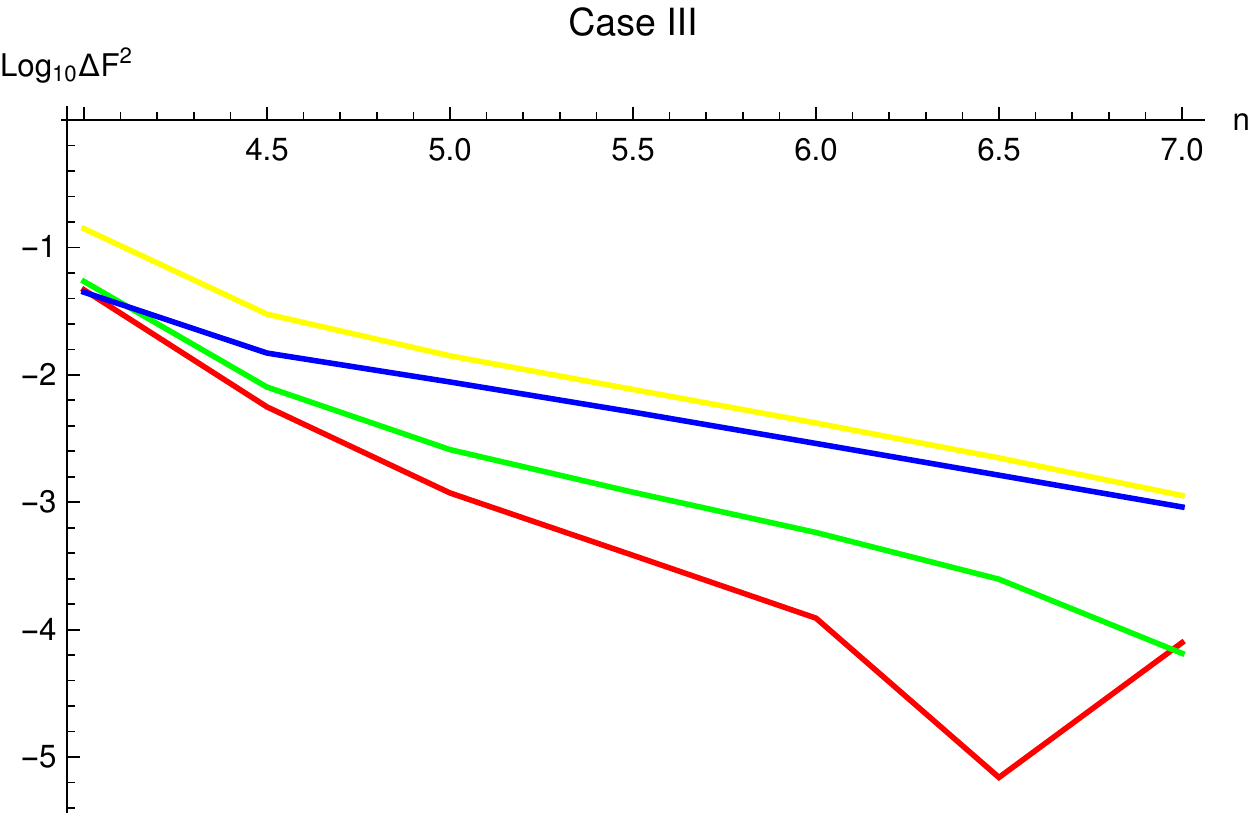}
\caption{\label{CaseIIIf2convergence} Same as Fig. \ref{CaseIf1convergence} except for $F_2^\rho$ and loop III.}
\end{figure}

We have integrated the forces using the cubature method as was done for the KT
string with a sequence for $m=0$ (no over-retardation), exactly causal Green function, for
$n=4-7$ and we adjust $\Delta=e^{-1.15129 n}$ (so $\Delta =10^{-2}$ at $n=4$ and $10^{-3}$ at
$n=6$) so that the limit is the idealized kink solution. The results are shown in
Figs.~\ref{CaseIIIf1convergence} and \ref{CaseIIIf2convergence}. Note that the errors in the [2D] results are far
smaller than the size of the
individual components that make up the [1D] calculation (the line integral,
the delta function and the kink contributions).

%%%%%%%%%%%%%%%%%%%%%%%%%%%%%%%%%%%%%%%%%%%%%%%%%%%%%%%%%%%%%%%%%%%%%%%%%%%%
\section{Parametrized fits near the cusp for the non-intersecting KT string}
\label{sec:VV-fits}

In this appendix we give quantitative numerical fits to the behaviour of the self-force in the
vicinity of the cusp for the non-intersecting KT string ($\alpha=1/2$, $\phi=0$) described in
Sec.~\ref{sec:VV-nonSI}. We investigated several approaches to fitting
the forces in the vicinity of the cusp.

Here, we begin with fitting
the integral part of ${\cal F}^\mu_{\rm conf}$ since the
delta-function part is known in analytic form. We sampled worldsheet
points about the cusp
extending from $2 \times 10^{-2}$ to $2 \times 10^{-8}$ in radius
and at fixed angle $\theta$ and found
\begin{eqnarray}
  \log |F^\mu| & = & A^\mu_1(\theta) \log x + A^\mu_2(\theta) x + A^\mu_3(\theta) \\
  x & = & \log r \\
  r & = & \sqrt{\delta \tau^2 + \delta \zeta^2}
\end{eqnarray}
fits each ray very well.
The ratio of the magnitude of $\log |F^\mu|$ to
the root mean square error for one fitted ray is typically $\sim 10^3$.
We observed that angle-dependent coefficients $A^\mu_i(\theta)$ varied
systematically and we selected simple parameterized
forms. For $F^t$ and $F^x$ the form
\begin{eqnarray}
  A^\mu_i & = & a^\mu_i \chi(\theta)^{b^\mu_i} \\
  \chi(\theta) & = & \lim_{r \to 0} \frac{\sqrt{-\gamma}}{r^2}
\end{eqnarray}
gave reasonable fits. There are 6 scalars needed to fit each of
the 2d function $F^t$ and $F^x$. For $F^y$ we found simple
angle-dependent coefficients $A^\mu_i(\theta)$ worked well:
\begin{eqnarray}
  A_i & = & a_i + b_i T \left( \frac{\theta-c}{2 \pi} \right)
\end{eqnarray}
where $T(x)$ is the triangle wave with amplitude $1$, period $1$,
$T(0)=0$ and $T'(0)=4$. There are 7 scalars to fit (note that $c$ is the
same fitted parameter for the 3 $A_i$). For $F^z$ the form is tolerably
well fit by
\begin{eqnarray}
  A_i & = & a_i + b_i U\left( \theta - \frac{\pi}{2} \right)
  S\left( \frac{\theta-c}{3 \pi/2} \right) - \\
  & & d_i U \left( \frac{\pi}{2} - \theta \right) \nonumber
\end{eqnarray}
where $U(x)$ is the step function ($0$ for $x<0$ and $1$ for $x \ge 0$),
$S(x)$ is the sawtooth wave that varies from $0$ to $1$ with
period $1$, $S(0+)=0$, $S(1-)=1$. There are 9 scalars to fit (here
$c=0.5$ is fixed and not varied). Results are summarized in
the Table below. The fits for $F^t$ and $F^x$ are close to $1/r$
because $A_2^t \sim A_2^x \sim -1$ is the dominant term (and
$a_2^t \sim a_2^x \sim -1$ and $b_2^t \sim b_2^x \sim 0$). The fit for $F^y$ is
close to $\log r$ (setting $A_2^y=0$ makes minimal difference
in $Q$, as shown in the Table). The fit for
$F^z$ requires both $A_1^z$ and $A_2^z$.
\begin{widetext}
~
\begin{table}[H]
  \begin{center}
    \begin{tabular}{|c|cc|cc|cc|c|}
  \hline
  $\mu$ & $a_1$ & $b_1$ & $a_2$ & $b_2$ & $a_3$ & $b_3$ & $Q$ \\
\hline
$t$ & $0.0512$ & $-0.0474$ & $-0.996$ & $-0.000301$ & $2.982$ & $-0.110$ & $123$ \\
$x$ & $0.0207$ & $-0.0497$ & $-0.998$ & $-0.000625$ & $3.027$ & $-0.108$ & $123$ \\
\hline
    \end{tabular}
    \bigskip

  \begin{tabular}{|c|cc|cc|cc|c|c|}
      \hline
  $\mu$ & $a_1$ & $b_1$ & $a_2$ & $b_2$ & $a_3$ & $b_3$ & $c$ & $Q$ \\
\hline
$y$ & $0.931$ & $0.651$ & $0.00119$ & $0.0349$ & $3.096$ & $-1.429$ & $0.314$ & $87$ \\
    & $0.942$ & $0.335$ & - & - & $3.084$ & $-1.089$ & $0.314$ & $82$ \\
\hline
  \end{tabular}
  \bigskip

  \begin{tabular}{|c|ccc|ccc|ccc|c|c|}
  \hline
  $\mu$ & $a_1$ & $b_1$ & $d_1$ & $a_2$ & $b_2$ & $d_2$ & $a_3$ & $b_3$ & $d_3$ & $c$ & $Q$ \\   \hline
$z$ & $1.053$ & $-0.961$ & $0.617$ & $0.0145$ & $-0.0524$ & $0.0396$ & $0.679$ & $2.495$ &
  $-1.422$ & $0.5$ & $87$ \\
    & $0.925$ & $-0.490$ & $0.262$ & - & - & - & $0.815$ & $1.992$ &
  $-1.041$ & $0.5$ & $30$ \\
\hline
\end{tabular}
  \end{center}
  \caption{Fits based on $\log |F| = A_1 \log x + A_2 x + A_3$
    where $x=\log r$ and $A_i(\theta)$ are
    specific forms described in the text. $Q$ is the ratio of the
    mean $\log |F|$ to the root mean square error of the fit.
  }\label{tab:VVcuspasymptotics1}
\end{table}
\end{widetext}

We have also fit four parameter, adhoc forms. These are generally
less good than the previous set and the
quality varies considerably. It is best in the quadrants without the
visible fold and it remains difficult to fit all quadrants together. The results (given in Table
\ref{tab:VVcuspasymptotics}) imply that the $t$- and $x$-components diverge as $1/r$, where $r
\equiv \sqrt{\tau^2 + \zeta^2}$ is the Euclidean distance on the worldsheet. A possible divergence
in the $y$- and $z$-components cannot be ruled out, but is certainly not as strong as the $t$- and
$x$-components.
\begin{widetext}
~
\begin{table}[H]
  \begin{center}
    \begin{tabular}{|cc|cccc|cc|}
  \hline
  $\mu$ & ${\rm sgn} \ \delta \tau \ {\rm sgn} \ \delta \zeta$ &
  a & b & c & d & $\epsilon$ & Q\\
  \hline
$t$ & {\rm All} & $ 3.95292$ & $ -0.414102$ & $ 0.151793$ & $ 0.0721235$ & $ 0.131773$ & $ 4.77971$ \\
  & {\rm ++} & $ 3.01226$ & $ -0.502883$ & $ 0.178088$ & $ 0.212065$ & $ 0.00341496$ & $ 163.388$ \\
  & {\rm +-} & $ 4.22776$ & $ -0.474782$ & $ -0.757188$ & $ 0.686015$ & $ 0.0569798$ & $ 11.6713$ \\
  & {\rm -+} & $ 3.24345$ & $ -0.468097$ & $ 0.395402$ & $ 0.427249$ & $ 0.118002$ & $ 4.78373$ \\
  & {\rm --} & $ 2.69632$ & $ -0.502038$ & $ -0.0722535$ & $ -0.418941$ & $ 0.0189629$ & $ 28.1211$ \\
\hline
$x$ & {\rm All} & $ 3.97366$ & $ -0.412535$ & $ 0.151562$ & $ 0.0719106$ & $ 0.131271$ & $ 4.78016$ \\
  & {\rm ++} & $ 3.04136$ & $ -0.500672$ & $ 0.176313$ & $ 0.20999$ & $ 0.00282476$ & $ 196.672$ \\
  & {\rm +-} & $ 4.24515$ & $ -0.473347$ & $ -0.755537$ & $ 0.684603$ & $ 0.0567632$ & $ 11.6804$ \\
  & {\rm -+} & $ 3.26301$ & $ -0.466743$ & $ 0.395591$ & $ 0.425764$ & $ 0.11768$ & $ 4.78189$ \\
  & {\rm --} & $ 2.72609$ & $ -0.499806$ & $ -0.0708712$ & $ -0.416556$ & $ 0.0186554$ & $ 28.457$ \\
\hline
$y$ & {\rm All} & $ 4.20813$ & $ -0.0398377$ & $ -0.381031$ & $ 0.144944$ & $ 0.134169$ & $ 2.41009$ \\
  & {\rm ++} & $ 3.26127$ & $ -0.0948215$ & $ -0.0164481$ & $ 0.48741$ & $ 0.0189321$ & $ 7.52428$ \\
  & {\rm +-} & $ 3.61753$ & $ -0.0744807$ & $ -0.580823$ & $ -0.340387$ & $ 0.167632$ & $ 1.57078$ \\
  & {\rm -+} & $ 3.70482$ & $ -0.0686625$ & $ -0.518819$ & $ 0.296286$ & $ 0.019127$ & $ 8.42306$ \\
  & {\rm --} & $ 3.92864$ & $ -0.0768625$ & $ -0.17871$ & $ 0.10457$ & $ 0.0088657$ & $ 15.2576$ \\
\hline
$z$ & {\rm All} & $ 2.65579$ & $ -0.0161345$ & $ -0.344105$ & $ 0.0271806$ & $ 0.200512$ & $ 1.58037$ \\
  & {\rm ++} & $ 1.78559$ & $ -0.0666965$ & $ -0.072034$ & $ 0.460484$ & $ 0.00948379$ & $ 14.77$ \\
  & {\rm +-} & $ 1.91589$ & $ -0.0211916$ & $ -0.156757$ & $ -0.808392$ & $ 0.27011$ & $ 1.37301$ \\
  & {\rm -+} & $ 3.04493$ & $ -0.0625678$ & $ 0.234966$ & $ -0.80721$ & $ 0.0899292$ & $ 2.78705$ \\
  & {\rm --} & $ 2.76864$ & $ -0.0513231$ & $ 0.0852916$ & $ 0.348301$ & $ 0.0157434$ & $ 7.3724$ \\
\hline
\end{tabular}
  \end{center}
  \caption{Two dimensional fits to $\log | {\cal F}^\mu_{\rm conf} |$ near the cusp ($0.0013 <
          \{\tau|,|\zeta|\} < 0.0021$) using the form $a + b \log \sqrt{-\gamma} + c \cos \theta + d \sin
          \theta $ where $\cos \theta = \tau/r$, $\sin \theta = \zeta/r$ and $r = \sqrt{\tau^2 +
          \zeta^2}$. The fit goodness is quantified by $\epsilon$, the root mean square error between the
          fit and the actual data (both as logs), and $Q$, the ratio of the total variation in $\log
          {\cal F}^\mu_{\rm conf}$ divided by $\epsilon$. ``All'' means a single fit to the whole plane; $++$
          means the fit restricted to $\tau \ge 0$ and $\zeta \ge 0$ and similarly for $+-$, $-+$ and
          $--$. Asymptotically $F \propto 2b/\tau$ for $\zeta=0$ and $2b/\zeta$ for $\tau=0$.
  }\label{tab:VVcuspasymptotics}
  \end{table}
\end{widetext}

In addition to the full fits, we have made simpler one-dimensional fits along the coordinate axes.
Table \ref{tab:VVcusptauvaries} gives fits to $|\log {\cal F}^\mu_{\rm conf}|$ as a power law in $|\log
\tau|$ along the coordinate axis $\zeta=0$.
We have separately fit the total force and just the part of the force
that comes from the integral contribution.
For the $t$- and $x$-components of the total force the
scaling is clear and unambiguous. There is a measurable
difference in the amplitude of the force before and after the cusp formation even though the rate
of divergence is identical. (The delta function
contribution scales as $1/r$ for t- and x-components.)
The amplitude difference or Stokes phenonmenon is smaller
in the integral piece than in the total. The delta function
contribution is related to the choice of the retarded
Green function and the dichotomy of emission and absorption near the cusp. The $y$- and
$z$-components grow more slowly as the time to the cusp
shrinks and the asymptotics are not as well-determined.
Nonetheless, the similarity of total and integral quantities implies
that the delta function contribution which is constant in $r$ (for the y- and z-components)
is subdominant at the scales probed. Similarly, Table
\ref{tab:VVcuspzetavaries} considers the analogous situation for $|\log {\cal F}^\mu_{\rm conf}|$ as a
power law in $|\log \zeta|$ along the coordinate axis $\tau=0$. Again,
the scaling with distance to the cusp is unambiguous.
The $t$- and $x$-components diverge
like $|\zeta|^{-1}$ and have similar amplitudes. The Stokes-like phenomenon
seems to be weaker or absent in the total force than it is in the
integral contribution alone (contrary to the previous example).
The string segment behaves approximately symmetrically along its length. Again, the $y$- and
$z$-components grow more slowly as the cusp is approached.
\begin{table}[H]
  \begin{center}
    \begin{tabular}{|cc|ccc|ccc|}
  \hline
  $\mu$ & ${\rm sgn} \ \delta \tau$ & a & b & Q & a & b & Q \\
  & & \multicolumn{3}{c|}{total} & \multicolumn{3}{c|}{integral} \\
  \hline
  $t$ & $+$ & $3.56$ & $-1.00$ & $8.9 \times 10^3$ &
  $3.30$ & $-1.00$ & $6.8 \times 10^3$\\
  & $-$ & $3.08$ & $-1.00$ & $5.6 \times 10^3$ &
  $3.39$ & $-1.00$ & $7.7 \times 10^3$ \\
  \hline
  $x$ & $+$ & $3.56$ & $-1.00$ & $2 \times 10^4$ &
  $3.31$ & $-1.00$ & $1.6 \times 10^4$ \\
  & $-$ & $3.08$ & $-1.00$ & $1 \times 10^4$ &
  $3.40$ & $-1.00$ & $1.5 \times 10^4$ \\
  \hline
  $y$ & $+$ & $3.27$ & $-0.19$ & $60$ &
  $3.46$ & $-0.18$ & $65$ \\
  & $-$ & $4.43$ & $-0.11$ & $1.1 \times 10^2$ &
  $4.51$ & $-0.10$ & $1.1 \times 10^2$ \\
  \hline
  $z$ & $+$ & $1.71$ & $-0.14$ & $82$ &
  $2.50$ & $-0.09$ & $1.3 \times 10^2$ \\
  & $-$ & $2.94$ & $-0.07$ & $1.5 \times 10^2$ &
  $3.27$ & $-0.05$ & $1.9 \times 10^2$ \\
  \hline
\hline
\end{tabular}
  \end{center}
  \caption{One dimensional fits for the
    variation of $\log | {\cal F}^\mu_{\rm conf}|$ (total and continuous
    integral contribution) with $\tau$
    for $\zeta=0$. The range of
    the fit is  $0.0013 < |\delta \tau| < 0.0021$ and
    the form is $a + b \log | \tau |$.
    $Q$ is the ratio of the total variation in $\log {\cal F}^\mu_{\rm conf}$ divided by the
    RMS error between the fit and the actual data.
  }\label{tab:VVcusptauvaries}
  \end{table}
\begin{table}[H]
  \begin{center}
    \begin{tabular}{|cc|ccc|ccc|}
  \hline
  $\mu$ & ${\rm sgn} \ \delta \zeta$ & a & c & Q & a & c & Q \\
  & & \multicolumn{3}{c|}{total} & \multicolumn{3}{c|}{integral} \\
  \hline
  $t$ & $+$ & $2.78$ & $-1.00$ & $4 \times 10^3$ &
  $2.56$ & $-1.00$ & $3.3 \times 10^3$\\
  & $-$ & $2.70$ & $-1.00$ & $4 \times 10^3$ &
  $2.89$ & $-1.00$ & $4.8 \times 10^3$\\
  \hline
  $x$ & $+$ & $2.79$ & $-1.00$ & $9 \times 10^3$ &
  $2.58$ & $-1.00$ & $7 \times 10^3$\\
  & $-$ & $2.71$ & $-1.00$ & $8 \times 10^3$ &
  $2.90$ & $-1.00$ & $10^4$ \\
  \hline
  $y$ & $+$ & $4.00$ & $-0.14$ & $84$ &
  $4.02$ & $-0.13$ & $85$\\
  & $-$ & $3.77$ & $-0.15$ & $76$ &
  $3.80$ & $-0.15$ & $77$\\
  \hline
  $z$ & $+$ & $2.36$ & $-0.10$ & $120$ &
  $2.48$ & $-0.09$ & $125$\\
  & $-$ & $2.35$ & $-0.10$ & $110$ &
  $2.47$ & $-0.09$ & $120$\\
  \hline
\hline
\end{tabular}
  \end{center}
  \caption{One dimensional fits for the variation of $\log | {\cal F}^\mu_{\rm conf}|$
     (total and continuous
    integral contribution) with $\zeta$ at
                $\tau=0$. The range of the fit is
                $0.0013 < |\zeta| < 0.0021$ and the form is $a + c \log | \zeta |$. $Q$ is the ratio of the
                total variation in $\log {\cal F}^\mu_{\rm conf}$ divided by the RMS error between the fit and the
                actual data.
  }\label{tab:VVcuspzetavaries}
  \end{table}

\bibliographystyle{apsrev4-1}
\bibliography{references}

\end{document}